\documentclass[review]{fcs}
\usepackage{microtype}
\usepackage{graphicx}
\usepackage{subfigure}
\usepackage{subcaption}
\usepackage{forest}
\usetikzlibrary{arrows.meta,shapes,positioning,trees}
\title{A Survey on Learning from Graphs with Heterophily: Recent Advances and Future Directions}
\author[1]{Chenghua GONG}
\author[1]{Yao CHENG}
\author[1]{Jianxiang YU}
\author[1]{Can XU}
\author[2]{Caihua SHAN}
\author[3]{Siqiang LUO}
\author*[1]{Xiang LI}
\address[1]{School of Data Science and Engineering, East China Normal University, Shanghai, China}
\address[2]{Microsoft Research Asia, Shanghai, China}
\address[3]{School of Computer Science and Engineering, Nanyang Technological University, Singapore}
\fcssetup{
  received       = {month dd, yyyy},
  accepted       = {month dd, yyyy},
  corr-email     = {xxx@xxx.xxx},
} 

\tikzset{
    basic/.style = {draw, font=\rmfamily\fontsize{5pt}{6pt}\selectfont, rectangle},
    root/.style = {basic, thin, rounded corners=2pt,  align=center, text width=1.25cm},
    a/.style = {basic, thin, rounded corners=2pt, 
    align=center, text width=2.4cm},
    b/.style = {basic, thin, rounded corners=2pt, 
    align=center, text width=2.3cm},
    c/.style = {basic, thin, rounded corners=2pt, 
    align=center, text width=2.2cm},
    d/.style = {basic, thin, rounded corners=2pt,
    align=left, text width=7.35cm},
    e/.style = {basic, thin, rounded corners=2pt,
    align=left, text width=4.3cm},
    edge from parent/.style={draw=black, edge from parent fork right}
    }

\begin{abstract}
    Graphs are structured data that models complex relations between real-world entities. 
    Heterophilic graphs, where linked nodes are prone to be with different labels or dissimilar features, have recently attracted significant attention and found many real-world applications.
    Meanwhile, increasing efforts have been made to advance learning from graphs with heterophily.
    Various 
    graph heterophily measures,
    benchmark datasets, 
    and learning paradigms 
    are emerging rapidly.
    In this survey, 
    we comprehensively review existing works on learning from graphs with heterophily. 
    First, we overview over 500 publications, 
    of which more than 340 are directly related to heterophilic graphs.
    After that, 
    we survey existing metrics of graph heterophily and 
    list recent benchmark datasets.
    Further, we systematically categorize existing methods based on a hierarchical taxonomy 
    including GNN models, learning paradigms and practical applications.
    In addition,
    broader topics related to graph heterophily are also included.
    Finally, we discuss the primary challenges of existing studies and highlight promising avenues for future research. 
\end{abstract}


\begin{document}

    \section{Introduction}
\begin{figure}[!ht]
    \centering
    \subfigure[A citation network.]{
    \includegraphics[width=0.445\linewidth]{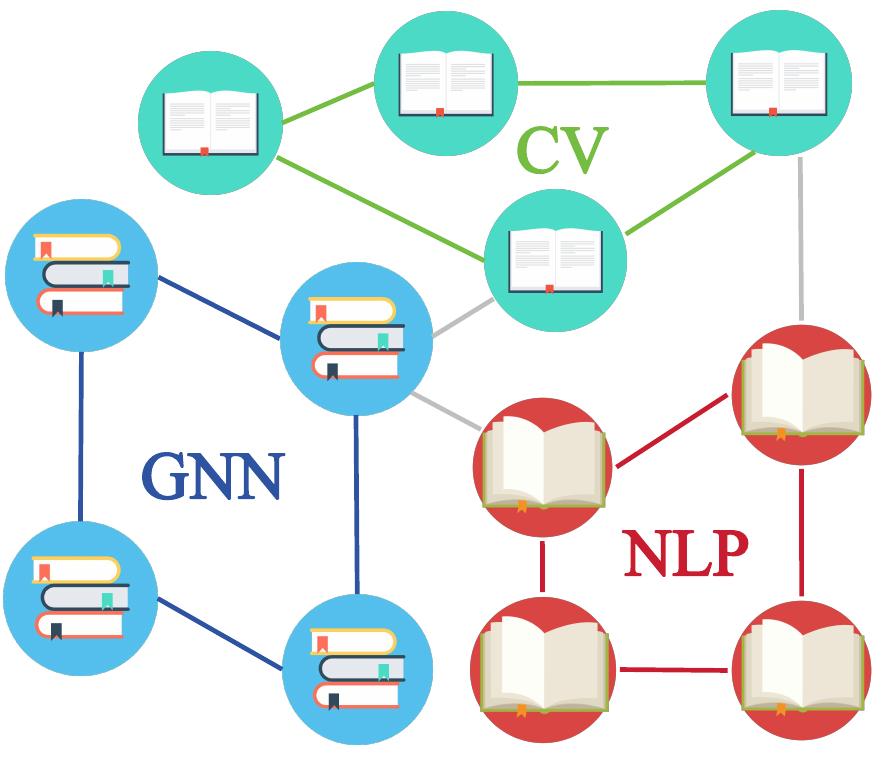}
    \label{homograph}
    }
    \subfigure[A social network with bots.]{
    \includegraphics[width=0.48\linewidth]{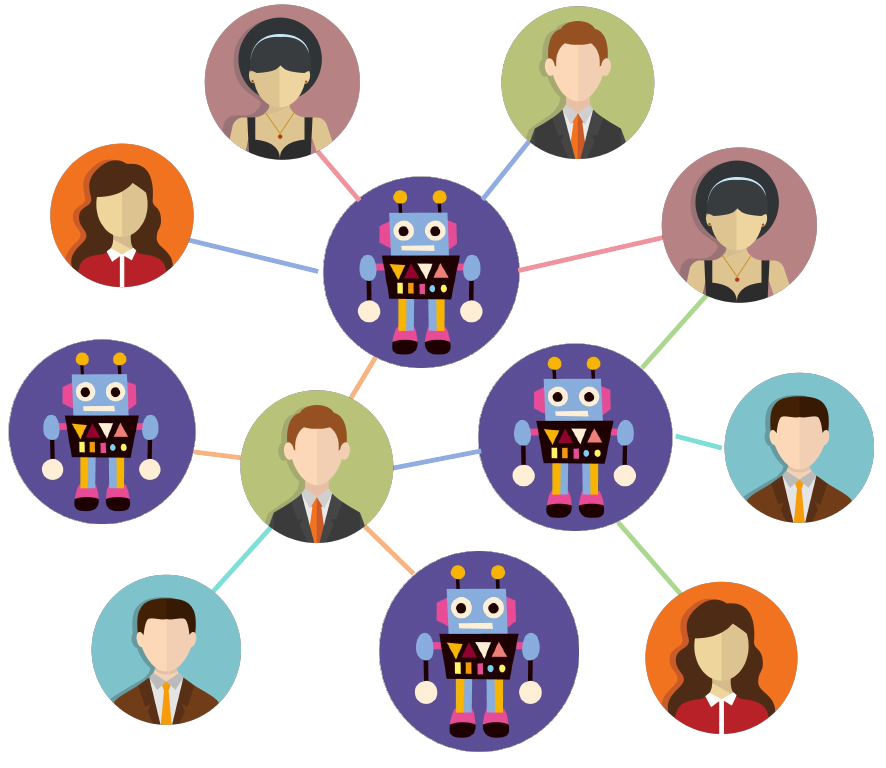}
    \label{heterograph}
    }
    \subfigure[A brain network with community structure.]{
    \includegraphics[width=0.96\linewidth]{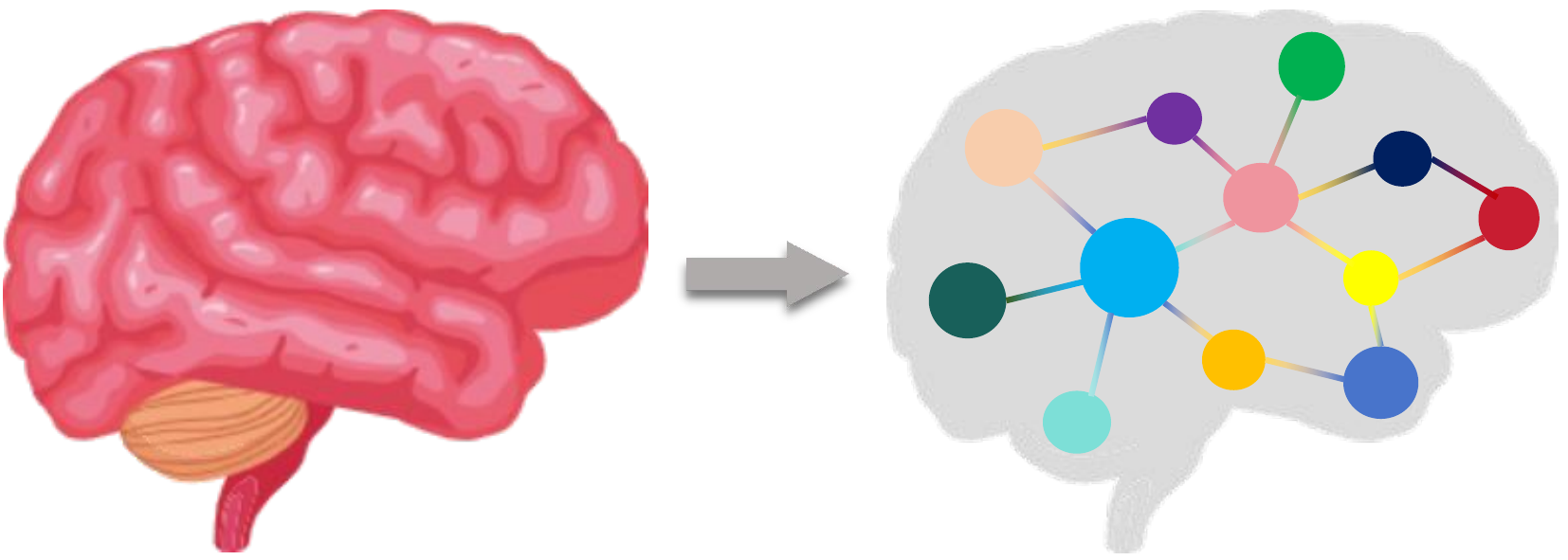}
    \label{brain}
    }
    \caption{The toy examples of homophilic and heterophilic graphs in real-world applications.}
    \end{figure}

    Graph-structured data is ubiquitous in the real world, which models entities as nodes and the complex relationships between entities as edges.
    Some graphs exhibit homophily, where linked nodes tend to have the same label or similar features, such as citation networks, friendship networks and political networks. 
    As shown in Figure~\ref{homograph}, the citation relations between papers show typical homophily, because papers are more likely to cite other papers within the same research field.
    In other cases, there also exist many graphs with heterophily, where the nodes with different labels or dissimilar features are more likely to be connected.

    \noindent \textbf{Heterophily in real-world applications.}
    Graphs with heterophily have found various practical applications, which verifies the significance of the research topic. 
    For example, social bots have been widely employed to spread rumors, sow panic, and even manipulate elections, causing serious negative social impacts~\cite{shao2018spread}.
    In Figure~\ref{heterograph}, we present a social network with automated bots, where social bots tend to establish connections with users instead of other bots. 
    Due to the significant difference in characteristics and behaviors between social bots and normal users, this network exhibits the typical graph heterophily. 
    Moreover, we can view the human brain as a complex network in Figure~\ref{brain}, with different regions regarded as nodes and the connections between regions seen as edges. 
    Considering that each region supports different physiological and psychological functions, they exhibit distinct structures and features~\cite{lynn2019physics}. 
    Therefore, the brain network is far from being homophilic but rather heterophilic.
    Moving to the urban computing~\cite{zheng2014urban}, the city is usually modeled as an urban graph where nodes are urban objects such as functional regions and edges are physical or social dependencies such as human mobility, traffic flow, and geographical data. 
    Taking the urban graph constructed with human mobility as an example, graph heterophily usually exists in an urban graph as end nodes of an edge could be of different functionalities, such as residential area and workplace, respectively.
    In summary, the heterophily inherent in graphs is widespread across various application scenarios and is closely related to our human bodies, daily lives, and even the cities and societies we inhabit.

    \noindent \textbf{Problems when facing graph heterophily.}
    Recently, many Graph Neural Networks (GNNs)~\cite{kipf2016semi,hamilton2017inductive,velivckovic2017graph} have be proposed and achieved great success when dealing with the graph-structured data. 
    Traditional GNNs implicitly assume that graphs are homophlic and follow the message passing mechanism, where each node aggregates messages from neighbors to update its representation. However, this paradigm performs poorly when facing graph heterophily, 
    and the degradation can be attributed to nodes incorrectly aggregating the information from neighbors.
    More precisely, message passing mechanism fails to discriminate local uninformative nodes and explore global informative nodes under the heterophily setting.  
    For the former, simply aggregating neighborhood information without discrimination can easily introduce noise, 
    resulting in ineffective node representations. 
    For the latter, message passing mechanism is naturally constrained by the local topology and fails to reach distant but informative nodes. 
    To deeply dive into graph heterophily, 
    various model architectures, learning paradigms, learning topics and practical applications have emerged recently.

    \noindent \textbf{Difference from existing surveys.}
    Due to the increasing popularity, we have recently witnessed some relevant surveys on this topic~\cite{zheng2022graph,zhu2023heterophily,luan2024graph}. 
    The early surveys~\cite{zheng2022graph,zhu2023heterophily} are limited in scale, and solely focus on the GNN models while neglecting other learning topics such as learning paradigms and practical applications.
    The recent handbook~\cite{luan2024graph} comprehensively collects existing works related to heterophilic graphs, 
    but it simply lists for easy reference but lacks a systematic categorization. 
    Therefore, there is an urgent need for a more comprehensive and systematic review in this field to summarize recent progresses and envision future directions.
    
    In this survey,
    we start with the introduction of metrics of graph heterophily and benchmark datasets, elaborating on the sources and characteristics of each dataset in detail. 
    Then, we 
    categorize existing heterophilic GNN models and beyond.
    After that,
    we introduce existing studies on 
    self-supervised learning and prompt learning, 
    and also provide 
    more related topics 
    to broaden the scope of research.
    Finally, we summarize practical applications related to graph heterophily, and give an outlook on the future development of this field.    
    Our main contributions can be summarized as follows:
    
    \noindent \textbf{Comprehensive Review.} 
    To our knowledge, this survey is currently 
    the most comprehensive one in the area of heterophilic graph learning.
    
    \noindent \textbf{Systematic Taxonomy.} 
    This survey introduces a systematic taxonomy that categorizes existing works from diverse learning aspects. 
    
    \noindent \textbf{Prospective Future.} 
    We also reveal the challenges faced by existing works and present insightful future directions.

    \section{Preliminaries}
    \label{sec2}
    
    \subsection{Notations}
    Let $\mathcal{G}=(\mathcal{V}, \mathcal{E})$ denote a graph with a set of nodes $\mathcal{V}$ and a set of edges $\mathcal{E}$, where $N=|\mathcal{V}|$ is the number of nodes. 
    The adjacency matrix of $\mathcal{G}$ is denoted as $\mathbf{A} = [a_{ij}] \in \{0,1\}^{N \times N}$, where $a_{ij}=1$ if there exists an edge $e_{ij} = (v_i, v_j)$. 
    The degree matrix $\mathbf{D}$ is a diagonal matrix with each diagonal element $d_i=\sum_{i=1}^N a_{ij}$ being the degree of node $v_i$. 
    The neighbor set of node $v$ is denoted as $\mathcal{N}(v) = \{ v_j : (v_i, v_j)\in \mathcal{E}\}$. 
    The node feature matrix is denoted as $\mathbf{X}$, where the $i$-th row $x_i$ is the feature vector of node $v_i$. 
    The node representation matrix is denoted by $\mathbf{H}$, where the $h_i$ is the representation of node $v_i$.
    The label matrix is denoted by $\mathbf{Y} \in \mathbb{R}^{N \times C}$, where $C$ is the number of classes and the $i$-th row $y_i$ is the one-hot encoded label vector of $v_i$. 
    For nodes $v_i, v_j \in \mathcal{V}$, if $y_i=y_j$, they are viewed as intra-class nodes; 
    otherwise,
    they are inter-class nodes. 
    Equally, an edge $e_{ij} \in \mathcal{E}$ is taken as an intra-class edge if $y_i = y_j$, 
    or an inter-class edge if $y_i \neq y_j$.

    \subsection{Message Passing Framework}
    The message passing framework~\cite{gilmer2017neural} stands for a broad category of GNN architectures, 
    where each node aggregates information from neighboring nodes and then combining the aggregated information. 
    The process can be formulated as:
    \begin{equation}
    \label{mpp}
    h_{i}^{(l)} = \texttt{COM} \left( h_{i}^{(l-1)}, \texttt{AGG}\{h_j^{(l-1)}: v_j \in \mathcal{N}(v) \} \right), 
    \end{equation}
    where $0 \leq l \leq L$ and $L$ is the number of GNN layers, $h_i^{(0)}=x_i$ and $h_i^{(l)} (1 \leq l \leq L)$ denotes the node representation of $v_i$ at the $l$-th layer.
    The choice of aggregation function \texttt{AGG}$(\cdot)$ is flexible (e.g, mean, sum, max pooling), and the combination function \texttt{COM}$(\cdot)$ in each layer can also be customized. 
    GNNs based on the message passing mechanism have achieved significant advancements, 
    but they still encounter challenges 
    such as over-smoothing~\cite{li2018deeper}, over-squashing~\cite{alon2020bottleneck}, limited propagation~\cite{corso2020principal}, and the graph heterophily issues ~\cite{zhu2020beyond}. 

    \subsection{Graph Transformer}
    Recently, Transformer~\cite{vaswani2017attention} has rapidly advanced and revolutionized the field of NLP~\cite{devlin2018bert,brown2020language,patwardhan2023transformers} and CV~\cite{dosovitskiy2020image,liu2021swin,han2022survey,khan2022transformers}. 
    Inspired by that, 
    Graph Transformers (GTs) has emerged as a prominent approach in graph learning~\cite{min2022transformer,shehzad2024graph}.
    Here, we briefly introduce the core components of GTs: Self-Attention Mechanism, Positional and Structural Encodings.

    \noindent \textbf{Self-Attention Mechanism.} 
    Self-attention is the fundamental mechanism of GTs.
    Given the node feature matrix $\mathbf{X}$, a GT layer first projects $\mathbf{X}$ into the Query, Key, and Value matrices:
    \begin{equation}
        \mathbf{Q} = \mathbf{X}\mathbf{W}_Q, \mathbf{K} = \mathbf{X}\mathbf{W}_K, \mathbf{V} = \mathbf{X}\mathbf{W}_V,
    \end{equation}
    where $\mathbf{W}_Q, \mathbf{W}_K, \mathbf{W}_V$ are three trainable weight matrices.
    After that, the node representation $\mathbf{H}$ with a single self-attention head is computed as follows:
    \begin{equation}
    \label{gts}
        \mathbf{S} = \frac{\mathbf{Q} \mathbf{K}^\mathsf{T}}{\sqrt{d}},
        \mathbf{H} = \texttt{softmax}(\mathbf{S})\mathbf{V},
    \end{equation}
    where $d$ denotes the dimensionality of the attention head, and $\mathbf{S} \in \mathbb{R}^{N \times N}$ denotes the self-attention matrix. 
    Subsequently, the extension to multi-head attention can be directly implemented by concatenating multiple heads. 
    Moreover, the MLP module, residual connections, and normalization operations are followed to produce the final output. 
    The main difference between GTs and the message passing framework is that GTs completely ignores the graph topology and treats the input as a fully-connected graph. 
    Learning global attention of GTs on graphs can to some extent alleviate the heterophily issue.

    \noindent \textbf{Positional and Structural Encodings.} 
    To incorporate the critical graph structural information into GTs, various Positional Encodings (PE) and Structural Encodings (SE) have been proposed. 
    SEs focus exclusively on topological information on local, relative, or global levels, such as node degree, triangle and cycle counting, and the context of subgraph. 
    More advanced methods include DSE~\cite{ying2021transformers}, RWSE~\cite{dwivedi2021graph} and TCSE~\cite{bouritsas2022improving}.
    Different from SEs, 
    PEs focus on perceiving the relative node positions towards other nodes and absolute node positions within the entire graph. 
    For example, LapPE~\cite{kreuzer2021rethinking}, RWPE~\cite{ma2023graph}, JaccardPE~\cite{zhang2020adaptive} are all PEs. Whether SEs and PEs in GTs can help solve the graph heterophily issue still remains unsolved.
    
    \subsection{Graph Laplacian and Filters}
    Graph Laplacian~\cite{chung1997spectral} is defined as $\mathbf{L}=\mathbf{D}-\mathbf{A}$, and its normalized version is $\widetilde{\mathbf{L}}=\mathbf{I} - \mathbf{D}^{-\frac{1}{2}}\mathbf{A}\mathbf{D}^{-\frac{1}{2}}$.
    Its eigen-decomposition gives $\mathbf{L} = \mathbf{U} \mathbf{\Lambda} \mathbf{U}^\mathsf{T}$, where $\mathbf{U}$ is the eigenvector matrix, also called graph fourier basis. 
    Given a graph signal $\mathbf{x}$, the graph fourier transform of $\mathbf{x}$ is defned as $\hat{\mathbf{x}} = \mathbf{U}^\mathsf{T} \mathbf{x}$, and the inverse transform is $\mathbf{x} = \mathbf{U}^\mathsf{T} \hat{\mathbf{x}}$.
    The eigenvalue matrix $\mathbf{\Lambda} = Diag(\lambda_1, \lambda_2,..., \lambda_{N})$ with  $0 \leq \lambda_1 \leq \lambda_2,..., \lambda_N \leq 2$ and $\lambda_i$ is also called as frequency~\cite{luan2020complete}. 
    Here, 
    smaller $\lambda_i$ corresponds to low-frequency signals (smooth information), while larger $\lambda_i$ corresponds to high-frequency signals (non-smooth information)~\cite{dakovic2019local}. 
    According to the spectral graph theory, the graph convolution operation can be expressed as:
    \begin{equation}
        g*\mathbf{x} = \mathbf{U} g(\mathbf{\Lambda}) \mathbf{U}^\mathsf{T} = \sum_{i=0}^{N}g(\lambda_i)\mathbf{u}_i \mathbf{u}_i^\mathsf{T} \mathbf{x}
    \end{equation}
    where $g(\mathbf{\Lambda}) = Diag(g(\lambda_1),...,g(\lambda_N))$ denotes the diagonal matrix when applied to the graph spectrum and $g:[0,2]\rightarrow \mathbb{R}$ is the spectral graph filter to re-weight different frequencies. 
    
    In practice, performing eigen-decomposition directly is infeasible on large-scale graphs in terms of time complexity. 
    Hence, it is common to utilize the polynomial approximation with regard to $\widetilde{\mathbf{L}}$ as an approximate filter $g(\widetilde{\mathbf{L}})$~\cite{liao2024benchmarking}:
    \begin{equation}
        g*\mathbf{x} = \mathbf{U} g(\mathbf{\Lambda}) \mathbf{U}^\mathsf{T} \approx 
        g(\widetilde{\mathbf{L}}) \cdot \mathbf{x}.
    \end{equation}
    Such estimation is typically considered for arbitrary smooth signals, while non-smooth signals are more important under graph heterophily. 
    Therefore, various high-pass and low-pass filters are utilized to capture the complex patterns on graphs.
    The common low-pass filter $\mathbf{F}_{LP}$ is built based on the affinity matrix $\mathbf{F}_{LP} = \widetilde{\mathbf{A}} = \mathbf{D}^{-\frac{1}{2}}\mathbf{A}\mathbf{D}^{-\frac{1}{2}}$
    , while the corresponding high-pass filters $\mathbf{F}_{HP}$ is built from the normalized Laplacian matrix $\mathbf{F}_{HP} = \widetilde{\mathbf{L}} = \mathbf{I} - \mathbf{D}^{-\frac{1}{2}}\mathbf{A}\mathbf{D}^{-\frac{1}{2}}$~\cite{luan2020complete}.
    
    \subsection{Learning Paradigms}
    Currently, the mainstream learning paradigms in graph learning include three main categories: supervised learning, self-supervised learning, and the recently proposed prompt learning.
    
    \noindent \textbf{Supervised Learning.} 
    Based on supervision signals,
    there have been proposed some effective GNNs specially designed for graph heterophily. 
    Given a GNN model, when applied to the task of node classification, it is essentially semi-supervised because information aggregation in GNN leverages unlabeled nodes in the test set; when the GNN is used for graph classification, it becomes supervised due to the ignorance of test set. Therefore, for brevity, we merge semi-supervised learning into supervised learning on heterophilic graphs. 


    \noindent \textbf{Self-supervised Learning.} 
    Despite the remarkable success of supervised learning, the heavy reliance on labels brings high annotation cost, weak model robustness, and the over-fitting problem.
    To this end, Self-Supervised Learning (SSL)~\cite{liu2021self,liu2022graph} on graphs designs a series of pretext tasks, leverages the input itself as the supervision to learn informative representations from unlabeled data.
    
    \noindent \textbf{Prompt Learning.} 
    Built on the recent development of SSL and pre-training techniques~\cite{hu2019strategies}, prompt learning is an emerging paradigm primarily applied to the fine-tuning and optimization of NLP pre-trained models~\cite{liu2023pre}. 
    The core idea of prompt learning is to bridge the gap between pretexts and downstream tasks through a unified task template, thereby fully leveraging the pre-trained model. 
    By avoiding the costly fine-tuning, prompt learning can achieve great results with a small number of parameters in the few-shot setup.
    Currently, this paradigm has been explored to a certain extent in graphs~\cite{sun2023graph,long2024towards}.
    
    \begin{figure*}[ht]
    \centering
    \subfigure[Number of papers published per year.]{
    \includegraphics[width=0.36\linewidth]{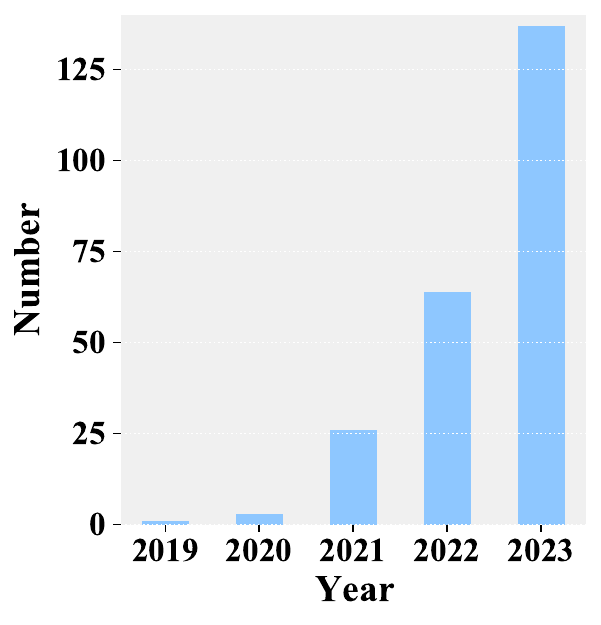}
    \label{num}
    }
    \subfigure[The distribution of sources for the collected papers.]{
    \includegraphics[width=0.50\linewidth]{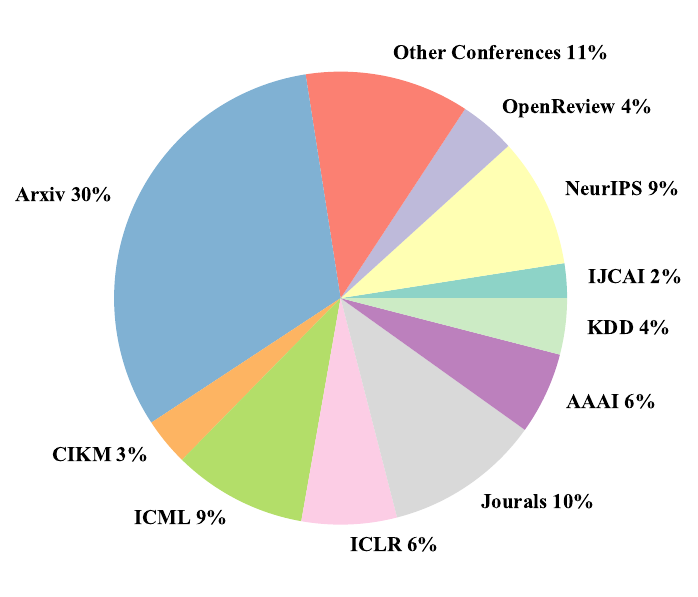}
    \label{piechart}
    }
    \caption{The statistics of collected papers related to learning from graphs with heterophily.}
    \label{Statistics}
    \end{figure*}
    
    \section{Overview}
    \label{sec3}
    This section first provides a brief literature overview, followed by the introduction of organizational structure of our survey. 
    
    \subsection{Literature Overview}
    In this survey, we collect over 500 papers, of which more than 340 are directly related to heterophilic graphs, and pay much  attention to those published in top conferences or journals. 
    We also include latest studies released on OpenReview and Arxiv.
    All the resources related to this survey are given at our GitHub repository\footnote{https://github.com/gongchenghua/Papers-Graphs-with-Heterophily}.
    In Figure~\ref{Statistics}, 
    we present the statistics of collected papers.
    First, we compile the annual publication statistics of papers over the past five years.
    As can be seen in Figure~\ref{num}, 
    the number of papers released per year shows a significant growth trend in the past five years, indicating the enormous potential of this direction.
    Moreover, the distribution of sources for the collected papers is given in Figure~\ref{piechart}.  
    It is worth noting that more than half of the collected papers have been published in top conferences or journals, indicating the reliability of our survey.
    \begin{figure}[h]
    \includegraphics[width=0.93\linewidth]{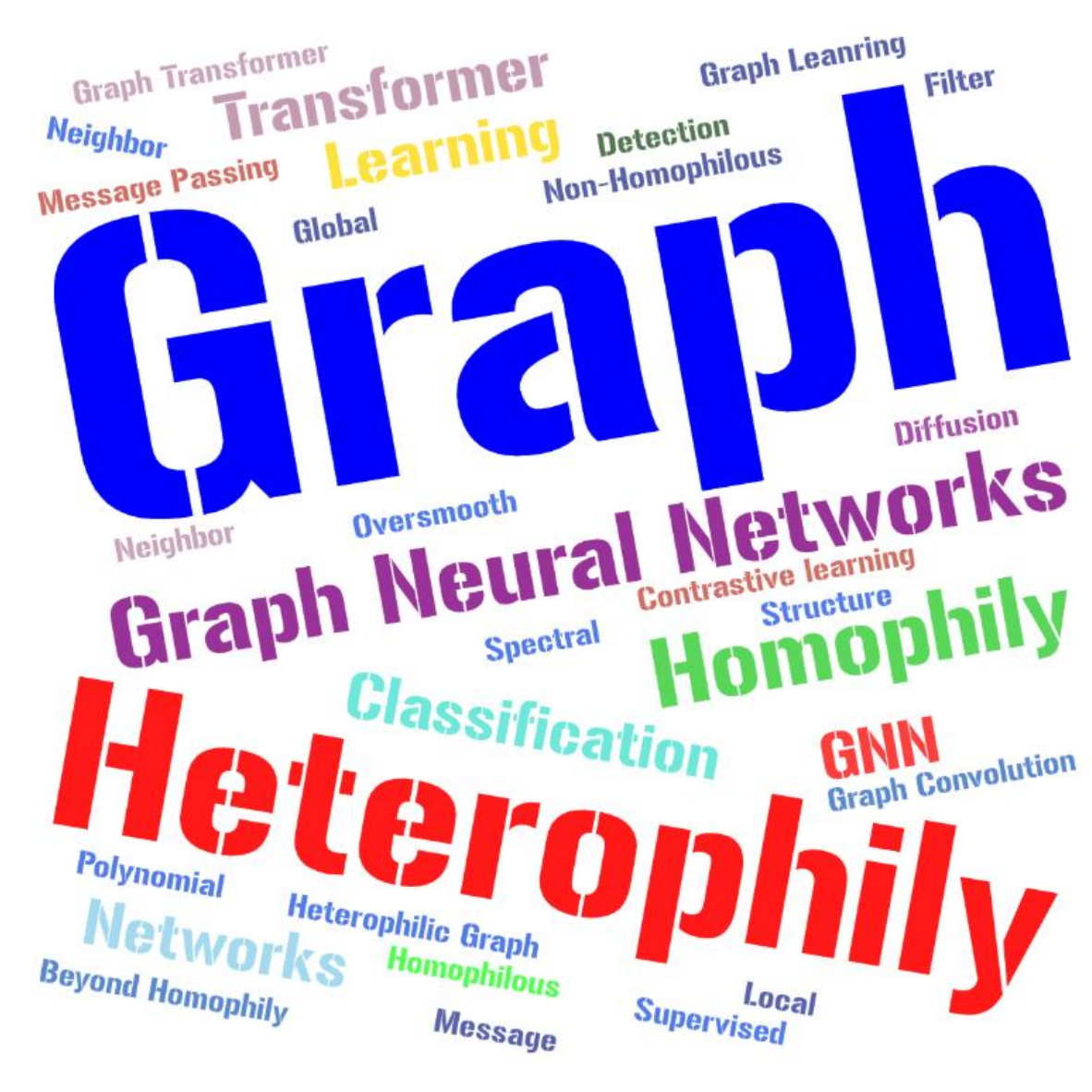}
    \caption{The word cloud composed of high-frequency words from the titles of collected articles in this survey.}
    \label{keywords}
    \end{figure}
    Further, we analyze the general content of collected papers, and present the most frequently occurring words of the titles using the word cloud in Figure~\ref{keywords}. 
    Notably, 
    these keywords are closely related to the topic of this survey, which center around learning from graphs with heterophily.
    
    \begin{figure*}[ht]
    \centering
    \includegraphics[width=0.98\textwidth]{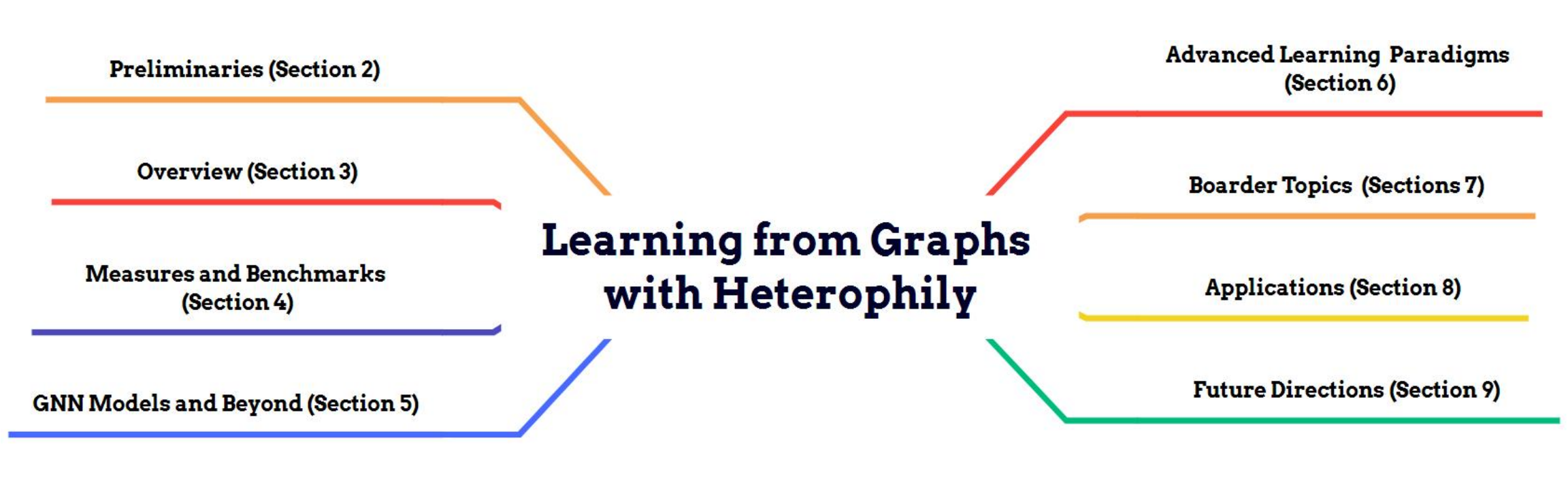}
    \caption{The overall organizational structure of this survey.}
    \label{overview}
    \end{figure*}
    
    \subsection{Organizational Structure}
    We present the organizational structure of this survey in Figure~\ref{overview}.
    In Section~\ref{sec2}, we provide the notions and preliminaries, and we present an overview of the survey in Section ~\ref{sec3}.
    In Section~\ref{sec4}, we introduce representative metrics of graph heterophily, and provide a detailed description of current benchmark datasets.
    In Section~\ref{sec5}, we summarize the representative GNN models and beyond related to graph heterophily, and provide a detailed grouping of these models.
    Apart from the supervised learning paradigm, we also summarize other popular learning paradigms on heterophilic graphs in Section~\ref{sec6}, including self-supervised learning and prompt learning.
    Apart from the remarkable GNN models and learning paradigms, some extensible topics
    are also mentioned in Section~\ref{sec7}.
    In Section~\ref{sec8}, we discuss the practical applications of heterophilic graphs, and further provide unique insights and prospects for future explorations in Section~\ref{sec9}.

    \section{Measures and Benchmarks}
    \label{sec4}
    The success of graph learning depends on the high-quality datasets. 
    To evaluate proposed models, various heterophilic datasets have been released~\cite{rozemberczki2021multi,pei2020geom,lim2021new,lim2021large,platonov2023critical,luan2024heterophily}. 
    Meanwhile, a variety of metrics for graph heterophily have been given to further characterize these datasets. 
    This section first discusses graph heterophily measures, and then provides a comprehensive review on benchmark datasets.
    
    \subsection{Measuring Graph Heterophily}
    Graph heterophily refers to the phenomenon that connected nodes tend to share different features or labels.
    Understanding this concept and establishing relevant metrics is crucial for further study. 
    Here, we introduce the most representative metrics for measuring heterophily, typically presented in terms of the level of homophily.
    For example, node homophily~\cite{pei2020geom} and edge homophily~\cite{zhu2020beyond} are most commonly used metrics.
    Specifically, the node homophily measures the homophily at the node level, where the homophily degree for each node is computed as the proportion of the neighbors sharing the same label. 
    Then the node homophily $\mathcal{H}_{node}$ is defined as the average homophily degree of all the nodes on graphs: 
    \begin{equation}
     \mathcal{H}_{node} = \frac{1}{|\mathcal{V}|} \sum\limits_{v \in \mathcal{V}}  \frac{|\{ u \in \mathcal{N}(v): y_v = y_u\}|}{|\mathcal{N}(v)|},
    \end{equation}
    where $y_v, y_u$ denote the labels of nodes $v$ and $u$, respectively. 
    It is worth noting that $\mathcal{H}_{node}$ only reflects the graph homophily within 1-hop neighbors. 
    Recent researches~\cite{xiao2024simple,cavallo20222} extend this definition to the $k$-hop neighbors to measure high-order homophily:
    \begin{equation}
     \mathcal{H}_{high-order} = \frac{|\{ u \in \mathcal{N}_k(v): y_v = y_u\}|}{|\mathcal{N}_k(v)|},
    \end{equation}
    where $\mathcal{N}_k(v)$ denotes the set of $k$-hop neighborhood of node $v$.
    The edge homophily $\mathcal{H}_{edge}$ measures the graph homophily at the edge level, and is defined as the fraction of edges connecting nodes with the same label:
    \begin{equation}
     \mathcal{H}_{edge} = \frac{|\{ (v, u) \in \mathcal{E}: y_v = y_u\}|}{|\mathcal{E}|}.
    \end{equation} 
    Both node and edge homophily range from $[0,1]$, and the high homophily indicates low heterophily, and vice versa. 
    While widely used, these two simple metrics are highly sensitive to the number of classes, leading to limited utility~\cite{lim2021new}. 
    To mitigate the class imbalance issue by treating all classes equally, another metric referred to as class homophily~\cite{lim2021new} is defined as:
    \begin{equation}
    \mathcal{H}_{class} = \frac{1}{C-1} \sum\limits_{k=1}^{C}\left[\mathcal{H}_k-\frac{|C_k|}{|\mathcal{V}|}\right]_+, 
    \end{equation}
    where $\left[x\right]_+ = max\{x, 0\}$, $C$ is the number of classes, $C_k$ is the set of nodes in class $k$, and $\mathcal{H}_k$ is the class-wise homophily metric:
    \begin{equation}
    \mathcal{H}_k = \frac{\sum_{v: y_v=c} |\{ u \in \mathcal{N}(v): y_v = y_u\}|}{\sum_{v: y_v=c}|\mathcal{N}(v)|}. 
    \end{equation}
    Although $\mathcal{H}_{class}$ can measure heterophily more fairly, there are still existing some issues~\cite{platonov2024characterizing}. 
    For example, the class homophily neglects the variation of node degrees when correcting the fraction of intra-class edges by its expected value.
    Combined with a traditional graph measure, referred to as assortativity coefficinent~\cite{newman2003mixing}, a more advanced measure named adjusted homophily~\cite{platonov2024characterizing} is defined as:
    \begin{equation}
     \mathcal{H}_{adj} = \frac{\mathcal{H}_{edge}-\sum_{k=1}^C \mathcal{D}_k^2/(2|\mathcal{E}|)^2}{1-\sum_{k=1}^C \mathcal{D}_k^2/(2|\mathcal{E}|)^2},
    \end{equation}
    where $\mathcal{D}_k$ is the sum of degrees of all the nodes with label $k$. 
    The adjusted homophily $\mathcal{H}_{adj}$ is shown to be comparable across different datasets with varying numbers of classes~\cite{platonov2023critical}.
    In addition to the representative metrics detailed above, various statistic-based metrics for graph heterophily are continuously emerging and are becoming dominant~\cite{li2023restructuring,gong2023neighborhood,luan2022revisiting,yang2021diverse,jin2022raw,lee2024feature}.
    Recently, some studies use classifier-based\cite{luan2024graph} and unsupervised-based~\cite{ojha2024affinity} approaches to measure graph heterophily.
    Zheng et al.~\cite{zheng2024missing} further attempt to disentangle the heterophily from label, structural, and feature aspects, and provide a comprehensive review of existing heterophily metrics.
    While various proposed metrics can help people characterize the graph heterophily, many recent studies have empirically demonstrated that the performance of GNNs does not always align with the heterophily~\cite{luan2022revisiting,luan2024graph,ma2021homophily}, and how heterophily influences GNNs still remains unclear and controversial.
    
    \begin{table*}[h]
    \centering
    \caption{The statistics of current benchmark datasets for heterophilic graphs.}\label{benchmark}
    \tiny
    \begin{tabular}{ccccccccc}
    \toprule
    Dataset &  \# Nodes & \# Edges & \# Features & \# Classes &  $\mathcal{H}_{node}$ &  $\mathcal{H}_{edge}$ & Sources & Literature\\
    \midrule
    Cornell & 183 & 295 & 1,703 & 5 & 0.20 & 0.30 & WebKB & \cite{pei2020geom}\\
    Texas & 183 & 309 & 1,703 & 5 & 0.06 & 0.06 & WebKB & \cite{pei2020geom}\\
    Wisconsin & 251 & 499 & 1,703 & 5 & 0.10 & 0.17 & WebKB & \cite{pei2020geom}\\
    Chameleon & 2,277 & 36,101 & 2,325 & 5 & 0.25 & 0.23 & Wikipedia & \cite{pei2020geom}\\
    Squirrel & 5,201 & 217,073 & 2,089 & 5 & 0.22 &  0.22 & Wikipedia & \cite{pei2020geom}\\
    Actor & 7,600 & 33,544 & 931 & 5 & 0.20 & 0.22 & Actor & \cite{pei2020geom}\\
    \midrule
    Penn94 & 41,554 & 1,362,229 & 5 & 2 & 0.48 & 0.47 & Social Media & \cite{lim2021large}\\
    Pokec & 1,632,803 & 30,622,564 & 65 & 2 & 0.39 & 0.44 & Social Media & \cite{lim2021large}\\
    Genius & 421,961 & 984,979 & 12 & 2 & 0.10 & 0.62 & Social Media & \cite{lim2021large}\\
    Twitch-Gamers & 168,114 & 6,797,557 & 7 & 2 & 0.56 & 0.55 & Social Media & \cite{lim2021large}\\
    Deezer-Europe & 28,281 & 92,752 & 31,241 & 2 & 0.53 & 0.53 & Social Media & \cite{lim2021new}\\
    Blog-Catalog & 5,196 & 171,743 & 8,189 & 6 & 0.39 & 0.40 & Social Media & \cite{huang2017label}\\
    Flickr & 7,575 & 239,738 & 12,047 & 9 & 0.24 & 0.24 & Social Media & \cite{huang2017label}\\
    ArXiv-Year & 169,343 & 1,166,243 & 128 & 5 & 0.28 & 0.22 & Citation & \cite{lim2021large}\\
    Snap-Patents & 2,923,922 & 139,975,788 & 269 & 5 & 0.19 & 0.07 & Citation & \cite{lim2021large}\\
    Wiki &  1,925,342 & 303,434,860 & 600 & 5 & - & 0.39 & Wikipedia & \cite{lim2021large}\\
    Wiki-Cooc &  10,000 & 2,243,042 & 100 & 5 & 0.18 & 0.34 & Wikipedia & \cite{zhiyao2024opengsl}\\
    \midrule 
    Roman-Empire & 22,662 & 32,927 & 300 & 18 & 0.05 & 0.05 & Wikipedia & \cite{platonov2023critical}\\
    Amazon-Ratings & 24,492 & 93,050 & 300 & 5 & 0.38 & 0.38 & E-commerce & \cite{platonov2023critical}\\
    Minesweeper & 10,000 & 39,402 & 7 & 2 & 0.68 & 0.68 & Digital Game & \cite{platonov2023critical}\\
    Tolokers & 11,758 & 519,000 & 10 & 2 & 0.63 & 0.59 & Crowdsource & \cite{platonov2023critical}\\
    Questions & 48,921 & 153,540 & 301 & 2 & 0.90 & 0.84 & Q$\&$A Platform & \cite{platonov2023critical}\\
    \bottomrule
    \end{tabular}
    \end{table*}
    
    \subsection{Benchmark Datasets}
    To boost learning from graphs with heterophily, there is an urgent need for trustworthy and high-quality benchmarks. We further categorize existing benchmarks into three types: basic, large-scale and advanced benchmarks. 
    Note that all 23 datasets focus on the node classification task, and detailed statistics are presented in Table~\ref{benchmark}.
    
    \noindent \textbf{Basic Benchmark.} 
    Inspired by studies in the field of complex networks~\cite{tang2009social,garcia2016using,rozemberczki2021multi}, Pei et al.~\cite{pei2020geom} summarize the first batch of benchmark datasets for heterophilic graphs. 
    To date, most studies have evaluated their models on these  datasets, which we refer to as basic benchmark. 
    Basic benchmark includes 6 datasets: $\mathtt{Cornell}$, $\mathtt{Texas}$, $\mathtt{Wisconsin}$, $\mathtt{Chameleon}$, $\mathtt{Squirrel}$ and $\mathtt{Actor}$. 
    Next, we categorize them based on their sources and provide detailed descriptions.
    
    \noindent \emph{-WebKB.} WebKB is a webpage dataset collected from computer science departments by Carnegie Mellon University. $\mathtt{Cornell}$, $\mathtt{Texas}$, $\mathtt{Wisconsin}$ are three sub-datasets of WebKB~\cite{garcia2016using}, where nodes represent web pages, and edges represent hyperlinks between web pages.

    \noindent \emph{-Wikipedia.} $\mathtt{Chameleon}$ and $\mathtt{Squirrel}$ are two page-page networks on specific topics collected from Wikipedia~\cite{rozemberczki2021multi}, which is a free, online encyclopedia that anyone can edit. In these datasets, nodes represent web pages and edges are mutual links between web pages. 

    \noindent \emph{-Actor.} $\mathtt{Actor}$, also called $\mathtt{Film}$, is the actor-only induced subgraph of a ``film-director-actor-writer'' network~\cite{tang2009social}. Each node represents an actor, and the edge between two nodes denotes their co-occurrence on the same Wikipedia page. 

    Despite the wide use, 
    the drawbacks of the basic benchmark are the small scale and limited domain of the datasets, which are insufficient to evaluate heterophily-specific models adequately.
    
    \noindent \textbf{Large-scale Benchmark.} 
    Aside from the overfitting risks caused by the small-scale data~\cite{dwivedi2023benchmarking}, the evaluation on the basic benchmark is plagued by high variance across different data splits~\cite{zhu2020beyond}. 
    To this end, a series of large-scale datasets from diverse domains are collected and released in~\cite{lim2021new,lim2021large}, forming the large-scale benchmark:
    $\mathtt{ArXiv-Year}$, $\mathtt{Snap-Patents}$, $\mathtt{Wiki}$, $\mathtt{Penn94}$, $\mathtt{Pokec}$, $\mathtt{Genius}$,  $\mathtt{Twitch-Gamers}$ and $\mathtt{Deezer-Europe}$.
    In addition to the datasets mentioned above, there are also some representative large-scale datasets, such as $\mathtt{Wiki-Cooc}$, $\mathtt{BlogCatalog}$ and $\mathtt{Flickr}$.
    Since these datasets show the significant heterophily, we also include them in the large-scale benchmark.
     
    \noindent \emph{-Social Media.} $\mathtt{Penn94}$~\cite{traud2012social} is a social network from the Facebook of university students, where nodes represent students and edges represent the friendships. 
    Similarly, $\mathtt{Pokec}$~\cite{jure2014snap} is the friendship network of a Slovak online social network, where nodes represent users and edges represent directed friendship relations. 
    $\mathtt{Genius}$~\cite{lim2021expertise} is a subset of the social network on a web site for crowdsourced annotations of song lyrics. The nodes are users, and edges connect users that follow each other on the website. 
    $\mathtt{Twitch-Gamers}$~\cite{rozemberczki2021twitch} is a network of relationships between accounts on the livestreaming platform Twitch. Each node represents a Twitch account, and edges exist between accounts sharing mutual followers. $\mathtt{Deezer-Europe}$~\cite{rozemberczki2020characteristic} is a social network of users on Deezer from European countries, where edges represent mutual follower relationships.
    $\mathtt{BlogCatalog}$~\cite{huang2017label} is a also social network created from an online community where bloggers can follow each other.
    $\mathtt{Flickr}$~\cite{huang2017label} is a dataset created from an online website sharing images and videos where users can follow each other, forming a social network.
    
    \noindent \emph{-Citation.} $\mathtt{ArXiv-Year}$~\cite{hu2020open} is the Ogbn-ArXiv citation network labeled by the posted year, instead of subject areas. The nodes are ArXiv papers, and directed edges connect a paper to other papers that it cites. 
    $\mathtt{Snap-Patents}$~\cite{leskovec2005graphs,jure2014snap} is a dataset of utility patents granted between 1963 to 1999 in the US, where each node is a patent, and edges connect patents that cite each other.

    \noindent \emph{-Wikipedia.} $\mathtt{Wiki}$~\cite{lim2021large} is a dataset of Wikipedia articles, where nodes represent pages and edges represent links between them. 
    $\mathtt{Wiki-Cooc}$~\cite{zhiyao2024opengsl} is a dataset based on the English Wikipedia, where nodes denote unique words and edges connect frequently co-occurring words.

    Motivated by the large-scale benchmark, research papers on learning from graph heterophily start to flourish. 
    To better evaluate models and verify whether they capture heterophily patterns, 
    we urgently need more advanced, high-quality datasets.

    \noindent \textbf{Advanced Benchmark.} 
    Recently, a critical study has suggested existing benchmarks for heterophily-specific evaluation have serious drawbacks, rendering results based on them unreliable~\cite{platonov2023critical}.
    The fatal drawback is the presence of duplicate nodes in $\mathtt{Chameleon}$ and $\mathtt{Squirrel}$, causing train-test data leakage. 
    Experiments have shown that removing duplicate nodes strongly affects the performance of many heterophily-specific GNNs.  
    In addition, $\mathtt{Cornell}$, $\mathtt{Texas}$, $\mathtt{Wisconsin}$ are not innocent either. 
    These datasets have very imbalanced classes.
    For example, 
    $\mathtt{Texas}$ has a class that consists of only one node, which makes this class for training and evaluation meaningless.
    Aware of these issues, several new datasets with different nature and with diverse structural properties are released, collectively forming the advanced benchmark~\cite{platonov2023critical}. 
    The advanced benchmark includes 5 datasets for evaluation: $\mathtt{Roman-Empire}$, $\mathtt{Amazon-Ratings}$, $\mathtt{Minesweeper}$, $\mathtt{Tolokers}$ and $\mathtt{Questions}$.

    \noindent \emph{-Wikipedia.} 
    $\mathtt{Roman-Empire}$ is based on the one of the longest articles on Wikipedia, Roman Empire. 
    Each node corresponds to one (non-unique) word in the text and two nodes are connected with an edge if subjected to the proposed conditions.

    \noindent \emph{-E-commerce.} 
    $\mathtt{Amazon-Ratings}$ is based on the Amazon product co-purchasing network metadata dataset from SNAP~\cite{jure2014snap}. 
    Nodes are products (books, CDs, DVDs, video tapes), and edges connect products that are frequently bought together.
    
    \noindent \emph{-Digital Game.} $\mathtt{Minesweeper}$ is a synthetic network based on the Minesweeper game. It is a 100 $\times$ 100 grid where node is connected to eight neighbors, except for nodes at the edge of the grid. 20$\%$ of the nodes are randomly selected as mines and the task is to predict which nodes are mines.
    
    \noindent \emph{-Crowdsource.}
    $\mathtt{Tolokers}$ is based on data from the Toloka crowdsourcing platform~\cite{lhoest2021datasets}. 
    The nodes represent tolokers (workers) in at least one of 13 selected projects. An edge connects two tolokers if they are in the same task.

    \noindent \emph{-Q$\&$A Platform.} $\mathtt{Questions}$ is based on the data from a question-answering website Yandex Q. The nodes are users, and an edge connects two nodes if one user answered the other user’s question during a one-year time interval. 
   
    Evidently, we can find that the advanced benchmark not only has unique graph heterophily characteristics, 
    but also covers a broader scope, which is more closely related to our daily lives.
    For further advancement of this field, we advocate for the release of high-quality open-source datasets from a broader spectrum of fields. 
    We also recommend further cross-pollination  of heterophilic datasets with NLP, CV, and other fields, enabling models to truly address real-world application challenges.

    \subsection{Reassessment under Graph Heterophily}
    Due to aforementioned flaws in benchmarks, some studies has proposed to comprehensively reassess existing models.
    Platonov et al.~\cite{platonov2023critical} first uncover the fatal problems within popular datasets and reveal that the previous evaluation is unreliable. 
    They further propose the advanced benchmark for reassessment and show that standard GNNs generally outperform heterophily-specific methods.
    Through extensive empirical analysis, Luo et al.~\cite{luo2024classic} investigate the influence of various GNN configurations such as dropout~\cite{srivastava2014dropout,shu2022understanding}, normalization~\cite{cai2021graphnorm},  residual connections~\cite{he2016deep,li2019deepgcns}, network depth~\cite{li2018deeper,li2019deepgcns,li2020deepergcn}, and jumping knowledge~\cite{xu2018representation} on the node classification task. 
    Experiments show that classic GNNs (GCN, GAT, and GraphSAGE) with these configurations can achieve great performance, matching or even exceeding that of recent Graph Transformers across most heterophilic and homophilic datasets with slight hyper-parameter tuning.
    Beyond classic GNNs, Liao et al.~\cite{liao2024benchmarking} extensively benchmark spectral GNNs with a focus on the frequency perspective. 
    They propose that the majority of existing spectral models can be divided into three categories: fixed filter, variable filter and filter bank. 
    All three types can achieve satisfactory performance on homophilic graphs, while variable filters and filter bank models excel under graph heterophily. 
    To identify the real challenging subsets of heterophilic datasets, Luan et al.~\cite{luan2024heterophilic,luan2024heterophily} benchmark the commonly used heterophilic datasets and classify them into benign, malignant and ambiguous heterophily datasets. 
    Trough model reassessment, they point out that good models for heterophily should perform significantly better than baselines under graph heterophily, especially on malignant and ambiguous heterophilic datasets, and perform at least as good as baselines on homophilic graphs.
    We call on everyone to use comprehensive benchmarks and fair evaluation to objectively compare the performance of proposed models, as this can provide a foundation for subsequent model, and further promoting the development of learning from graphs with heterophily.
    
    \section{GNN Models and Beyond}
    \label{sec5}
\begin{figure*}[htbp]
    \centering
    \begin{forest}
    for tree={
    grow'=east,
    anchor=west,
    parent anchor=east,
    child anchor=west,
    edge path={
    \noexpand\path[\forestoption{edge},-, >={latex}] 
    (!u.parent anchor) -- +(5pt,0pt) |- (.child anchor)
    \forestoption{edge label};
    }
    }
    [GNN Models and Beyond, root
    [Spectral Graph Filters, a    
        [Fixed Filter, b
            [{GCNII~\cite{chen2020simple}
             }, d
            ]  
        ]
        [Variable Filter, b
            [{GPR-GNN~\cite{chien2020adaptive}, ASGC~\cite{chanpuriya2022simplified}, ChebNetII~\cite{he2022convolutional}, ClenshawGCN~\cite{guo2023clenshaw},
            BernNet~\cite{he2021bernnet},
            LegendreNet~\cite{chen2023improved}, JacobiConv~\cite{wang2022powerful}, FavardGNN~\cite{guo2023graph}
             }, d
            ]  
        ]
        [Filter Bank, b
            [{FB-GNN~\cite{luan2020complete},
              ACM-GNN~\cite{luan2022revisiting},
              GSCNet~\cite{du2024graph},
              G$^{2}$CN~\cite{li2022g},
              FiGURe~\cite{ekbote2024figure}
             }, d
            ]  
        ]
        [Further Exploration, b
            [{PyGNN~\cite{geng2023pyramid},
              PD-GNN~\cite{lu2024flexible},
              PC-Conv~\cite{li2024pc},
              NewtonNet~\cite{xu2023shape},
              AdaptKry~\cite{huang2024optimizing},
              UniFilter~\cite{huang2024universal},
              Node-MoE~\cite{han2024node}
             }, d
            ]  
        ]
    ]
    [Utilizing High-order Neighbors, a
        [Multi-hop View, b
            [{MixHop~\cite{abu2019mixhop},
              H$_2$GCN~\cite{zhu2020beyond},
              2NCS~\cite{cavallo20222},
              U-GCN~\cite{jin2021universal},
              FSGNN~\cite{maurya2022simplifying},
              HP-GNN~\cite{xu2022hp},
              PowerEmbed~\cite{huang2022local},
              MIGNN~\cite{choi2023node}
             }, d
            ]  
        ]
        [Tree-structure View, b
            [{TDGNN~\cite{wang2021tree},
              Ordered-GNN~\cite{song2023ordered}
             }, d
            ]  
        ]
        [Path-based View, b
            [{PathNet~\cite{sun2022beyond},
              RAW-GNN~\cite{jin2022raw},
              PathMLP~\cite{xie2023pathmlp}
             }, d
            ]  
        ]
    ]
    [Exploring Global Homophily, a
        [Pre-computed Extension, b
            [Feature Similarity, c
                [{SimP-GCN~\cite{jin2021node},
                  U-GCN~\cite{jin2021universal}
                }, e
                ]
            ]
            [Structure Similarity, c
                [{Geom-GCN~\cite{pei2020geom},
                  WRGNN~\cite{suresh2021breaking}
                }, e
                ]
            ]
            [Mixtures, c
                [{NSGCN~\cite{ai2024neighbors},
                  NDGCN~\cite{liu2025beyond},
                  MVGFN~\cite{wang2023improving}
                }, e
                ]
            ]
        ]
        [Affinity Learning, b
            [Latent Space Similarity, c
                [{NL-GNN~\cite{liu2021non},
                  GPNN~\cite{yang2022graph},
                  SNGNN~\cite{zou2023similarity},
                  KNN-GNN~\cite{li2024knn},
                  DC-GNN~\cite{dong2024differentiable},
                  Deformable-GCN~\cite{park2022deformable}
                }, e
                ]
            ]
            [Compatibility Matrix, c
                [{CPGNN~\cite{zhu2021graph},
                  CLP~\cite{zhong2022simplifying},
                  CMGNN~\cite{zheng2024revisiting}
                }, e
                ]
            ]
            [Decoupled Scheme, c
                [{GloGNN~\cite{li2022finding},
                  HOG-GCN~\cite{wang2022powerful},
                  BM-GCN~\cite{he2022block},
                  LG-GNN~\cite{yu2024lggnn},
                  SIMGA~\cite{liu2023simga},
                  INGNN~\cite{liu2022uplifting}
                }, e
                ]
            ]
        ]   
    ]
    [Discriminative Message Passing, a
        [Signed Message Passing, b
            [{FAGCN~\cite{bo2021beyond},
              SAGNN~\cite{wu2023signed},
              SADE-GCN~\cite{lai2023self},
              GGCN~\cite{yan2022two},
              Choi et al.~\cite{choi2023signed},
              M2M-GNN~\cite{liang2024sign}
             }, d
            ]  
        ]
        [Directed Message Passing, b
            [{AMUD~\cite{sun2024breaking},
              GNNDLD~\cite{chaudhary2024gnndld},
              Dir-GNN~\cite{rossi2024edge},
              Holonets~\cite{koke2023holonets},
              CGCN~\cite{zhuo2024commute}
             }, d
            ]  
        ]
        [Gating Mechanism, b
            [Neighborhood Gating, c
                [{GBK-GNN~\cite{du2022gbk},
                  NH-GCN~\cite{gong2023neighborhood},
                  HES~\cite{wang2024heterophilic}
                }, e
                ]
            ]
            [Layer Gating, c
                [{PriPro~\cite{cheng2023prioritized},
                  GNN-AS~\cite{deng2024learning}
                }, e
                ]
            ]
            [Attribute Gating, c
                [{DMP~\cite{yang2021diverse},
                  HA-GAT~\cite{wang2024heterophily}
                }, e
                ]
            ]
            [Advanced Gating, c
                [{G$^2$~\cite{rusch2022gradient},
                  Co-GNN~\cite{finkelshtein2023cooperative}
                }, e
                ]
            ]
        ]
    ]
    [Graph Transformers, a
        [Polynomial View, b
            [{PolyFormer~\cite{ma2024polyformer},
              PolyNormer~\cite{deng2024polynormer}
             }, d
            ]  
        ]
        [Signed Attention, b
            [{SignGT~\cite{chen2023signgt},
              SIGformer~\cite{chen2024sigformer}
             }, d
            ]  
        ]
        [Mitigate Over-globalization, b
            [{Coarformer~\cite{kuang2024transformer},
              Gapformer~\cite{liu2023gapformer},
              LGMformer~\cite{li2024learning},
              VCR-Graphormer~\cite{fu2024vcr},
              CobFormer~\cite{xing2024less}
             }, d
            ]  
        ]
        [Token Sequence, b
            [{ANS-GT~\cite{zhang2022hierarchical},
              NTFormer~\cite{chen2024ntformer}
             }, d
            ]  
        ]
        [PEs and SEs, b
            [Positional Encodings, c
                [{DGT~\cite{park2022deformable},
                  MpFormer~\cite{limpformer} 
                }, e
                ]
            ]
            [Structural Encodings, c
                [{AGT~\cite{ma2023rethinking}
                }, e
                ]
            ]
            [Other Explorations, c
                [{Muller et al.~\cite{muller2023attending},
                  SpecFormer~\cite{bo2023specformer},
                  DeGTA~\cite{wang2024graph}
                }, e
                ]
            ]
        ]
    ]
    [Neural Diffusion Process, a
        [Non-smooth Diffusion, b
            [{GIND~\cite{chen2022optimization},
              PDE-GCN~\cite{eliasof2021pde},
              GraphCON~\cite{rusch2022graph}
             }, d
            ]  
        ]
        [Diffusion with External Forces, b
            [{CDE~\cite{zhao2023graph},
              ACMP~\cite{wang2022acmp},
              GREAD~\cite{choi2023gread},
              ADR-GNN~\cite{eliasof2023adr},
              FLODE~\cite{maskey2024fractional}
             }, d
            ]  
        ]
        [Diffusion with Modulators, b
            [{G$^2$~\cite{rusch2022gradient},
              EGLD~\cite{zhang2024unleashing},
              MHKG~\cite{shao2023unifying},
              A-GGN~\cite{gravina2022anti}
             }, d
            ]  
        ]
        [Other Explorations, b
            [{NSD~\cite{bodnar2022neural},
              Conn-NSD~\cite{barbero2022sheaf},
              QDC~\cite{markovich2023qdc},
              Giovanni et al.~\cite{di2022understanding},
              PCGCN~\cite{zhang2023steering},
              FGND~\cite{wan2024flexible},
              HiD-Net~\cite{li2024generalized}
             }, d
            ]  
        ]
    ]
    ]
    \end{forest}
    \caption{Taxonomy of heterophilic GNNs and beyond with representative examples.}
    \label{models}
    \end{figure*}
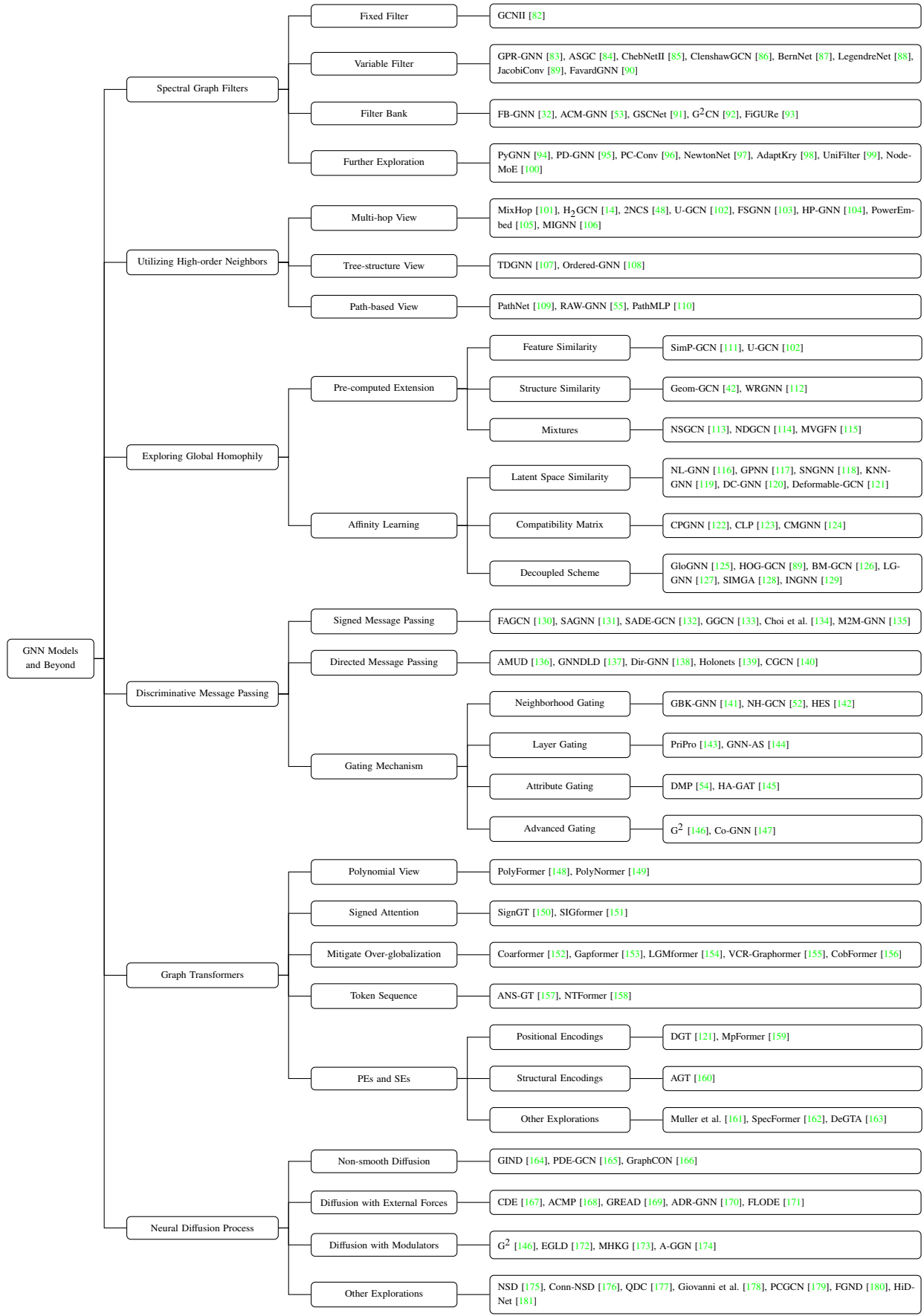

    As can be seen in Figure~\ref{models}, 
    graph heterophily has led to the emergence of numerous 
    heterophilic GNNs. 
    In this section, 
    we categorize the most representative GNN models related to graph heterophily and conduct a series of analyses and discussions.
    
    \subsection{Spectral Graph Filters}
    Based on the theory of graph signal processing~\cite{ortega2018graph}, spectral graph filters have been integrated into the GNNs to enhance the expressive power.
    Most GNNs smooth representations of connected nodes, which is equivalent to performing low-pass filtering~\cite{nt2019revisiting}. 
    For example, GCN~\cite{kipf2016semi} utilizes the self-loop operation to enhance the low-pass filter, and its approximate filter $g(\widetilde{\mathbf{L}})$ can be expressed as:
    \begin{equation}
        g(\widetilde{\mathbf{L}}) =\mathbf{I} + \mathbf{F}_{LP} = \mathbf{I} + \widetilde{\mathbf{A}} = 2\mathbf{I} - \widetilde{\mathbf{L}}.
    \end{equation}
    However, the high-frequency signals that capture the dissimilarity between nodes are often neglected, restricting the expressive power of heterophily. 
    To this end, many studies attempt to design advanced spectral filters with the aim of capturing the heterophily pattern on graphs.
    Here, we follow Liao et al.~\cite{liao2024benchmarking} and categorize existing spectral methods into fixed filter, variable filter and filter bank.

    \noindent \textbf{Fixed Filter.} 
    For the first type, the basis and parameters of graph filters are both constant, resulting in fixed filters $g(\widetilde{\mathbf{L}})$.
    Inspired by the homophily scenario~\cite{gasteiger2018predict,li2019label,gasteiger2019diffusion}, 
    GCNII~\cite{chen2020simple} extends the vanilla GCN under graph heterophily with two simple yet effective techniques: Initial residual and Identity mapping, to preserve ego or local information. 
    The spectral interpretation of the polynomial approximation is:
    \begin{equation}
        g(\widetilde{\mathbf{L}}) = \sum_{k=0}^K \alpha(1-\alpha)^k (\mathbf{I}-\widetilde{\mathbf{L}})^k,
    \end{equation}
    where $K$ is the spectral filter order, $\alpha \in [0,1]$ is the coefficient for balancing neighbor propagation and ego preservation, where a larger $a$ results in stronger node identity and weaker neighboring impact, and vice versa. 
    Due to the preset parameters, methods based on fixed filters still face the problem of insufficient expressive power when dealing with graph heterophily.

    \noindent \textbf{Variable Filter.} 
    Compared to fixed filters, variable filters extend the approximate filter to a variable version $g(\widetilde{\mathbf{L}},\theta)$, rendering a better capability to capture high-frequency signals. 
    Inspired by the generalized PageRank computation~\cite{li2019optimizing,gasteiger2018predict},
    GPR-GNN~\cite{chien2020adaptive} produces the spectral filter formulated as:
    \begin{equation}
        g(\widetilde{\mathbf{L}},\theta) = \sum_{k=0}^K \theta^k (\mathbf{I}-\widetilde{\mathbf{L}})^k,
    \end{equation}
    where $\theta$ is the learnable weight to the fit the heterophily pattern. 
    ASGC~\cite{chanpuriya2022simplified} adopts the same principle of GPR-GNN and provides a more simplified version.
    ChebNet~\cite{defferrard2016convolutional} is also a typical example of using the variable filter, which utilizes the Chebyshev polynomial as basis:
    \begin{equation}
        g(\widetilde{\mathbf{L}},\theta) = \sum_{k=0}^K \theta^k T^{(k)}(\widetilde{\mathbf{L}}).
    \end{equation}
    Each term $T^{(k)}(\widetilde{\mathbf{L}})$ can be expressed recursively:   
    \begin{align}
        T^{(k)}(\widetilde{\mathbf{L}}) = 2\widetilde{\mathbf{L}}T^{(k-1)} &(\widetilde{\mathbf{L}})- T^{(k-2)}(\widetilde{\mathbf{L}}), \notag \\ 
        T^{(1)}(\widetilde{\mathbf{L}}) = \widetilde{\mathbf{L}}, &T^{(0)}(\widetilde{\mathbf{L}}) = \mathbf{I}. 
    \end{align}
    Based on ChebNet, ChebNetII~\cite{he2022convolutional} employs the Chebyshev interpolation~\cite{gil2007numerical} to enhance the filter: 
    \begin{equation}
        g(\widetilde{\mathbf{L}},\theta) =
        \frac{2}{K+1} \sum_{k=0}^K \sum_{\kappa =0}^K
        \theta_\kappa  T^{(k)}(x_\kappa)
        T^{(k)}(\widetilde{\mathbf{L}}),
    \end{equation}
    where $T^{(k)}(x_\kappa)$, $T^{(k)}(\widetilde{\mathbf{L}})$ follow the Chebyshev basis, and 
    $x_\kappa= \cos \frac{(\kappa+1/2)\pi}{K+1}$ are the Chebyshev nodes of $T^{K+1}$~\cite{he2022convolutional}. Through this extension, ChebNetII exhibits superiority than ChebNet on heterophilic graphs.
    ClenshawGCN~\cite{guo2023clenshaw} applies Clenshaw algorithm to incorporate explicit residual connections. 
    The form of its spectral filter is similar to chebNet, but with some differences in the iteration:
    \begin{align}
        T^{(k)}(\widetilde{\mathbf{L}}) = 2\widetilde{\mathbf{L}}T^{(k-1)} &(\widetilde{\mathbf{L}})- T^{(k-2)}(\widetilde{\mathbf{L}}), \notag \\ 
        T^{(1)}(\widetilde{\mathbf{L}}) = 2\widetilde{\mathbf{L}}, &T^{(0)}(\widetilde{\mathbf{L}}) = \mathbf{I}. 
    \end{align}
    To avoid oversimplified or ill-posed learnt weights, BernNet~\cite{he2021bernnet} substitutes the generalized PageRank with Bernstein polynomial: 
    \begin{equation}
        g(\widetilde{\mathbf{L}},\theta) = \sum_{k=0}^K \frac{\theta_k}{2^K} T^{(k)}(\widetilde{\mathbf{L}}), 
        T^{(k)} = \binom{K}{k}
        (2\mathbf{I}-\widetilde{\mathbf{L}})^{K-k}(\widetilde{\mathbf{L}})^k.
    \end{equation}
    BernNet can approximate complex filters such as band-rejection and comb filters, and provide better interpretability for graph heterophily.
    In addition to the above models, other models such as LegendreNet~\cite{chen2023improved}, JacobiConv~\cite{wang2022powerful}, FavardGNN~\cite{guo2023graph} are dedicated to designing more advanced variable filters to enhance the expressive power for heterophily.
    For more details, please refer to ~\cite{liao2024benchmarking}.
    
    \noindent \textbf{Filter Bank.} 
    Some studies question whether a single filter can adequately capture complex graph signals on graphs. 
    Hence, GNNs with filter bank integrate multiple fixed or variable filters to enhance the expressive power. 
    Given that the number of filters is $Q$, the integration can be formulated as:
    \begin{equation}
        \hat{g}(\widetilde{\mathbf{L}},\gamma) = \bigoplus^Q_{q=1} \gamma^q \cdot g_q(\widetilde{\mathbf{L}}),
    \end{equation}
    where $\bigoplus$ denotes an arbitrary combination such as sum or concatenation, and each filter (fixed or variable filter) $g_q(\widetilde{\mathbf{L}})$ is assigned with a weight parameter $\gamma^q$.
    FB-GNN~\cite{luan2020complete} first introduces the concept of filter bank for combining multiple filters under heterophily. 
    It designs a dual-channel scheme with simple low-pass and high-pass filters to learn the smooth and non-smooth components together:
    \begin{equation}
        \hat{g}(\widetilde{\mathbf{L}}; \gamma) = \gamma_1 \mathbf{F}_{LP} + \gamma_2 \mathbf{F}_{HP} = \gamma_1 \widetilde{\mathbf{L}} + \gamma_2 (\mathbf{I} - \widetilde{\mathbf{L}}),
    \end{equation}
    where $\gamma_1, \gamma_2 \in [0,1]$ is the learnable scalar parameters for weighted sum. 
    ACM-GNN~\cite{luan2022revisiting} extends FBGNN to three filters, including an all-pass filter to maintain node identity:
    \begin{equation}
        \hat{g}(\widetilde{\mathbf{L}}; \gamma) = \gamma_1 \widetilde{\mathbf{L}} + \gamma_2 (\mathbf{I} - \widetilde{\mathbf{L}}) + \gamma_3 \mathbf{I}.
    \end{equation}
    FAGCN~\cite{bo2021beyond} integrates two spectral filters with bias to capture both low- and high-frequency signals:
    \begin{equation}
        \hat{g}(\widetilde{\mathbf{L}}; \gamma) = \gamma_1 ((\beta+1)\mathbf{I} - \widetilde{\mathbf{L}}) + \gamma_2 ((\beta-1)\mathbf{I} + \widetilde{\mathbf{L}}),
    \end{equation}
    where $\beta \in [0,1]$ is the scaling coefficient.
    GSCNet~\cite{du2024graph} proposes a novel simple basis that decouples the positive and negative activation:
    \begin{equation}
        \hat{g}(\widetilde{\mathbf{L}}; \gamma) = \sum_{i=0}^{K_1} \gamma_i (2\mathbf{I}-\widetilde{\mathbf{L}})^i+ \sum_{j=0}^{K_2} \gamma_j (\widetilde{\mathbf{L}})^j,
    \end{equation}
    where the ratios of activation can be adjusted by hyper-parameters $K_1$ and $K_2$.
    G$^{2}$CN~\cite{li2022g} proposes the Gaussian filters with sufficient flexibility and better performance:
    \begin{align}
        \hat{g}(\widetilde{\mathbf{L}}, &\gamma) = \gamma_1 \sum_{k=1}^{\lfloor K/2 \rfloor} \theta_{1,k} T_1^{(k)} +
        \gamma_2 \sum_{k=1}^{\lfloor K/2 \rfloor} \theta_{2,k} T_2^{(k)},  \notag \\
        &T_1^{(k)} = ((1+\beta_1)\mathbf{I} - \widetilde{\mathbf{L}}), \theta_{1,k} = \frac{\alpha_1^k}{k!}, \notag \\
        &T_2^{(k)} = ((1-\beta_2)\mathbf{I} - \widetilde{\mathbf{L}}), \theta_{2,k} = \frac{\alpha_2^k}{k!},
    \end{align}
    where $\alpha, \beta \in [0,1]$ is the decay coefficient and scaling coefficient.
    Typically, FiGURe~\cite{ekbote2024figure} utilizes the filter-level parameters to control each component of filter bank: 
    \begin{equation}
        \hat{g}(\widetilde{\mathbf{L}}; \gamma) = \sum^Q_{q=1} \gamma^q \cdot g_q(\widetilde{\mathbf{L}}).
    \end{equation}
    Various filters, including Chebyshev, Bernstein and other filters, can be used to compose the filter bank of FiGURe. 
    It is worth noting that variable filters and filter bank models outperform fixed filter under heterophily~\cite{liao2024benchmarking}. 
    Therefore, a comprehensive consideration of model capability and computational cost is required when selecting a suitable filter for learning distinct graph signals.

    \noindent \textbf{Further Explorations.}
    In addition to the filter design, there are also some further explorations of spectral graph filter.
    For example, PyGNN~\cite{geng2023pyramid} attempts to improve the performance of spectral filters using sampling techniques, while PD-GNN~\cite{lu2024flexible} introduces a new parameterized Laplacian matrices, offering more flexibility in controlling the diffusion distance than graph Laplacian.
    Inspired by graph heat equation~\cite{weber2008analysis}, PC-Conv~\cite{li2024pc} introduces the Possion-Charlier polynomial~\cite{kroeker1977wiener} to make an exact numerical approximation.
    NewtonNet~\cite{xu2023shape} establishes the connections between spectral frequencies and graph heterophily, and further integrates spectral filter with Newton Interpolation~\cite{hildebrand1987introduction}.
    Huang et al.~\cite{huang2024optimizing} propose to optimize the polynomial filters in terms of Krylov subspace~\cite{liesen2013krylov}, and further reveal how universal polynomial bases enhance spectral GNNs~\cite{huang2024universal}.
    Motivated by Mixture of Experts~\cite{cai2024survey}, Node-MoE~\cite{han2024node} proposes that node-wise filtering can achieve linear separability for all nodes, and further leverages a Mixture of Experts framework for spectral GNNs.
    Besides Node-MoE, both DSF~\cite{guo2023graph} and NFGNN~\cite{zheng2023node} are dedicated to designing adaptive node-specific filtering.
    
    Despite remarkable success, existing GNNs based spectral graph filters still face two major issues: the polynomial limitation and the transductive limitation~\cite{xuslog}. 
    The co-existence of homophily and heterophily patterns on graphs, as well as scalability on large graphs, remain challenges that spectral GNNs urgently need to address.
    
    \subsection{Utilizing High-order Neighbors}
    Under graph heterophily setting, simply weighted averaging the representations of neighbors~\cite{kipf2016semi,velivckovic2017graph} in message passing will inevitably introduce noise, potentially resulting in low-quality node representations~\cite{zhu2020beyond}.
    
    \noindent \textbf{Multi-hop View.}
    To mitigate the negative impact of heterophilic neighbors, a natural idea is to utilize informative nodes within multi-hops.
    Formally, the k-hop neighbor set of nodes can be defined as:
    \begin{equation}
        \mathcal{N}_k(v) = \{ u: d(u,v) = k \},
    \end{equation}
    where $d(u,v)$ measures the shortest distance between nodes $u$ and $v$. 
    MixHop~\cite{abu2019mixhop} demonstrates that mixing information from multi-hop neighbors can provide a wider class of representations. 
    Hence, Mixhop extends the message passing mechanism within 1-hop to multi-hops:
    \begin{align}
        r_{i,k}^{(l)} = \texttt{AGG}(&\{h_j^{(l-1)}: v_j \in \bigcup_{q=0}^{k} \mathcal{N}_q(v)\}), (k=0,1,2), \notag \\
        &h_{i}^{(l)} = \texttt{CONCAT} (r_{i,0}^{(l)}, r_{i,1}^{(l)},r_{i,2}^{(l)}),
    \end{align}
    where $\texttt{CONCAT}$ denotes the column-wise combination. The arbitrary linear combinations of columns can be learned to model more complex patterns on graphs, such as the heterophily pattern.
    Similar to Mixhop, H$_2$GCN~\cite{zhu2020beyond} considers 2-hop message passing, and further utilizes the combination of intermediate representations:
    \begin{align}
        r_{i,k}^{(l)} = \texttt{AGG}&(\{h_j^{(l-1)}: v_j \in \mathcal{N}_k(v)\}), (k=0,1,2), \notag \\
        &h_{i}^{(l)} = \texttt{CONCAT} (r_{i,0}^{(l)}, r_{i,1}^{(l)},r_{i,2}^{(l)}), \notag \\
        &h_i =  \texttt{CONCAT}(h_i^{(0)}, h_i^{(1)}, h_i^{(2)}).
    \end{align}
    To ``utilizing'' instead of ``mixing'' multi-hop neighbors, both Mixhop and H$_2$GCN utilize the $\texttt{CONCAT}$ operation to separate the ego and neighbor representation of nodes. 
    Furthermore, H$_2$GCN particularly emphasizes the importance of 2-hop neighbors, and demonstrates that the 2-hop neighborhood for a node is always homophily-dominant.
    Cavallo et al.~\cite{cavallo20222} also find that the 2-hop Neighbor Class Similarity (2NCS) correlates with GNN performance more strongly.
    U-GCN~\cite{jin2021universal} reaches the similar conclusion and discovers that 1-hop, kNN and 2-hop neighbors are more suitable as neighborhoods in network with complete homophily, randomness and complete heterophily, respectively. 
    To extract information from the three kinds neighborhood simultaneously, U-GCN performs a multi-type convolution based on the attention mechanism. 
    Moreover, FSGNN~\cite{maurya2022simplifying} employs the softmax operation to regularize the message from different hops, and HP-GNN~\cite{xu2022hp} designs a memory unit to retain information from multi-hop neighbors.
    PowerEmbed~\cite{huang2022local} projects multi-hop neighborhoods into a low-dimensional space for further concatenation, and MIGNN~\cite{choi2023node} utilizes the mutual information to model the dependence between nodes within k-hop neighborhoods.

    \noindent \textbf{Tree-structure View.}
    Another line of utilizing high-order neighbors is to model the neighborhood as a tree structure. 
    For example, TDGNN~\cite{wang2021tree} models the high-order neighbor interactions through tree decomposition. 
    Viewing higher-order neighborhoods as tree structures provides new insights into addressing heterophily. 
    However, most works overlook the importance of orders of tree structures.
    To this end, Ordered-GNN~\cite{song2023ordered} integrates the inductive bias from rooted-tree hierarchy and encodes the neighborhood information at some orders. 
    The ordered mechanism prevents the mixing of node features within hops and possesses stronger expressive power, leading to superior performance on both heterophilic and homophilic graphs.

    \noindent \textbf{Path-based View.}
    In addition to tree-structure view, high-order neighborhoods can also be transformed into multiple paths.
    PathNet~\cite{sun2022beyond} first sheds light on the path-level patterns which can explicitly reflect rich semantic and structural information. 
    This work proposes a novel path aggregation strategy with a maximal entropy path sampler, and designs a structure-aware recurrent cell to perform path-level aggregation. 
    RAW-GNN~\cite{jin2022raw} employs the Breadth-First Search (BFS) and Depth-First Search (DFS) views to capture graph homophily and heterophily, respectively. 
    After obtaining BFS and DFS paths, RAW-GNN utilizes the RNN-based aggregators to encode each path and learns the importance of each path through an attention mechanism. 
    Moreover, PathMLP~\cite{xie2023pathmlp} designs a similarity-based path sampling strategy, encodes path-level messages through simple transformation and concatenation, and effectively learns node representations through adaptive path aggregation under graph heterophily.
    
    However, the choice of high-order neighborhood size remains a tricky issue; a large neighborhood will incur high computational costs, while a smaller one may not guarantee solving the heterophily issue. 
    Although empirical results suggest that the 2-hop neighborhood may be a suitable choice~\cite{zhu2020beyond,cavallo20222,jin2021universal}, this cannot ensure the applicability in complex real-world scenarios.
    
    \subsection{Exploring Global Homophily}
    Since graph heterophily in the neighborhood has a negative impact on GNNs, another approach is to go beyond local neighborhood and explore potential connections globally.
    Generally, we can measure the similarity between nodes and further introduces a new potential neighbor set $\mathcal{N}_p$:
    \begin{equation}
        \mathcal{N}_p = \{ u: sim(u,v) > \rho \},
    \end{equation}
    where $\rho$ is a threshold, and $sim(\cdot)$ is a similarity function that can be implemented based on feature, structure or other metrics.
    In this way, we can extend the neighbor set involved in Equation~\ref{mpp} from a global perspective to mitigate the heterophily issue.
    
    \noindent \textbf{Pre-computed Extension.}
    One straightforward approach is to extend the neighbor set during the pre-computation stage.

    \noindent \emph{-Feature Similarity.}
    Exploring homophily globally based on node feature similarity is the most commonly used method.
    Here, $sim(\cdot)$ is implemented by simple dot product or cosine similarity, and the feature similarity matrix can be pre-computed as:
    \begin{equation}
        S_{ij} = cos(x_i, x_j) = \frac{x_i^\mathsf{T} x_j}{\Vert x_i \Vert \ \Vert x_j \Vert}.
    \end{equation}
    We can choose k Nearest Neighbors for each node to construct the kNN feature graph as the potential neighbor set. 
    For example, SimP-GCN~\cite{jin2021node} and U-GCN~\cite{jin2021universal} adaptively combine information from the original graph and kNN feature graph, and find this simple method achieves good performance under graph heterophily setting. 
    
    \noindent \emph{-Structure Similarity.}
    Structure-based methods, on the other hand, place greater emphasis on utilizing graph structure to discover potential neighbors. 
    Typically, Geom-GCN~\cite{pei2020geom} uses three graph structural metrics: Isomap~\cite{tenenbaum2000global}, Poincare~\cite{nickel2017poincare}, and Struc2Vec~\cite{ribeiro2017struc2vec}, to implement $sim(\cdot)$, and maps nodes to the latent space for geometric relation mining. 
    In addition to inherent neighbors of input graph, neighbors that conform to the defined geometric relationships, i.e., potential neighbors $\mathcal{N}_p$, also participate in the message passing. 
    Moreover, WRGNN~\cite{suresh2021breaking} takes the degree sequence of neighbor nodes as the similarity metric to construct a weighted multi-relational graph. 
    Message passing on this computation graph breaks the limit of local assortative and facilitates global integration.
    
    \noindent \emph{-Mixtures.}
    It is also feasible to simultaneously utilize feature and structural similarity for neighbor extension.
    For example, NSGCN~\cite{ai2024neighbors} computes the common neighbors distribution and feature similarity for extension, while NDGCN~\cite{liu2025beyond} further considers the higher-order neighborhood distribution.
    MVGFN~\cite{wang2023improving} defines the semantic kNN neighbors based on feature similarity and the structural kNN neighbors based on graph embedding algorithms, such as diffusion process~\cite{gasteiger2019diffusion} and node2vec~\cite{grover2016node2vec}. 
    Combined with 1-hop and 2-hop neighbors, MVGFN aggregates hybrid information from these four neighbor sets and introduces a multi-view graph fusion framework.
    
    \noindent \textbf{Affinity Learning.}
    While pre-computed extension based on node features or structure is available, exploring global homophily through learning methods offers more flexibility. 
    Generally, this type of methods learn an affinity matrix to model global relationships and guide message passing, rather than relying on fixed similarity computations.
    
    \noindent \emph{-Latent Space Similarity.}
    Considering the computation cost of global similarity, an intuitive idea is to map nodes to the latent space to explore global homophily.
    For example, NL-GNN~\cite{liu2021non} directly utilizes the GNNs to embed nodes, and GPNN~\cite{yang2022graph} leverages the Pointer Networks~\cite{vinyals2015pointer} to obtain node embeddings. 
    After that, both of them compute the global affinity matrix based on node embeddings through attention, and perform non-local aggregation to capture long-range dependencies from distant nodes. 
    Further, SNGNN~\cite{zou2023similarity} updates the affinity matrix at each layer for propagation, which dynamically characterizes the global interactions in each layer.
    KNN-GNN~\cite{li2024knn} first separates the ego and neighbor representations of nodes, and then finds kNN neighbors in latent space for non-local aggregation. 
    Since capturing the global homophily in the latent space is convenient, 
    a similar approach is also applied in DC-GNN~\cite{dong2024differentiable} and Deformable-GCN~\cite{park2022deformable} to combat the graph heterophily issue.
    
    \noindent \emph{-Compatibility Matrix.}
    Apart from exploring global homophily in the latent space, some studies introduce the compatibility matrix~\cite{zhu2021graph} to boost homophily mining.
    Compatibility matrix models the connection probability of nodes between each pair of classes, which can be transformed into an affinity matrix to guide message passing.
    CPGNN~\cite{zhu2021graph} first incorporates the compatibility matrix for modeling the heterophily level in the graphs and designs the compatibility-guided propagation.
    CLP~\cite{zhong2022simplifying} first learns the class compatibility matrix and then generalises the label propagation algorithm, weighted by the class compatibility, to accommodate arbitrary heterophily assumptions.
    Zheng et al.~\cite{zheng2024revisiting} revisits the message passing under graph heterophily, reformulates them into an unified framework and reveals that the success of heterophily-specific GNNs is attributed to implicitly enhancing the compatibility matrix among classes. 
    Inspired by this, 
    they introduce the Compatibility Matrix-aware Graph Neural Network (CMGNN) to explicitly leverage and improve the compatibility matrix. 
    Moreover, LRGNN~\cite{liang2024predicting} extends the compatibility matrix to a signed global relationship matrix,  and reformulates the prediction of relationship matrix into a robust low-rank matrix approximation problem.

    \noindent \emph{-Decoupled Scheme.}
    Another representative approach is to model the affinity matrix through decoupling the graph structure and node feature views.
    Typically, GloGNN~\cite{li2022finding} uses MLPs to map the node feature and adjacency matrix into feature and topology views separately, and further fuses them with a term weight.
    After the decoupling, GloGNN aggregates information from global nodes and designs an acceleration strategy to avoid the quadratic time complexity.
    Moreover, HOG-GCN~\cite{wang2022powerful}, BM-GCN~\cite{he2022block} and LG-GNN~\cite{yu2024lggnn} share the similar idea with GloGNN. 
    HOG-GCN applies the label propagation to enhance the topology view, BM-GCN introduces the block modeling-guided propagation and LG-GNN integrates the SimRank~\cite{jeh2002simrank} and ListNet loss~\cite{cao2007learning} to enhance the two views.
    Further, SIMGA~\cite{liu2023simga} and INGNN~\cite{liu2022uplifting} theoretically proves that this decoupled strategy is effective in discovering global relationship and grouping similar nodes under graph heterophily.
    
    We acknowledge that global computation significantly enhances the model performance on heterophilic graphs. However, the unavoidable trade-off is even higher computational costs.
    Whether global neighbors are truly necessary, and whether exploring global homophily will bring negative effects, are questions worth investigation.
      
    \subsection{Discriminative Message Passing}
    Besides high-order neighbors or global homophily, heterophilic information within the local neighborhood is also worth consideration. 
    We can adopt the discriminative message passing mechanism to retain useful information and distinguish noisy messages, such as signed message passing, directional message passing, and gating mechanisms.
    
    \noindent \textbf{Signed Message Passing.}
    Generally, the aggregation weights in message passing are mostly set as positive, which leads to an inability to capture heterophilic patterns. 
    In light of that, FAGCN~\cite{bo2021beyond} proposes the signed message strategy and allows the aggregation weights in message passing to be negative.
    Further, SAGNN~\cite{wu2023signed} and SADE-GCN~\cite{lai2023self} regularize the aggregation weights and restrict the range to $[-1, 1]$, while GGCN~\cite{yan2022two} approximates the sign function for message passing with cosine similarity to separate the contribution of similar neighbors from that of dissimilar neighbors.
    From the theoretical perspective, Choi et al.~\cite{choi2023signed} points that signed messages escalate the inconsistency between neighbors and increase the uncertainty in predictions. 
    Based on the observations, they propose to utilize calibration for signed GNNs to reduce uncertainty and enhance the quality.
    Further, Liang et al.~\cite{liang2024sign} proposes that signed message passing have two limitations: undesirable representation update for multi-hop neighbors and vulnerability against over-smoothing issues. 
    To address these issues, they propose the Multi-set to Multi-set GNN (M2M-GNN), and the core idea is to ensure that heterophilic node representations are not intertwined. 

    \noindent \textbf{Directed Message Passing.}
    The directionality of edges is often overlooked in graph learning, yet recent studies have shown that directed message passing can alleviate heterophily issues.
    AMUD~\cite{sun2024breaking} investigates the impact of directed topology in homophily and heterophily and offers modeling guidance for natural digraphs from a statistical perspective, maximizing the benefits for subsequent graph learning.
    GNNDLD~\cite{chaudhary2024gnndld} integrates the edge directionality into the heterophily-specific models, and highlights that this simple concept can significantly boost performance.
    Rossi et al.~\cite{rossi2024edge} observe that the directionality of a graph substantially increases its effective homophily in heterophilic settings, while posing negligible or no impact on homophilic settings. 
    Inspired by this, they introduce the Directed Graph Neural Network (Dir-GNN) to separate aggregations for directed edges, and prove that Dir-GNN is strictly more expressive than vanilla message passing.
    Combining spectral graph theory, Koke et al.~\cite{koke2023holonets} integrate spectral graph filters with digraphs to improve performance on heterophilic graphs. They extend the spectral convolution to directed graphs and provide a detailed analysis from the frequency perspective.
    CGCN~\cite{zhuo2024commute} introduces a new Laplacian formulation for digraphs, along with a rapid computational method for determining commute times, effectively leveraging the path asymmetry to address graph heterophily.
    
    \noindent \textbf{Gating Mechanism.}
    Designing gating mechanisms for heterophily in message passing is highly flexible, and can be applied within node neighborhoods, between model layers, and even at the attribute level.
    
    \noindent \emph{-Neighborhood Gating.}
    Typically, GBK-GNN~\cite{du2022gbk} demonstrates that feature transformation with a single kernel fails to classify the heterophily pattern, and proposes a gated bi-kernel mechanism with two kernels to model graph homophily and heterophily, respectively. 
    Through gate selection for neighbor aggregation and bi-kernel feature transformation, GBK-GNN can theoretically enhance the discriminative ability for heterophily.
    Similarly, NH-GCN~\cite{gong2023neighborhood} introduces the Neighborhood Homophily (NH) and designs the NH-based gating mechanism to separate neighbors into two channels with channel-specific weights.
    Moreover, HES~\cite{wang2024heterophilic} proposes the Snowflake Hypothesis~\cite{wang2023snowflake} underpinning the concept of ``one node, one receptive field'', and establishes the gates for nodes to filter out the heterophilic information in the neighborhood. 

    \noindent \emph{-Layer Gating.}
    The gating mechanism can be used not only to distinguish heterophilic neighbors, 
    but also to customize personalized aggregation between GNN layers.
    For example, PriPro~\cite{cheng2023prioritized} designs the gating mechanism between GNN layers to adaptively integrate or discard inter-layer information for each node, implicitly leveraging higher-order relations on graphs to address the heterophily issue.
    Following PriPro, the similar idea of customizing personalized receptive field for each node has also been applied in GNN-AS~\cite{deng2024learning}.

    \noindent \emph{-Attribute Gating.}
    At a finer granularity, DMP~\cite{yang2021diverse} sets dimension-level gates on the node representation and calibrates the propagation of each attribute, thereby enhancing the discriminative ability for heterophily. 
    Moreover, HA-GAT~\cite{wang2024heterophily} designs the edge heterophily preference matrix to enhance the heterophily-aware aggregation at the attribute level.
    
    \noindent \emph{-Advanced Gating.}
    Meanwhile, we also notice that some studies have employed more advanced gating mechanisms.
    Considering that graph learning may benefit from different rates of propagation, Rusch et al.~\cite{rusch2022gradient} propose a novel multi-rate message gating scheme called G$^2$ that leverages graph gradients to ameliorate the heterophily issue. 
    G$^2$ is a flexible framework which can plug existing GNNs, and can be interpreted as a discretization of a dynamical system governed by nonlinear differential equations.
    Inspired by social interactions, Co-GNN~\cite{finkelshtein2023cooperative} proposes a novel cooperative framework where each node is viewed as a dynamic player. 
    Co-GNN utilizes an advanced gating mechanisms to endow nodes with four states: Standard, Listen, Broadcast and Isolate, allowing nodes to choose whether to receive or broadcast messages. 
    This flexible paradigm possesses greater expressive power of homophily and heterophily, and allows the exploration of typical topology while learning. 

    Despite the emergence of various GNNs with advanced propagation, they still adhere to the message passing paradigm and inevitably encounter issues such as over-smoothing, over-squashing and graph heterophily. 
    Whether further progress can be made on the graph heterophily issue partly depends on the ability to innovate GNN architectures.

    \subsection{Graph Transformers}
    Due to the powerful global attention mechanism, Graph Transformer (GT), this advanced architecture is considered to have inherent advantages in addressing the graph heterophily issue.

    \noindent \textbf{Polynomial View.}
    Spectral GNNs utilize various polynomial bases to approximate graph convolution and demonstrate powerful capabilities on heterophilic graphs. 
    Inspired by that, PolyFormer~\cite{ma2024polyformer} incorporates the spectral information through polynomial approximation into GTs to enhance the expressive power.
    This framework defines the polynomial token to perform node-wise filtering based on various polynomial bases. 
    Subsequently, PolyFormer designs the global attention mechanism for polynomial tokens to balance node-specific and global patterns.  
    Further, PolyNormer~\cite{deng2024polynormer} introduces the first polynomial-expressive GT, where graph topology and node features are integrated into polynomial coefficients separately. 
    With a linear local-to-global attention scheme, PolyNormer can learn high-degree equivariant polynomials and perform well under both homophily and heterophily settings.

    \noindent \textbf{Signed Attention.}
    Due to the inherent positive attention, the self-attention mechanism in GTs fails to effectively capture the high-frequency signals, thereby making learning from graph heterophily inefficient.
    To this end, SignGT~\cite{chen2023signgt} extends the self-attention calculation to include signed attention, thereby preserving different frequency information of graphs. 
    Moreover, SignGT utilizes the multi-hop topology to maintain the local structural bias from a global perspective.
    The same concept has also been extended to the field of recommender systems, improved as SIGformer~\cite{chen2024sigformer}.
    
    \noindent \textbf{Mitigate Over-globalization.}
    While global attention partially addresses the graph heterophily issue, the massive distant nodes inevitably divert a significant portion of attentions, regardless of actual relevance.
    Motivated by this, Coarformer~\cite{kuang2024transformer} downsamples the input graph by grouping nodes into a less number of virtual super-nodes. 
    Then, a Transformer-based module is proposed to explore the pairwise interactions super-nodes and capture the coarse but long-range dependencies. 
    Similarly, Gapformer~\cite{liu2023gapformer} introduces graph pooling to mitigate the influx of irrelevance, LGMformer~\cite{li2024learning} mitigates the over-globalization via common K-Means, and VCR-Graphormer~\cite{fu2024vcr} specifically encodes the heterophily information and rewires graphs through virtual connections and super-nodes.
    Xing et al.~\cite{xing2024less} explore whether global attention always benefits GTs and reveal the over-globalizing problem in GTs. 
    They propose CobFormer to prevent the over-globalizing phenomenon while keeping the ability to extract valuable information from distant interactions. 

    \noindent \textbf{Token Sequence.}
    This category of GTs considers sampling a token sequence for each node as the model input, implicitly leveraging the information from multi-hop neighbors.
    For example, ANS-GT~\cite{zhang2022hierarchical} samples the node sequence from 1-hop, 2-hop and KNN neighbors or based on PageRank, and then formulates the optimization of sampling strategies as an adversary bandit problem. 
    NTFormer~\cite{chen2024ntformer} introduces two types of token sequences, node-wise and neighborhood-wise token sequences, from attribute and topology perspectives, and incorporates a Transformer-based backbone with the adaptive fusion module to learn the final node representations.
    
    \noindent \textbf{PEs and SEs.}
    Due to the inherent global mechanism which treats inputs as fully connected graphs, GTs urgently require PEs and SEs to supplement structure information, especially under heterophily.
    
    \noindent \emph{-Positional Encodings.}
    Inspired by Katz index~\cite{katz1953new}, DGT~\cite{park2022deformable} designs the learnable PE named Katz PE to improve the expressive power of GTs by incorporating structural and semantic similarity between nodes.
    MpFormer~\cite{limpformer} also introduces a novel PE technique called HeterPos. HeterPos introduces the shortest path distance to define relative positions and captures feature distinctions between ego-nodes and neighbors, facilitating the incorporation of heterophilic information into the GTs.

    \noindent \emph{-Structural Encodings.}
    Apart from PEs, AGT~\cite{ma2023rethinking} highlights the necessity of SEs and introduces the learnable centrality encoding and kernelized local structure encoding from the node centrality and subgraph views.
    This framework fills the gap of SEs for GTs in learning from heterophilic graphs.

    \noindent \emph{-Other Explorations.}
    To investigate whether PEs and SEs can effectively address heterophily, Muller et al.~\cite{muller2023attending} benchmark multiple model variants with PEs and SEs on heterophilic datasets.
    They conclude that PEs and SEs can lead to large gains in model performance, while global attention mechanism only provides small or no improvements at all.
    Noting that eigenvalues of graph Laplacian convey rich information, SpecFormer~\cite{bo2023specformer} proposes the novel Eigenvalue Encoding based on the spectral theory.
    Moreover, DeGTA~\cite{wang2024graph} identifies that existing GTs primarily face two challenges: multi-view chaos and local-global chaos and proposes a novel decoupled model. 
    To address these issues, it provides clear definitions of positional, structural, and attribute information, as well as local and global interaction, and rationally integrates all the information in a rational manner.

    Although the Transformer-based architecture has achieved significant success in the domain of NLP, there remain questions about whether this remarkable success can be  effectively transferred to graph-structure data. 
    Furthermore, GTs inevitably require significant computational resources. 
    Although techniques have been developed to reduce the computational complexity, they still encounter the scalability challenges when processing large-scale graph data.

    \subsection{Neural Diffusion Process}
    Intuitively, the process of heat diffusion has a natural correspondence with message passing mechanism of GNNs.
    Recent studies have revealed the connection between diffusion dynamics and message passing~\cite{chamberlain2021grand}. 
    A considerable number of neural diffusion models have emerged, leveraging continuous dynamics formulations to mitigate the known pitfalls of GNNs, such as over-smoothing and heterophily issues, inspired by physics-informed insights~\cite{han2023continuous}.
    Here, we focus on neural diffusion-based methods to address graph heterophily.
    
    \noindent \textbf{Non-smooth Diffusion.}
    Previous works~\cite{chamberlain2021grand,chamberlain2021beltrami,thorpe2022grand++} have sought to extend the isotropic diffusion in GCN~\cite{kipf2016semi} to anisotropic diffusion, aiming to enhance the expressive power.
    However, the underlying mechanism of heat diffusion is still susceptible to the challenge of graph heterophily.
    Therefore, GIND~\cite{chen2022optimization} extends the linear isotropic diffusion to a more expressive nonlinear diffusion mechanism, thereby avoiding the aggregation of noisy information by excluding dissimilar neighbors.
    PDE-GCN~\cite{eliasof2021pde} employs the oscillation, a non-smooth dynamical system, to enhance the neural diffusion process. 
    Similarly, GraphCON~\cite{rusch2022graph} considers a more general oscillatory dynamics which combines a damped oscillating process with a coupling function. Approaches based on non-smooth dynamics can, to some extent, alleviate the over-smoothing problem, and also exhibit superior performance in solving the heterophily issue.
    
    \noindent \textbf{Diffusion with External Forces.}
    However, single dynamic processes, such as diffusion or oscillation, may still have limitations when dealing with heterophily.
    In fact, we can enhance the dynamic processes by explicitly  introducing external forces, such as convection, advection, and reaction, to adapt better to graph heterophily.
    For example, CDE~\cite{zhao2023graph} incorporates the principle of heterophily by modeling the flow of information on nodes through the convection-diffusion equation.  
    Building upon this, the homophilic ``diffusion'' and the heterophilic ``convection'' are effectively combined to capture complex patterns in graphs.
    ACMP~\cite{wang2022acmp} further enhances the diffusion process by incorporating the Allen-Cahn reaction term~\cite{allen1979microscopic}, thus forming a reaction-diffusion process. 
    The reaction process in ACMP is implemented through the repulsive force between nodes, which can be interpreted as modeling a negative diffusion coefficient to explicitly model heterophily. 
    GREAD~\cite{choi2023gread} proposes a more general reaction-diffusion dynamics framework, integrating a variety of dynamical processes, with the reaction term selected from diverse domains. 
    For instance, Fisher reaction~\cite{fisher1937wave} is used to describe the spreading of biological populations; Zeldovich reaction~\cite{gilding2004travelling} is a generalized equation that describes the phenomena in combustion theory; Allen-Cahn reaction~\cite{allen1979microscopic} is also included in the framework of GREAD.
    Based on GREAD, ADR-GNN~\cite{eliasof2023adr} incorporates an explicit advection term into the reaction-diffusion process, thus forming the advection-diffusion-reaction process. 
    Moreover, FLODE~\cite{maskey2024fractional} proposes the fractional heat equation for modeling anomalous diffusion processes. 
    With the capability to capture non-local interactions, the framework incorporating the fractional Laplacian is able to establish long-range dependencies, making it well-suited for heterophilic graphs.
    In summary, these methods aim to introduce external force terms into the diffusion process, conceptually akin to a high-pass filter, with the goal of inducing a sharpening effect to address the heterophily issue.

    \noindent \textbf{Diffusion with Modulators.}
    Different from imposing external forces, 
    modulating the diffusion process, such as using a gating mechanism, can to some extent alleviate the heterophily issues~\cite{han2023continuous}.
    For example, G$^2$~\cite{rusch2022gradient} proposes that controlling the speed of diffusion can counteract the over-smoothing and address graph heterophily. 
    Hence, this framework explicitly introduces a gating function to modulate the neural diffusion. 
    The concept of gating has been similarly explored in EGLD~\cite{zhang2024unleashing}. 
    EGLD introduces a dual-channel neural diffusion framework by conducting low-pass and high-pass filtering, and further introduces a dimension-level gate to coordinate the obtained representations. 
    In contrast to dimension-level gating,
    MHKG~\cite{shao2023unifying} introduces a filter-level gating strategy for modulation based on the reverse heat kernel. 
    Specifically, MHKG integrates high-pass and low-pass filters into the heat diffusion process, and controls the smoothing and sharpening of signals through weights of filter gates. 
    Moreover, A-GGN~\cite{gravina2022anti} imposes stability and conservation constraints to the diffusion process through anti-symmetric weight matrices, thereby enhancing the capability to capture long-range dependencies between nodes. 
    
    \noindent \textbf{Other Explorations.}
    In addition to the aforementioned methods, we briefly outline other relevant works below.
    From the perspective of higher-order geometries, some studies~\cite{bodnar2022neural,barbero2022sheaf} address heterophily through sheaf Laplacian~\cite{hansen2019toward} diffusion.
    Inspired by quantum diffusion, QDC~\cite{markovich2023qdc} proposes a novel graph convolution kernel and improves message passing with quantum mechanical kernel.
    Giovanni et al.~\cite{di2022understanding} attempt to understand graph convolution via energies, and Zhang et al.~\cite{zhang2023steering} explore steering GNNs with pinning control~\cite{yu2013synchronization}.
    Park et al.~\cite{parkmitigating} find that the reverse process of diffusion produces more distinguishable representations for heterophilic graphs.
    FGND~\cite{wan2024flexible} combines latent class representation learning with inherent graph topology to reconstruct the diffusion matrix during neural diffusion.
    From the macro perspective, a general diffusion framework is proposed in~\cite{li2024generalized}, which formally establishes the relationship between the diffusion process with many classic GNNs.
    Currently, various GNNs based on neural diffusion begin to flourish, with more and more studies taking graph heterophily into account. 
    However, pitfalls of model stability and robustness, gradient vanishing and explosion are still obstacles to overcome~\cite{han2023continuous}.
    
    \section{Advanced Learning Paradigms}
    \label{sec6}
    With the advancement of deep learning techniques, real-world demands are driving the emergence of paradigms that go beyond supervised learning.
    This section elaborates self-supervised and prompt learning paradigms, which have been attracting increasing popularity across various research fields.

    \subsection{Self-Supervised Learning}
    As listed above, learning from heterophily in a supervised manner has achieved remarkable success. 
    However, supervised learning requires a sufficient amount of labeled data for training, leading to high annotation costs and time consumption.
    Recently, Self-Supervised Learning (SSL)~\cite{jaiswal2020survey,jing2020self,liu2021self,gui2024survey} has emerged as a new paradigm for fully utilizing unlabeled samples, and there is a growing trend to extend such paradigm to graphs with heterophily.
    
    \noindent \textbf{Contrastive Learning.} 
    Graph Contrastive Learning (GCL) stands out as one of the most extensively studied SSL methods currently.
    The core idea of GCL is to bring positive samples closer and push negative samples further apart in the latent space.
    Most GCL methods follow the homophily assumption, either explicitly or implicitly~\cite{guo2023architecture}, limiting the applicability to heterophilic scenarios.
    Therefore, some GCL studies have been proposed for learning from graphs with heterophily. 

    \noindent \emph{-Spectral Filters.}
    Given the success of spectral filters in supervised learning, HLCL~\cite{yanggraph} attempts to extend this powerful tool to GCL.
    This framework first identifies a homophilic and a heterophilic subgraph based on feature similarity. 
    For each subgraph, it generates two augmented views and applies a high-pass filter to the heterophilic subgraph and a low-pass filter to the homophilic subgraph, respectively.
    After performing contrast in dual views, HLCL uses the low-pass filtered representations as the final output.
    GREET~\cite{liu2023beyond} employs an edge discriminator to distinguish heterophilic and homophilous edges, and then constructs the dual views through low-pass and high-pass filters. 
    To jointly learn edge distinction and node representations, it introduces an alternating training strategy for iterative optimization.
    Moreover, PolyGCL~\cite{chen2024polygcl} explores the powerful expressive capabilities of various polynomial spectral filters in GCL, and further integrates the low-frequency information with high-frequency information through linear combination.

    \noindent \emph{-Data Augmentation.} 
    Data augmentation has been proven to be an effective approach to enhance the performance of GCL~\cite{velivckovic2018deep,you2020graph,zhu2020deep,guo2023architecture}.
    However, simple random augmentation on heterophilic graphs does not effectively enhance GCL, necessitating more advanced augmentation strategies. 
    The representative method SimGCL~\cite{liu2024simgcl} directly computes the feature similarity and local feature assortativity~\cite{yang2021diverse} to perform pre-computed augmentation, while HGRL~\cite{chen2022towards} introduces structural learning augmentation based on feature similarity.
    Moreover, HeterGCL~\cite{wanghetergcl} proposes that random structure augmentation can lead to the destruction of the topology, and introduces an adaptive aggregation strategy to establish connections from high-order neighbors, explore structural information with an adaptive local-to-global contrastive loss.
    Further, GASSER~\cite{yang2024gauss} injects perturbations to specific frequencies in the spectral domain, with edge perturbation selectively guided by spectral hints. 
    The augmentation strategy is adaptive and controllable, as well as heuristically fitting the homophily ratios and spectrum of the graph.
    However, there are methods that question the necessity of data augmentation. 
    For example, SP-GCL~\cite{wang2022single}, AF-GCL~\cite{wang2022augmentation} and GraphACL~\cite{xiao2024simple} do not advocate the use of data augmentation in GCL and propose augmentation-free GCL architectures. 
    Whether data augmentation is necessary and how to conduct data augmentation under heterophily are topics worth exploration.
    
    \noindent \emph{-Multi-view Contrast.} 
    Due to complex patterns of graph heterophily, constructing multiple views in GCL to capture different hierarchical aspects of graphs becomes one of the solutions.
    For example, Khan et al.~\cite{khan2023contrastive} employ diffusion wavelets~\cite{coifman2006diffusion} to create augmented-view graphs, and utilize a multi-view contrast for learning invariant representations. 
    The diffusion wavelet filters can capture the band-pass response of graph signals, and explicitly highlight the higher-order information encoded within the graph.
    Realizing the importance of node attributes in heterophilic graphs, MUSE~\cite{yuan2023muse} constructs semantic and contextual views to capture the information of the ego node and its neighborhood. 
    Instead of simply combining multiple views, MUSE fuses the representations of dual views via a fusion controller. Emphasizing the semantic information of nodes significantly enhances the performance on heterophilic graphs.
    
    \noindent \emph{-Capture Monophily.} 
    Monophily is a commonly observed phenomenon in real-world graphs, for example, the attributes of a node’s friends are likely to be similar to the attributes of that node’s other friends~\cite{altenburger2018monophily}. 
    Intuitively speaking, monophily essentially describes the similarities between two-hop neighbors. 
    Inspired by this, GraphACL~\cite{xiao2024simple} presents a graph asymmetric contrastive framework, and proves that the asymmetric design can capture one-hop neighborhood context and two-hop monophily similarities. 
    Based on GraphACL, a more efficient version named GraphECL~\cite{xiaoefficient} has been proposed to enable fast inference for both homophilic and heterophilic graphs.
    S3GCL~\cite{wans3gcl} introduces a cosine-parameterized Chebyshev polynomial filters to enhance GCL, and establishes a cross-pass GCL objective between full-pass MLP and biased-pass graph filters.
    The principle of S3GCL is similar to GraphACL, both of them treat the neighboring nodes as positive pairs, eliminate the need for random augmentation, and capture the monophily on graphs beyond the homophily assumption.
    
    \noindent \emph{-Homophily Enhanchment.}
    Since graph heterophily can to some extent restrict the performance of GCL, we consider enhancing homophily to address this issue.
    NeCo~\cite{he2023contrastive} first demonstrates that the presence of intra-class edges impacts the performance of GCL. To address this, it integrates the positive neighbor sampling of GCL and and the homophily discrimination of GNNs into the same framework. 
    By removing inter-class edges and enhancing homophily during training, the performance of GCL on both heterophilic and homophilic graphs can be improved.
    Recognizing that homophily affects GCL, 
    HomoGCL~\cite{li2023homogcl} introduces soft clustering to discover potential positive and negative samples from neighbors.
    Moreover, HEATS~\cite{zhuo2024improving} and ROSEN~\cite{zhuo2024graph} both learn an affinity matrix in an unsupervised manner to capture global homophily beyond local affinity. 
    Therefore, how to measure heterophily or homophily on graphs without supervision will be an interesting topic.

    \noindent \emph{-Alleviate Ambiguity.}
    Research indicates that GNNs can result in ambiguous node representations due to the neighborhood aggregation, particularly in heterophilic graphs~\cite{luan2022revisiting, yan2022two}.
    DisamGCL~\cite{zhao2024disambiguated} first combines the issue of graph heterophily with the ambiguity problem of GNNs, and proposes that disambiguation can improve the performance of GNNs under heterophilic scenarios. 
    This framework introduces a memory cell to efficiently identify ambiguous nodes, and disambiguate these nodes by augmenting the learning process with a contrastive learning objective.  
    
    \noindent \textbf{Graph Auto-Encoders.}
    One objective of Graph Auto-Encoder (GAE) is to obtain low-dimensional embeddings through graph reconstruction for subsequent tasks. 
    Existing GAEs are mostly designed to reconstruct the direct links~\cite{kipf2016variational,li2023s}, so models trained in this way implicitly follow the homophily assumption and perform poorly when graphs exhibit heterophily.
    Since the pretext of link reconstruction is not reasonable under heterophily, SELENE~\cite{zhong2022unsupervised}, MVGE~\cite{lin2023multi}, and AGCN~\cite{li2024redundancy}  reconstruct node attributes and network structure in parallel to build heterophily-adapted GAEs.
    Moreover, PairE~\cite{li2022graph} introduces aggregated feature and assortativity versus reconstruction as extra pretexts to retain high-frequency signals between nodes.
    NWR-GAE~\cite{tang2022graph} proposes that the over-simplification of link reconstruction leads to a significant loss of information, thereby providing suboptimal performance in downstream tasks. 
    Hence, they propose a novel graph decoder to reconstruct the neighborhood information regarding both proximity and structure through Neighborhood Wasserstein Reconstruction (NWR). 
    By considering the proximity, structure and feature information, NWR-GAE distinguishes itself from other GAEs by performing well in the presence of heterophily.
    Recently, a large number of studies have attempted to combine GAE with GCL to construct more powerful SSL models~\cite{tian2024ugmae,luo2024masked,fang2024masked,yang2023cmgae}. 
    However, these efforts are largely limited to homophilic scenarios and overlook the existence of heterophily. 
    How to naturally integrate learning from heterophilic graphs using both SSL approaches is a direction worth exploring. 
    
    \noindent \textbf{Decoupled Learning.}
    Xiao et al.~\cite{xiao2022decoupled} propose that the semantic structure of graphs can be decoupled into latent variables that capture different aspects of node similarities, including attribute, label, and link similarity. 
    To leverage the heterophilic patterns for SSL, they introduce a Decoupled Self-Supervised Learning (DSSL) framework.
    DSSL imitates the generative process of nodes and links through latent variable modeling of the semantic structure, decoupling different underlying semantics between different neighborhoods into the SSL process.
    By decoupling the local diverse neighborhood context, DSSL does not rely on graph augmentations and downstream labels, and can easily handle the heterophily patterns on graphs.
    
    \subsection{Prompt Learning}
    Originating from the field of NLP, the ``pre-training, prompt-tuning'' paradigm~\cite{liu2023pre} reformulates various downstream tasks into an unified template of the pretext, and designs specific prompts for downstream adaptation. 
    Prompt learning fully unleashes the potential of pre-trained models, where adjusting only a few parameters can achieve excellent results even in few-shot settings. 
    Inspired by the success of prompt learning in NLP~\cite{brown2020language,wei2021finetuned} and CV~\cite{jia2021scaling,jia2022visual}, the graph domain has begun to shift the focus towards the “pre-training, prompting” paradigm~\cite{sun2023graph,long2024towards}.
    
    \noindent \textbf{Recent Advance.}  
    Since prompt learning in the graph domain is still in its early stage of development, we first survey the existing advances.
    
    \noindent \emph{-Unified Frameworks.}
    The core of prompt learning is the unified task framework, which is used to prevent ``negative transfer''~\cite{wang2021afec} between the pre-training pretext and downstream tasks.
    GPPT~\cite{sun2022gppt} first introduces the concept of prompt learning into graph learning, and presents an unified prompt framework based on link prediction.
    Following GPPT, GraphPrompt~\cite{liu2023graphprompt,yu2024generalized} and Prodigy~\cite{huang2024prodigy} both utilize the unified task template based on the subgraph similarity calculation.
    From another perspective, SGL-PT~\cite{zhu2023sgl} chooses to package various graph tasks as node generation,
    while ProG~\cite{sun2023all,zi2024prog} reformulates tasks at different levels to graph level and introduces meta-learning to boost multi-task learning. 
    Moreover, OFA~\cite{liu2023one} attempts to provide a general solution for building and training a foundation GNN model with in-context learning ability across different domains.

    \noindent \emph{-Prompt Strategies.}
    In addition to the unified framework, there are also studies dedicated to designing prompt strategies that are specifically tailored to graph data.
    For example, GPF~\cite{fang2024universal} injects learnable perturbations into the feature space to adapt the pre-trained model to a certain downstream task.
    VNT~\cite{tan2023virtual} inserts a set of virtual nodes into the graphs as prompt, while ProG~\cite{sun2023all,zi2024prog} and SUPT~\cite{lee2024subgraph} introduce virtual prompt graphs.
    GraphPrompt~\cite{liu2023graphprompt,yu2024generalized} utilizes the prompt token to efficiently adjust the model output, and MultiGPrompt~\cite{yu2024multigprompt} introduces multi-task pre-training to enhance prompt learning, further designing multiple pretext tokens to avoid mutual negative influence.
    In addition, other works, including GSPF~\cite{jiang2024unified}, IGAP~\cite{yan2024inductive}, ULTRA-DP~\cite{chen2023ultra}, and TGPT~\cite{wang2024novel} are all attempting to design prompts from more diverse views to match downstream tasks.

    \noindent \emph{-Various Extensions.}
    Recognizing the powerful capabilities of graph prompt learning, related extensions or variants have emerged in both the graph domain and other fields. 
    Considering the complex heterogeneous relations on graphs,
    HetGPT~\cite{ma2024hetgpt} and HGPrompt~\cite{yu2024hgprompt} are dedicated to harnessing the power of prompt tuning in pre-trained heterogeneous GNNs.
    To depict the dynamic relations between objects in graphs, DyGPrompt~\cite{yu2024dygprompt} presents a novel pre-training and prompting framework for dynamic graph modeling.
    Krait~\cite{song2024krait} reveals that backdoor disguise can benign graph prompts, and CrossBA~\cite{lyu2024cross} conducts investigations to assess the feasibility of backdoor attacks in cross-context graph prompt.
    Not only limited to the graph domain, graph prompting techniques are making significant progress in fields such as drug prediction~\cite{wang2024ddiprompt}, text-attribute graphs~\cite{liu2023one,ye2023natural,duan2024g,fang2024gaugllm,jiang2024killing}, urban computing~\cite{jin2024urban} and recommendation systems~\cite{zhang2024gpt4rec}.
    
    \noindent \textbf{Prompt Learning on Heterophilic Graphs.} 
    Despite the flourishing development of graph prompt learning, there is a lack of work focusing on the impact of graph heterophily.
    Self-Pro~\cite{gong2024self} first pays attention to the heterophily issue in graph prompt learning. 
    To accommodate both homophilic and heterophilic graphs, Self-Pro introduces asymmetric graph contrastive learning for pre-training, and unifies the pretext and downstream tasks to avoid negative knowledge transfer~\cite{wang2021afec}. 
    Through the self-adapter and semantic prompt injection, Self-Pro can perform well under the few-shot setting without introducing additional parameters. 
    Moreover, ProNoG~\cite{yu2024non} revisits existing pre-training methods on heterophilic graphs and introduces some non-homophily pre-training methods.
    For downstream adaptation, it proposes the condition-net~\cite{zhou2022conditional} to generate a series of prompts conditioned on various heterophilic patterns.
    From the structural perspective, PSP~\cite{ge2023enhancing} introduces virtual nodes in the prompting phase, enabling downstream tasks to benefit from topological patterns. 
    By decoupling node attributes from structure~\cite{lim2021new} during the pre-training phase, PSP can to some extent address the graph heterophily issue.

    \begin{figure*}[ht]
    \centering
    \begin{forest}
    for tree={
    grow'=east,
    anchor=west,
    parent anchor=east,
    child anchor=west,
    edge path={
    \noexpand\path[\forestoption{edge},-, >={latex}] 
    (!u.parent anchor) -- +(5pt,0pt) |- (.child anchor)
    \forestoption{edge label};
    }
    }
    [Broader Topics, root
    [Diversified Learning Tasks, a
        [Node Clustering, b
            [Spectral Filters, c
                [{CGC~\cite{xie2023contrastive},
                  AHGFC~\cite{wen2024homophily}
                }, e
                ]
            ]
            [Advanced Reconstruction, c
                [{DGCN~\cite{pan2023beyond},
                  SELENE~\cite{zhong2022unsupervised},
                  PFGC~\cite{xie2024provable},
                  PLCSR~\cite{zhu2024boosting}
                }, e
                ]
            ]
            [Homophily Enhancement, c
                [{HoLe~\cite{gu2023homophily}
                }, e
                ]
            ]
        ]
        [Link Prediction, b
            [{DisenLink~\cite{zhou2022link},
             GRAFF-LP~\cite{di2024link}
             }, d
            ] 
        ] 
        [Graph Classification, b
            [{IHGNN~\cite{yang2022incorporating}
             }, d
            ] 
        ] 
    ]
    [Model Scalability, a
        [Message Passing Framework, b
            [{LINKX~\cite{lim2021large},
              GLINX~\cite{papachristou2022glinkx},
              LD$^{2}$~\cite{liao2024ld2},
              HopGNN~\cite{chen2023node},
              AGS-GNN~\cite{das2024ags}
                }, d
            ]
        ]
        [Graph Transformer, b
            [{NodeFormer~\cite{wu2022nodeformer},
              DifFormer~\cite{wu2023difformer},
              SGFormer~\cite{wu2024simplifying},
              GOAT~\cite{kong2023goat},
              SpikeGraphormer~\cite{sun2024spikegraphormer}
             }, d
            ] 
        ]
    ]
    [Adversarial Attack and Robustness, a
        [Adversarial Attacks, b
            [{Zhu et al.~\cite{zhu2022does},
              Mid-GCN~\cite{huang2023robust},
              NSPGNN~\cite{zhu2024universally},
              EvenNet~\cite{lei2022evennet},
              LHS~\cite{qiu2024refining},
              PROSPECT~\cite{dengprospect}
                }, d
            ]
        ]
        [Resisting Noise, b
            [{R$^2$LP.~\cite{cheng2024resurrecting},
             }, d
            ] 
        ]
        [Model Privacy, b
            [{GPS~\cite{yuan2024unveiling},
              COSERA~\cite{wu2024provable},
              Mueller et al.~\cite{mueller2023privacy}
             }, d
            ] 
        ]
        [Model Fairness, b
            [{Loveland et al.~\cite{loveland2022graph}
             }, d
            ] 
        ]
    ]
    [Graph Structure Learning, a
        [Edge Discriminator, b
            [{DC-GNN~\cite{xue2024data},
             GNN-SATA~\cite{yang2024graph},
             ECG-GNN~\cite{deac2023evolving}
             }, d
            ] 
        ]
        [Dual Views, b
            [{GOAL~\cite{zheng2023finding},
              ATL~\cite{xu2023node},
             }, d
            ] 
        ]
        [Neighborhood Similarity, b
            [{DHGR~\cite{bi2024make},
              Choi et al.~\cite{choi2022finding}
             }, d
            ] 
        ]
        [Spectral Clustering, b
            [{GCN-SL~\cite{jiang2021gcn},
              Li et al.~\cite{li2023restructuring}
             }, d
            ] 
        ]
        [Probabilistic Modeling, b
            [{GEN~\cite{wang2021graph},
              L2A~\cite{wu2023learning}
             }, d
            ] 
        ]
        [Self-supervised Manner, b
            [{GPS~\cite{dong2023towards},
              HES-GSL~\cite{wu2023homophily},
              SUBLIME~\cite{liu2022towards},
              GSSC~\cite{wu2024learning}
             }, d
            ] 
        ]
        [Benchmarks, b
            [{OpenGSL~\cite{zhiyao2024opengsl},
              GSLB~\cite{li2024gslb}
             }, d
            ] 
        ]
    ]
    ]
    \end{forest}
    \caption{Taxonomy of broader topics related to graph heterophily with representative examples.}
    \label{topic}
    \end{figure*}
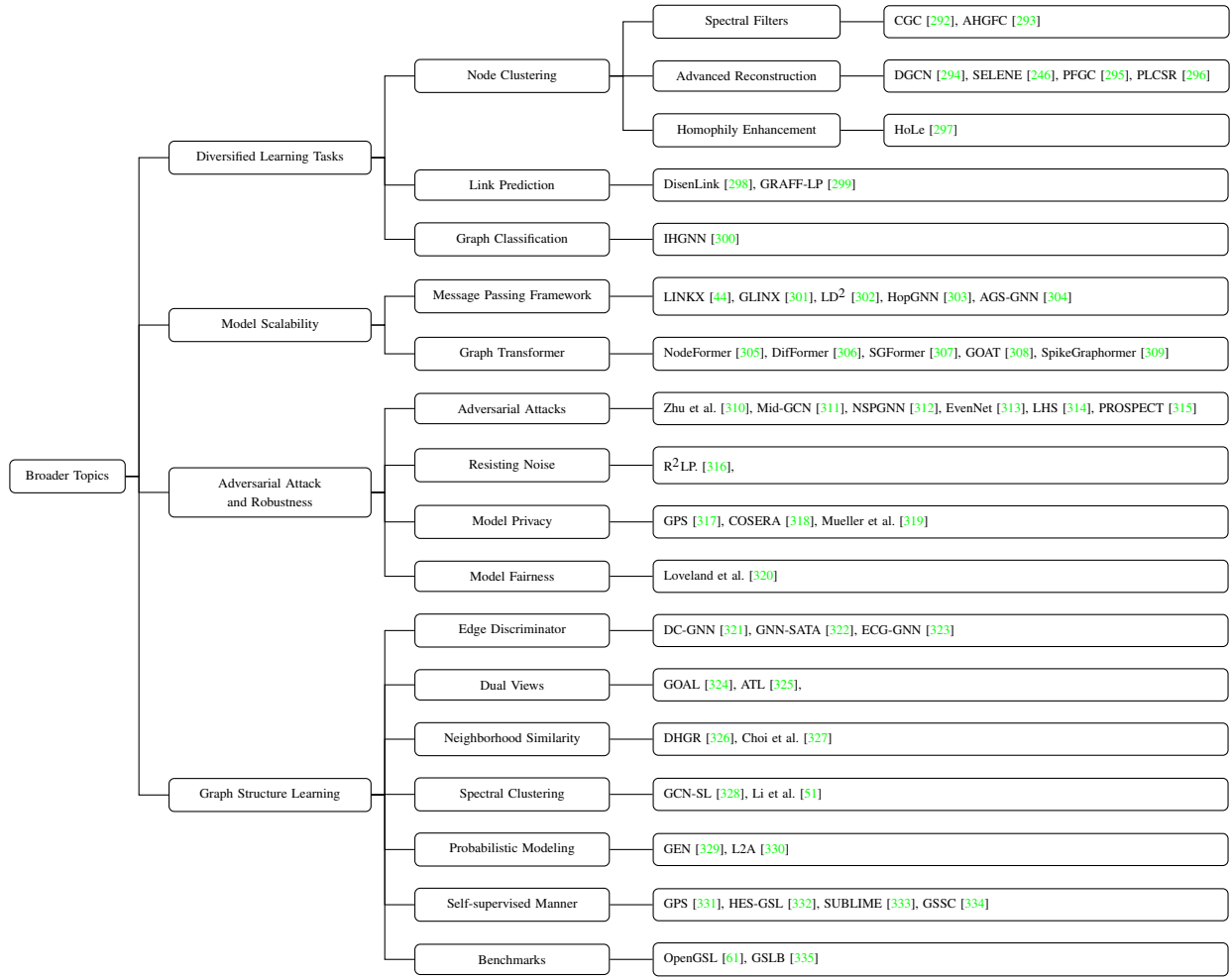
    
    \section{Broader Topics}
    \label{sec7}
    With the development of learning from heterophilic graphs, alongside notable GNN models and learning paradigms, there are also extensible topics worth attention, as shown in Figure~\ref{topic}.
    In this section, we aim to introduce broader topics that are also important components of heterophilic graph analysis.
    
    \subsection{Diversified Learning Tasks}
    The formats of graph tasks can be categorized into three classes: node-level, edge-level, and graph-level tasks. 
    The majority of heterophilic graph studies have been focusing on the node-level tasks, especially node classification. 
    As the heterophily issue has received increasing attention, researchers have begun to explore the potential of other graph tasks.
    
    \noindent \textbf{Node Clustering.} 
    Node clustering is the task of grouping similar nodes into the same category and dissimilar nodes into different categories in an unsupervised manner~\cite{wang2023overview}.
    Deep graph learning has facilitated the emergence of a large number of advanced graph clustering methods~\cite{tian2014learning,kipf2016variational,pan2021multi,zhao2021graph}. 
    Similar to node classification, the graph heterophily issue in node clustering has also drawn attention.

    \noindent \emph{-Spectral Filters.}
    Spectral graph filters have demonstrated their strong capabilities in heterophilic scenarios.
    To mine the heterophily patterns, CGC~\cite{xie2023contrastive} designs an adaptive filter that can automatically learn a suitable filter for node clustering under both homophilic and heterophilic settings, mining comprehensive information beyond low-frequency components. 
    Moreover, AHGFC~\cite{wen2024homophily} designs a hybrid graph filter based on joint aggregation with node features and adjacency relationships to make the low and high-frequency signals on graphs more distinguishable.
    
    \noindent \emph{-Advanced Reconstruction.}
    Under the unsupervised setting, most methods utilize the GAEs to obtain node embeddings for clustering.
    Considering the complex link patterns, 
    it is insufficient to only reconstruct the structural links. 
    Therefore, many methods consider employing advanced reconstruction techniques.
    For example, DGCN~\cite{pan2023beyond} utilizes dual encoders to separately map node features and graph structure into two low-dimensional spaces, and further fuses them for feature reconstruction.  
    SELENE~\cite{zhong2022unsupervised} restructures both node attributes and graph structure, and utilizes a dual-channel contrast between attributes and structure to enhance the discrimination of inter-class nodes. 
    After that, PFGC~\cite{xie2024provable} points out that potential homophily may exist in multi-hop neighborhood, so it suggests additionally reconstructing the high-order topology.
    Further, PLCSR~\cite{zhu2024boosting} integrates curriculum learning and contrastive learning to enhance GAEs for node clustering under graph heterophily.

    \noindent \emph{-Homophily Enhancement.}
    From a data-centric perspective, HoLe~\cite{gu2023homophily} observes that enhancing graph homophily can significantly improve node clustering. 
    To this end, HoLe proposes a structure homophily-enhanced method for node clustering. 
    By removing inter-class edges or adding intra-class edges, graph structure learning and node clustering will mutually reinforce and optimize each other. 

    \noindent \textbf{Link Prediction.} 
    The target of link prediction is to infer missing links or predict potential links based on the input graphs~\cite{lu2011link}. 
    GNN-based link prediction methods have achieved state-of-the-art performance~\cite{kipf2016variational,zhang2018link,li2023seegera,tan2023s2gae,li2023s}, but most of them adhere to the graph homophily assumption, ignoring potentially complex patterns. 
    Therefore, link prediction on heterophilic graphs is a promising direction that deserves further exploration.
    Zhou et al.~\cite{zhou2022link} first introduce link prediction into heterophilic graphs, and propose that connected nodes with low feature similarity may be similar in some latent factors.
    They propose the DisenLink framework to model the complex link patterns in heterophilic graphs by disentangled views, and learn disentangled representations by discovering edge factor with selection and factor-aware message passing. 
    Sharing the same topic, GRAFF-LP~\cite{di2024link} introduces GRAFF~\cite{di2022understanding}, a physics-inspired GNN, to enhance link prediction under heterophily. 
    Thanks to the incorporation of physics biases in message passing, these priors can improve the performance of link prediction.

    \noindent \textbf{Graph Classification.}
     In addition to node-level tasks, graph-level tasks are also closely related to heterophily. 
     Different from node-level tasks, graph classification requires adaptation to numerous graphs with varying homophily ratios. 
     Therefore, uniform aggregation and simple readout functions, 
     such as sum, ignore graph heterophily, leading to performance degradation~\cite{yang2022incorporating}.
     Inspire by H$2$GCN~\cite{zhu2020beyond}, IHGNN~\cite{yang2022incorporating} introduces the separation of the ego and neighbor representation of nodes and designs an adaptive aggregation strategy for different layers. 
     For graphs with varying node homophily ratios, IHGNN designs a graph-level readout function to reorganize nodes within each graph and align them across graphs. 
     To verify whether heterophily affects the performance of graph classification, Ding et al.~\cite{ding2023self} perform graph classification using spectral GNNs on molecular and protein datasets, and find that FAGCN~\cite{bo2021beyond} and ChebNet~\cite{defferrard2016convolutional} outperform vanilla GCN, indicating that high-frequency signals are also critical for graph classification.
    
    \subsection{Model Scalability}
    When dealing with graph heterophily, we inevitably encounter large-scale graphs, so there is an urgent need to enhance the model scalability. 
    Here, we categorize existing methods into two types based on the model architecture: Message Passing Framework and Graph Transformer.
    
    \noindent \textbf{Message Passing Framework.}
    The early model LINKX~\cite{lim2021large} separately embeds the adjacency matrix and node feature matrix using MLPs, and then combines the embeddings through concatenation.
    This simple method allows for straightforward mini-batch training and inference, demonstrating good performance on large-scale heterophilic datasets.
    Building on LINX, GLINX~\cite{papachristou2022glinkx} adds PEs and monophilous propagation to further improve the performance.
    LD$^{2}$~\cite{liao2024ld2} generates node embeddings from the adjacency matrix and node feature matrix in the pre-computation stage, and then applies multi-hop discriminative propagation with simple neural networks to learn node representation.
    Theoretical and empirical results shows that this scalable model has a time complexity $O(N)$ linear to the number of nodes. 
    HopGNN~\cite{chen2023node} enhances model scalability by pre-computing neighborhood information at each hop, then models interactions between multi-hops, fully utilizing high-order neighbors to enable generalization under heterophily.
    AGS-GNN~\cite{das2024ags} introduces an attribute-guided sampling strategy to scale models to large-scale graphs. Specifically, AGS-GNN pre-computes the sampling distribution, and then generates two subsets of neighbors based on feature-similarity and feature diversity. 
    Through neighbourhood pruning and group aggregation, AGS-GNN can perform well on both homophilic and heterophilic graphs,  achieving the desired scalability.
    
    \noindent \textbf{Graph Transformer.}
    One of the critical challenges for GTs is the scalability issue due to quadratic complexity, which is prohibitive for large-scale graphs.
    NodeFormer~\cite{wu2022nodeformer} introduces a kernelized Gumbel-Softmax operator to reduce the complexity of all-pair message passing to linear. 
    Thanks to efficiently propagation between arbitrary node pairs, NodeFormer demonstrates its promising potential for tackling heterophily, long-range dependencies and large-scale graphs.
    Inspired by the neural diffusion, DifFormer~\cite{wu2023difformer} attempts to elucidate the relationship between energy-driven diffusion and GTs. 
    This diffusion-based Transformer also needs to compute pairwise diffusivity with quadratic complexity, so an acceleration strategy through state updating is proposed to reduce the complexity to linear. 
    To further enhance the model scalability, SGFormer~\cite{wu2024simplifying} modifies the multi-head attention module in GTs to a single-layer and single-head attention module. 
    Combined with a simple local propagation, SGFormer retains the necessary expressiveness without PEs or SEs, data pre-processing, extra loss functions, or edge embeddings.
    Targeting at addressing the complexity issue, GOAT~\cite{kong2023goat} introduces dimension reduction based on the EMA K-Means algorithm, and proves that this approximate has bounded error compared to the global attention mechanism.
    Moreover, Spiking Neural Networks (SNNs)~\cite{ghosh2009spiking,tavanaei2019deep}, with event-driven and binary spikes properties, can perform energy-efficient computations. 
    In light of this, SpikeGraphormer~\cite{sun2024spikegraphormer} attempts to integrate SNNs into GTs, enabling all-pair node interactions on large-scale graphs with limited GPU memory.

    \subsection{Adversarial Attack and Robustness}
    Studying adversarial attacks~\cite{chakraborty2018adversarial} and improving model robustness~\cite{silva2020opportunities,xu2021robustness} can help us better understand the working principle of models, enhance the model performance and credibility when facing different environments. 
    Currently, the field of graph learning is also focusing on this topic, with heterophily being one of the primary focuses.

    \noindent \textbf{Adversarial Attacks.}
    Recent studies~\cite{zugner2018adversarial,dai2018adversarial} show that GNNs are sensitive to adversarial attacks: intentionally introduced minor changes in graph structure can lead to significant performance degradation.
    Zhu et al.~\cite{zhu2022does} first investigate the relationship between graph heterophily and GNN robustness against structural attacks. 
    They find that on homophilic graphs, effective structural attacks lead to increased heterophily, whereas on heterophilic graphs, attacks alter the homophily level contingent on node degrees. 
    Moreover, they propose that separating ego and neighbor representation can improve the robustness of GNNs against adversarial attacks.
    Combining with spectral graph theory,
    Huang et al.~\cite{huang2023robust} design a mid-pass filtering GCN model named Mid-GCN, leveraging the robustness of middle-frequency signals against adversarial attacks.
    NSPGNN~\cite{zhu2024universally} employs low-pass and high-pass filters on positive kNN and negative kNN graphs to enhance the robustness of GNNs.
    Lei et al.~\cite{lei2022evennet} point out that ignoring odd-hop neighbors can improve the robustness of GNNs and further present EvenNet, a simple yet effective spectral GNN based on the even-polynomial filter.
    Qiu et al.~\cite{qiu2024refining} further uncover that the predominant vulnerability on heterophilic graphs is caused by the structural out of-distribution (OOD) issue. 
    Therefore, they present LHS, a framework that strengthens GNNs against various attacks by refining latent homophilic structures over heterophilic graphs. 
    Equipped with the new adaptable structure, LHS can effectively mitigate the structural OOD threats over heterophilic graphs.
    Inspired by knowledge distillation, Deng et al.~\cite{dengprospect} introduce the MLP-to-GNN distillation framework for learning models robust against adversarial structure attacks on both homophilic and heterophilic graphs. 
    The 
    analysis indicates that Prospect-MLP can correct the wrong knowledge of Prospect-GNN regardless of homophily ratios, endowing the adversarial robustness and clear accuracy improvements.

    \noindent \textbf{Resisting Label Noise.}
    In addition to adversarial attacks, data noise can also significantly degrade the generalization ability of GNNs.
    Cheng et al.~\cite{cheng2024resurrecting} first study the impact of label noise in the context of arbitrary heterophily, and find that a high homophily rate can indeed mitigate the effect of graph label noise on GNN performance. 
    Motivated by this, they propose R$^2$LP to incorporate graph homophily reconstruction and noisy label rectification through label propagation. 
    Specifically, R$^2$LP iteratively performs graph reconstruction with homophily, label propagation for noisy label refinement and high-confidence sample selection over multi-rounds.
    
    \noindent \textbf{Model Privacy.}
    With the homophily property and message passing mechanism of GNNs, publicly available user information provides  adversaries with opportunities to infer private attributes, leading to privacy breaches~\cite{he2021node,liao2021information,hu2022learning}.
    Therefore, developing privacy-preserving GNN models to resist inference attacks is of significant importance.
    Yuan et al.~\cite{yuan2024unveiling} first investigate the problem of Graph Privacy Leakage via Structure (GPS), and introduce the heterophily metrics to quantify the various mechanisms contributing to privacy breach risks. 
    To counter privacy attacks, they also propose a graph data publishing method that employs learnable graph sampling, making the sampled graph suitable for publication. 
    Moreover, there are also some other recent privacy-preserving methods~\cite{wu2024provable,mueller2023privacy} that focus on the graph heterophily issue.
        
    \noindent \textbf{Model Fairness.}
    Unfortunately, the graph homophily assumption fails to account for local deviations where unfairness may impact certain groups, potentially amplifying unfairness~\cite{dai2021say,li2021dyadic}.
    Loveland et al.~\cite{loveland2022graph} are the first to consider graph heterophily to improve fair learning, as the connection between homophily and fairness has been underexplored.
    They demonstrate that adopting heterophily-specific GNNs can address the disassortative group labels and promote group fairness in graphs characterized by both heterophily and homophily.
    
    \subsection{Graph Structure Learning}
    In pursuit of an optimal graph structure for downstream tasks, recent studies have sparked efforts in Graph Structure Learning (GSL), which aims to learn an optimized structure and corresponding node representations jointly~\cite{zhu2021deep}.
    The heterophily introduces complex link patterns on graphs, posing significant challenges to GSL and attracting considerable attention.

    \noindent \textbf{Edge Discriminator.}
    Widely used in GSL, identifying heterophilic or noisy edges has been shown to enhance model performance~\cite{ye2021sparse}. 
    A very natural idea is to train a discriminatior to distinguish heterophilic edges.
    For example, DC-GNN~\cite{xue2024data} introduces a learnable edge classifier to transform the original heterophilic graph into its corresponding homophilic counterpart.
    GNN-SATA~\cite{yang2024graph} designs edge discriminator modules to dynamically remove or add edges to the adjacency matrix to make the model better fit with homophilic or heterophilic graphs.
    Moreover, ECG-GNN~\cite{deac2023evolving} builds the edge  discriminator based on pre-trained representations and selects k-nearest neighbors for each node to form a complementary graph.
    Downstream tasks and structural optimization can both benefit from parallel message passing in ECG-GNN on original and newly obtained structure.
     
    \noindent \textbf{Dual Views.}
    A dual-view model is also one of GSL approaches to address heterophily, by decoupling the original graph structure into homophily and heterophily views for further analysis.
    For example, GOAL~\cite{zheng2023finding} proposes to reconstruct graphs into heterophily and homophily views complementing each other for graph structure learning. 
    Specifically, this framework first groups intra-class nodes together, ranks them based on feature similarity, and further introduces a complemented aggregation strategy.
    Similarly, ATL~\cite{xu2023node} decomposes the given graph into two components and extracts complementary graph signals from these two components. 
    This dual structure learning framework can adaptively filter and modulate the graph signals, which is critical to address complex heterophilic patterns.

    \noindent \textbf{Neighborhood Similarity.}
    To better characterize heterophily in GSL, it is necessary to comprehensively consider the node neighborhood patterns, 
    resorting to neighborhood similarity for graph rewiring.
    For example, DHGR~\cite{bi2024make} introduces two metrics: Neighborhood Feature Distribution and Neighborhood Label Distribution, to identify edge polarity and further guide the process of graph rewiring.
    Similarly, Choi et al.~\cite{choi2022finding} propose to measure the similarity between nodes using their local subgraphs based on optimal transport, to better adapt to heterophilic graphs.
        
    \noindent \textbf{Spectral Clustering.}
    Since spectral clustering~\cite{nie2011spectral} can capture long-range dependencies on graphs, 
    some GSL methods resort to it for graph rewiring.
    For example, GCN-SL~\cite{jiang2021gcn} proposes an efficient version of spectral clustering to encode nodes, and further constructs the affinity matrix based on it. 
    By combining the affinity matrix with feature similarity, GCN-SL learns an optimized graph structure which benefits the downstream prediction task.
    Moreover, Li et al.~\cite{li2023restructuring} constructs the adjacency matrix based on the result of adaptive spectral clustering, with the aim of maximizing the proposed homophilic scores.

    \noindent \textbf{Probabilistic Modeling.}
    Based on Bayesian Inference to maximize the posterior probability, Wang et al.~\cite{wang2021graph} introduce a Graph structure Estimation Networks (GEN). 
    In addition to the observed links and node features, GEN also incorporates high-order neighborhood information to circumvent bias, and presents a model to jointly treat the above multi-view information as observations of the optimal graph. 
    Moreover, L2A~\cite{wu2023learning} performs the maximum likelihood estimation of GNNs and optimal graph structure learning with a variational inference framework to improve applicability to graphs with heterophily.
        
    \noindent \textbf{Self-supervised Manner.}
    Due to the independence from label dependency, self-supervised methods are also popular in GSL.
    GPS~\cite{dong2023towards} 
    estimates the edge likelihood based on self-supervised link prediction, and further rewires edges according to the uncertainty level. 
    HES-GSL~\cite{wu2023homophily} introduces the denoising auto-encoder~\cite{vincent2008extracting} for feature reconstruction to obtain node representations, and further conducts structural learning with homophily-enhanced self-supervision.
    SUBLIME~\cite{liu2022towards} leverages the contrastive learning to obtain node representations, and optimizes the original structure based on self-supervised learning.
    GSSC~\cite{wu2024learning} applies structural sparsification to remove potentially uninformative or heterophilic edges, and then performs structural self-contrasting in the sparsified neighborhood.
    
    \noindent \textbf{Benchmarks.}
    Recently, a set of GSL benchmarks have been released, attracting widespread attention.
    Representative GSL benchmarks OpenGSL~\cite{zhiyao2024opengsl} and GSLB~\cite{li2024gslb} 
    include heterophilic graphs and further discuss the performance of existing GSL methods on these benchmark datasets, indicating that graph heterophily is an inevitable topic for GSL.
    
    \section{Applications}
    \label{sec8}
    Graph heterophily is widely present in real-world graph structures and is receiving increasing attention in application-level research.
    In this section, we will delve into practical applications and provide detailed introductions.

    \subsection{Cyberspace Security}
    In the social cyberspace, graph heterophily has been a tricky issue for social network analysis. 
    Therefore, learning from heterophilic graphs has significant potential in cyberspace security, offering new insights and innovative approaches.
    
    \noindent \textbf{Anomaly Detection.}
    The rich relations between normal and abnormal objects can be modeled as graphs, giving rise to the Graph-based Anomaly Detection (GAD).  
    GAD also suffers from the graph heterophily issue, as anomalies are often submerged massive normal neighbors, and traditional GNNs uniformly smooth the neighboring nodes, undermining the discriminative message of anomalies.

    \noindent \emph{-Edge Discrimination.}
    Existing GAD methods suffer from heterophily induced by hidden anomalies connected to a considerable number of benign nodes.
    Therefore, SparseGAD~\cite{gong2023beyond} introduces a framework that sparsifies the graph structure to effectively reduce noise and collaboratively learns node representations.
    SparseGAD chooses to retain strong homophilic and heterophilic edges while removing other irrelevant edges for sparsification. Afterward, it performs heterophily-aware aggregation based on GPR-GNN~\cite{chien2020adaptive} to represent each node and categorize suspicious cases.
    To capture the discriminative information of anomalies, 
    TA-Detector~\cite{wen2024ta} introduces a trust classier to distinguish  between trust and distrust connections using label supervision, and
    GHRN~\cite{gao2023addressing} proposes a label-aware edge indicator to compute the post-aggregation similarity for heterophilic edge pruning.
    Moreover, TAM~\cite{qiao2024truncated} introduces an unsupervised anomaly scoring measure, local node affinity, to conduct edge discrimination and iteratively removes heterophilic edges based on this metric. 
    Apart from the edge pruning,  HedGe~\cite{zhang2024generation} introduces a metric called class homophily variance and emphasizes that generating potential homophilic edges based on this metric can improve the performance of GAD.

    \noindent \emph{-Spectral View.}
    Tang et al.~\cite{tang2022rethinking} first rethink GAD from the spectral view and observe that the existence of anomalies will lead to a ``right-shift'' phenomenon, where the spectral energy distribution concentrates less on low frequencies and more on high frequencies.
    Motivated by the observation, they propose the Beta Wavelet Graph Neural Network (BWGNN). 
    BWGNN employs spectral and spatial localized band-pass filters to better handle the ``right-shift'' phenomenon in GAD. 
    Further, AMNET~\cite{chai2022can} directly integrates high-pass and low-pass filters to establish a multi-frequency filter bank and utilizes the attention mechanism to combine them adaptively.

    \noindent \emph{-Self-supervised Manner.}
    Self-supervised learning methods, such as GCL or GAE, can also be applied to GAD to address the graph heterophily issue.
    For example, MELON~\cite{jin2024multi} explicitly models anomaly patterns and designs an enhanced data augmentation strategy by incorporating prior knowledge of various anomalies. 
    Following this, MELON proposes a dual-channel graph encoder with an edge discriminator and conducts multi-view discriminative edge heterophily contrastive learning.
    Roy et al.~\cite{roy2024gad} explore the feasibility of applying GAEs to GAD under heterophily, and find that existing GAEs excel at detecting cluster-type structural anomalies but struggle with more complex anomalies that do not conform to clusters. 
    To this end, they propose GAD-NR to extend NWR-GAE~\cite{tang2022graph}, which is specifically designed for graph heterophily, to the field of anomaly detection, leveraging its powerful modeling capability to identify abnormal nodes.

    \noindent \emph{-Distribution Shift.}
    Due to various time factors and the annotation preferences of human experts, 
    Gao et al.~\cite{gao2023alleviating} observe that the distribution of heterophily and homophily may shift across training and test data, a phenomenon referred to as structural distribution shift (SDS).
    They suggest that ignoring the SDS issue can lead to poor generalization in anomaly detection,
    and propose the Graph Decomposition Network (GDN) to address the issue with homophily guidance.
    
    \noindent \textbf{Fraud Detection.}
    Fraud detection~\cite{pourhabibi2020fraud}, widely used in financial, e-commerce, and insurance industries, aims to detect users exhibiting suspicious behaviors in the communication networks.
    Under the Graph-based Fraud Detection (GFD) setting, fraud nodes tend to interact with normal users and exhibit significantly abnormal characteristics, demonstrating typical graph heterophily.
    It is worth mentioning that GFD and GAD overlap to some extent; however, in this survey, we distinguish these two concepts and provide separate introductions.
    
    \noindent \emph{-Edge Discrimination.}
    Considering the heterophily patterns in fraud graphs, H$^2$-FDetector~\cite{shi2022h2} first identifies the homophilic and heterophilic connections with label supervision, and then customizes a special propagation strategy for heterophilic connections.
    Moreover, DRAG~\cite{kim2023dynamic} proposes to model the inherent heterophily in graphs through different types of relations, and further conducts relation-attentive aggregation at the edge level.

    \noindent \emph{-Group Aggregation.}
    Given that fraud graphs exhibit a blend of homophily and heterophily, it is advisable to explore advanced aggregation techniques to effectively address this complexity.  
    For example, GAGA~\cite{wang2023label} presents a Transformer-based method for fraud detection in multi-relation graphs. 
    To access distant neighbors and capture high-order information, GAGA segregates neighbor-labeled nodes into fraudulent, benign, and unlabeled groups and performs group aggregation over multi-hops.
    DGA-GNN~\cite{duan2024dga} employs decision tree binning encoding for feature transformation, and designs a dynamic grouping strategy to classify nodes into two distinct groups for hierarchical aggregation.
    Moreover, PMP~\cite{wang2023label} points out that the key to GFD lies not in excluding but in distinguishing inter-class neighbors. 
    Therefore, PMP utilizes label information to discriminate neighbors of different classes and customizes distinct group aggregation strategies.

    \noindent \emph{-Spectral View.}
    Another line of GFD involves addressing the graph heterophily issue from a spectral perspective.
    To capture appropriate graph signals, SplitGNN~\cite{wu2023splitgnn} analyzes the spectral distribution with varying degrees of heterophily and observes that the heterophily of fraud nodes causes the spectral energy to shift from low-frequency to high-frequency. 
    This framework employs an edge classifier to split the edges of original graph, facilitating more significant signal expressions across different frequency bands, and further adopts flexible band-pass spectral filters to learn node representations.
    Moreover, SEC-GFD~\cite{xu2024revisiting} first decomposes the spectrum of the graph signals, and then performs complicated message passing based on these frequency bands respectively to improve the performance of GFD.
    
    \noindent \textbf{Bot Detection.}
    Social bots are automated programs designed to simulate human activities on social media platforms. They have been widely employed to spread false information~\cite{shao2018spread}, manipulate elections~\cite{bergstrom2019information,deb2019perils,ferrara2017disinformation}, and thus pose cybersecurity risks and negative social impacts.
    These bots hide within social media platforms and tend to establish heterophilic connections with normal users, posing a challenge to Graph-based Bot Detection (GBD).
    
    \noindent \emph{-Edge Discrimination.}
    Interactions with real accounts results in social networks containing massive camouflaged, heterophilic and unreliable edges.
    Inspired by this, HOVER~\cite{ashmore2023hover} proposes that the key to GBD is to identify and reduce these edges for alleviating graph heterophily. 
    HOVER prunes the inter-class edges using heuristic criteria and further proposes an oversampling strategy for GBD.  
    SIRAN~\cite{zhou2023semi} combines relation discrimination with initial residual connections to reduce noise from neighbors, and enhance the capability of distinguishing different kinds of nodes in human-bot graphs with heterophily. 
    BothH~\cite{li2023multi} constructs a combination graph which consists of the original graph and feature similarity graph, and employs an edge classifier to discriminate between homophilic and heterophilic connections between nodes for message propagation.
    Instead of identifying edges to reduce the impact of heterophily, increasing homophily is also a feasible approach. 
    Motivated by this, HOFA~\cite{ye2023hofa} introduces homophily-oriented edge augmentation to mitigate the impact of heterophily, adding the homophilic edges based on the representation similarity.
    
    \noindent \emph{-Spectral View.}
    Rethinking GBD from the spectral view, existing methods tend to focus on the low-frequency information while neglecting the high-frequency information. 
    To address this, MSGS~\cite{shi2023muti} proposes a multi-scale architecture based on adaptive graph filters to intelligently exploit the low-frequency and high-frequency graph signals.
    
    \noindent \emph{-Contrastive Learning.}
    Due to the scarcity of labels, the popular self-supervised learning paradigm, graph contrastive learning, has been extended to the GBD scenario.
    BotSCL~\cite{wu2023heterophily} utilizes data augmentation to generate different graph views, and designs a channel-wise and attention-free encoder to overcome the heterophily issue. 
    Valuable label supervision is used to guide the encoder to aggregate class-specific information for GBD.

    \noindent \textbf{Rumor Detection.}
    Rumours on social media are an increasingly critical issue in terms of cybersecurity, potentially posing threats to societies~\cite{chen2022multi}.
    Thanks to deep graph learning, graph-based rumour detection has garnered significant attention recently~\cite{yang2021rumor,chen2022multi,yan2023graph}. 
    Real-world social networks exhibit low homophily, and heterophily in rumor graphs is more challenging due to the involvement of different modalities, such as users, posts, links, and hashtags. 
    To handle multi-modal heterophily, Nguyen et al.~\cite{nguyen2024portable} propose a Portable Heterophilic Graph Aggregation for Rumour detection On Social media (PHAROS). 
    Specifically, PHAROS generalizes direct relations into multi-hop aggregation and performs modality-aware aggregation. 
    By leveraging the Graph Transformer to explore global homophily, this framework encodes the rumour graph from three perspective: node features, graph topology, and node labels, and further reduces the training workload via the multi-head self-attention.
    
    \noindent \textbf{Crime Forecasting.}
    Since nearby regions typically exhibit similar socioeconomic characteristics, indicating similar crime patterns, recent solutions construct a distance-based region graph and utilize GNNs for crime forecasting.  
    However, this distance-based graph cannot fully capture the crime correlation between regions that are far apart but share similar crime patterns.
    Motivated by this, HAGEN~\cite{wang2022hagen} proposes a heterophily-aware constraint to regularize the optimization of the region graph to rewire the original graph. 
    The learned graph structure in HAGEN can reveal the dependencies between regions in crime occurrences and simultaneously capture the temporal patterns from historical crime records.
    
    \subsection{Recommender System}
    Recently, graph-based methods have become a popular paradigm to enhance the performance of recommender systems~\cite{he2020lightgcn,wang2019neural,sun2020framework,wu2021self}.
    Social networks are integrated into recommender systems based on the social homophily assumption~\cite{mcpherson2001birds}, wherein users tend to form connections with individuals who share similar interests. 
    The connections among like-minded users are harnessed to compensate for the information scarcity in the interaction graph, thereby providing more personalized recommendations. 
    TGIN~\cite{jiang2022triangle} first observes that user interests and click behaviors may exhibit heterophily on networks.
    Further, SHaRe~\cite{jiang2024challenging} calculates the preference-aware homophily ratios across real-world datasets and observes that user connections in social networks can be heterophilic. 
    To fully harness the potential of social connections, SHaRe adopts graph rewiring techniques to add highly homophilic relations and remove heterophilic relations. 
    This approach ensures that critical social relations are retained while introducing potential social relations that are beneficial for recommendations. 
    Further, considering the fairness of item side in graph-based recommendation, HetroFair~\cite{gholinejad2024heterophily} designs a fairness-aware attention mechanism to generate fair embeddings for users and items, and assigns distinct weights to different heterophilic features during the aggregation process.
    
    \subsection{Geographic Information}
    The advancement of GNNs offer novel insights into the research of geographic information, aiding in the better analysis of geographic spatial data and driving the development of geographic science.
    However, networks in geographic information exhibit more complex patterns, including the heterophily.

    \noindent \textbf{Urban Computing.}
    Urban graphs are widely applied in urban computing~\cite{zheng2014urban}, where nodes represent urban objects such as regions or points of interest, and edges denote urban dependencies like human mobility or road connections. 
    Heterophily is prevalent in urban graphs, reflecting the complex urban system where both similar and dissimilar urban objects can be interconnected.
    Therefore, SHGNN~\cite{xiao2023spatial} proposes a metric named the Spatial Diversity Score to uncover the spatial heterophily of urban graphs. 
    This framework employs a rotation-scaling module to cluster spatially close neighbors, and further processes each group with less internal diversity separately. 
    Subsequently, it introduces a heterophily-sensitive spatial interaction module to adaptively capture the complex patterns within different spatial groups.

    \noindent \textbf{Remote Sensing.}
    In the multimodal remote sensing context, each different modality or the combination results in distinct node types and heterophily interactions. 
    On remote sensing graphs~\cite{zoidi2015graph}, Label Propagation (LP) is essential to improve labeled data sparsity, enhance learning effectiveness, and support decision-making.
    However, traditional LP algorithms are based on the assumption of graph homophily and heterogeneity.
    To address the problem, Taelman et al.~\cite{taelman2021exploitation} design a novel LP method inspired by rules of ZooBP~\cite{eswaran2017zoobp} for multimodal remote sensing data. This method performs well on fully heterogeneous graphs and incorporates both homophily and heterophily interactions.

    \subsection{Computer Vison}
    The applications of graph learning in the field of CV are continuously expanding, providing new insights and solutions for understanding complex interactions between objects within scenes~\cite{han2022vision,peng2020learning,chen2019graph,yang2018graph}. 
    However, the relationships between objects in the visual domain are not necessarily homophilic.
    
    \noindent \textbf{Scene Generation.}
    The objective of Scene Graph Generation (SGG) is to detect objects and predict pairwise relations within an image~\cite{zhu2022scene,chang2021comprehensive}.
    Current SGG methods employ GNNs to model the relations and assume the homophily of scene graph while ignoring heterophily.
    Inspired by learning from heterophilic graphs, HL-Net~\cite{lin2022hl} presents a Heterophily Learning Network to comprehensively explore the homophily and heterophily between objects in scene graphs. 
    HL-Net first proposes an adaptive reweighting transformer, equivalent to general polynomial graph filtering~\cite{shuman2013emerging}, to deal with both high-frequency and low-frequency contexts. 
    Through a heterophily-aware messsage passing strategy, the heterophily and homophily interactions between objects in complicated visual scenes can be fully explored.
    Furthermore, KWGNN~\cite{chen2024kumaraswamy} rethinks the SGG from the spectral view and demonstrates that the spectral energy shifts towards the high-frequency part as heterophily in the scene graph increases. 
    Therefore, KWGNN adaptively generates band-pass filters inspired by kumaraswamy wavelet transform~\cite{kumaraswamy1980generalized} and integrates the filtering results to better accommodate varying levels of smoothness in scene graphs. 

    \noindent \textbf{Point Cloud Segmentation.}
    Point cloud segmentation is one of the crucial tasks in 3D computer vision, aiming to divide the target point cloud into different regions based on their attributes or functions.
    Recently, performing point cloud segmentation based on GNNs has become the mainstream trend~\cite{wang2019graph,lei2020spherical}. 
    In point cloud, it is inevitable that some regions exist nodes from multiple categories, indicating the presence of graph heterophiliy. 
    Traditional GNN methods overlook the crucial heterophilic information, leading to blurred boundaries in segmentation. 
    To adress this, Du et al.~\cite{du2024graph} model the point cloud as a homophilic-heterophilic graph and propose a graph regulation network to produce finer segmentation boundaries. 
    They first evaluate the extent of homophily between the nodes and then implements different weight strategies for homophilic and heterophilic relationships.
    After adaptively propagation, they design a prototype feature extraction module to mine the high-order homophily from the global prototype space. 
    It has been theoretically
    proven that this framework can constrain the similarity of representations between nodes according to the degree of heterophily.

    \noindent \textbf{3D Object Detection.}
    The objective of 3D Object Detection (3DOD) is to accurately locate 3D objects in point clouds. 
    Chen et al.~\cite{chen2024joint} point out that the relational knowledge in 3DOD should encompasses both homophily and heterophily. 
    Therefore, they propose a novel Joint Homophily and Heterophily Relational Knowledge Distillation method (H2RKD) for lidar-based 3D object detection, which simultaneously models both homophily and heterophily relations, enhancing the intra-object similarity and inter-object discrimination.
    
    \subsection{Biochemical Research}
    Graph learning has already achieved tremendous breakthroughs in the field of biochemistry~\cite{wang2022molecular,merchant2023scaling,fang2022geometry,li2024cgmega}, where modeling biochemical molecules as graph structures and conducting further analysis has become a mainstream learning paradigm. 
    However, recent research has found that graph heterophily remains a pressing problem that needs to be addressed in this domain.
    
    \noindent \textbf{Drug Discovery.} 
    Combination therapy~\cite{wu2022machine}, which involves the use of multiple drugs to improve clinical outcomes, has demonstrated advantages over monotherapy. 
    To avoid the costly high-throughput testing in drug combination studies, researchers have established drug-drug networks to explore possible combinations and accelerate the drug discovery process. 
    Chen et al.~\cite{cheng2019network} find that drug pairs with complementary exposure to the disease module tend to be effective combinations, which is consistent with the principle of non-overlapping pharmacology~\cite{jia2016overcoming}. 
    In other words, graph heterophily is also widely present in the drug-drug networks used for combination therapy.
    Furthermore, Chen et al.~\cite{chen2022drug} confirm that the drug-drug network exhibits heterophily and sparseness, limiting the learning effectiveness of tradition GNNs following the homophily assumption. 
    Therefore, they introduce a framework named DCMGCN to simultaneously optimize the drug combination prediction and drug representations. 
    Specifically, to address heterophily, DCMGCN expands the local neighborhood of drug nodes, searching globally for highly similar neighbors to enhance the learning performance.
    From drug-drug networks to drug–disease associations, Liu et al.~\cite{liu2024slgcn} find that the heterophily issue is also widespread in drug repositioning~\cite{xue2018review,jarada2020review,lotfi2018review}, leading to inefficient predictions. 
    They further propose a novel Structure-enhanced Line Graph Convolutional Network (SLGCN) for learning from drug–disease pairs. 
    SLGCN utilizes the transformation of line graphs to capture graph structural features, and assigns appropriate weights for homophily and heterophily structures in message passing with a gating mechanism.

    \noindent \textbf{Molecule Generation.} 
    Conditional molecule generation is of great significance for materials discovery and drug design, and its integration with deep learning is becoming increasingly seamless~\cite{kang2018conditional,yang2023cmgn,rigoni2020conditional}. 
    Existing methods often implicitly assume strong homophily, while overlooking the repulsions between dissimilar atoms, i.e., the heterophily in molecular structures. 
    In light of this, HTFlows~\cite{wangmolecule} proposes a novel flow-based method for conditional molecule generation. 
    By leveraging multiple interactive flows, the method effectively discerns both homophily and heterophily patterns between molecular entities, providing a more versatile representation of the intricate balance between molecular affinities and repulsions.
    
    \noindent \textbf{Neuroscience.}
    Modeling brain networks for analysis is crucial for the early diagnosis of neurodegenerative diseases, and deep learning methods for neuroscience and brain networks are rapidly advancing~\cite{wang2017multi,zhu2018dynamic,wang2023hypergraph,qu2023graph,xu2024data}.
    The brain networks exhibit typical heterophily, where dissimilar regions of interest physically connect, and the interplay between homophily and heterophily makes modeling and analysis challenging.
    Motivated by this, AGT~\cite{choneurodegenerative} introduces spectral node-wise filters based on wavelet transform~\cite{hammond2011wavelets,xu2019graph} to adaptively capture both localized homophily and heterophily patterns. 
    Considering the sequential variations in the progressive degeneration of brain networks, AGT
    proposes temporal regularization to control the distances between diagnostic groups of brain networks in the latent space, ensuring that the temporal dynamics along the groups are effectively captured.

    \subsection{Software Engineering}
    As GNNs are widely applied in the intelligent processes of software engineering, the issue of graph heterophily also arises.
    
    \noindent \textbf{Program Management.}
    Modeling software programs as graphs and employing graph learning techniques for management and analysis have been widely adopted in the software engineering domain. 
    To handle the heterophilic interactions between programs, HAGCN~\cite{xu2024heterophily} introduces the subtraction operation into GNNs to push apart dissimilar nodes in the representation space. 
    Meanwhile, HAGCN separately encodes each type of edges and uses a global relation-aware attention mechanism to aggregate the messages from different edge types for homophily-oriented enhancement. 
    
    \noindent \textbf{Defect Localization.}
    In the context of the booming global open-source ecosystem, a myriad of new developers register on GitHub, and millions of new code repositories are established. 
    Frequent code changes impose higher demands on Just-In-Time (JIT) defect localization~\cite{yan2020just,qiu2020jito,qiu2021deep}. 
    Through the analysis of graph heterophily, Zhang et al.~\cite{zhang2024just} observe that the code graphs in real-word scenarios exhibit very low homophily.
    Inspired by learning from heterophilic graphs, they utilize the classic heterophily-specific model, FAGCN~\cite{bo2021beyond}, to extract homophily and heterophily patterns from the code graph, and learn the graph representations through contrastive learning to enhance defective file prediction.
    
    \section{Future Directions}
    \label{sec9}
    After reviewing recent advances of learning from heterophilic graphs, several challenges and promising directions for further exploration remain. In this section, we will analyze these directions to provide insights for future research.
    
    \subsection{More Complex Scenarios}
    Currently, most studies on graph heterophily focus on static and homogeneous graphs, while neglecting the more intricate relationships that exist in real-world scenarios, such as graph heterogeneity, dynamic changes, and higher-order connections.  
    Addressing graph heterophily within these more complex contexts presents greater challenges but also offers deeper insights into understanding complex relationships in real-world settings.

    \noindent \textbf{Heterogeneous Graphs.}
    Heterogeneous graphs, also known as heterogeneous information networks, are network structures that comprise multiple types of nodes and edges~\cite{shi2016survey}. 
    The multitude of node and edge types in heterogeneous graphs poses great challenges to graph learning and has spurred the development of a series of Heterogeneous Graph Neural Networks~\cite{wang2019heterogeneous,li2017semi,li2021leveraging,li2016transductive,zhao2021heterogeneous,hu2019adversarial,ji2021heterogeneous}.
    Recent studies~\cite{guo2023homophily,li2023hetero} have noted the presence of heterophily in heterogeneous graphs and have attempted to measure this property. 
    Meanwhile, self-supervised learning on heterogeneous graphs has also taken the heterophily issue into account~\cite{shen2024heterophily}.
    Recently, the release of new benchmark datasets~\cite{lin2024heterophily} marks that this field will receive more widespread attention.

    \noindent \textbf{Temporal Graphs.}
    Temporal graphs formalize evolving graphs where nodes, edges and their features change dynamically, and GNNs are powerful tools for analyzing these graphs~\cite{longa2023graph,sahili2023spatio}. 
    For example, Greto~\cite{zhou2022greto} empirically and theoretically elucidates the existence of topology-task discordance and explains the failure of homophily-based GNNs on dynamic graphs. 
    Additionally, there is limited analysis of heterophily for temporal graphs, leaving ample room for future exploration.

    \noindent \textbf{Hypergraphs.}
    Hypergraphs are a generalization of graphs where an edge can connect to any number of vertices, allowing for the modeling of high-order interactions~\cite{antelmi2023survey}.
    Heterophily has been proved as a more common phenomenon in hypergraphs than in simple graphs~\cite{veldt2023combinatorial}.
    Wang et al.~\cite{wang2022equivariant} are the first to explore this issue and propose a hypergraph-based framework, demonstrating significant superiority in processing heterophily patterns in hypergraphs.
    Further follow-ups have been made in~\cite{nguyen2024sheaf, zou2024unig}, and we look forward to deeper research on the heterophily issue in hypergraphs.
    
    \subsection{Deeper Theoretical Insights}
    Undoubtedly, deeper theoretical insights can greatly inspire the development of learning from graphs with heterophily. Therefore, we organize the related theories and provide some promising perspectives of theoretical understanding.
    
    \noindent \textbf{Model Performance.}
    Intuitively, the presence of heterophily can lead to poor performance of GNNs, while strong homophily tends to result in better model performance. 
    However, Ma et al.~\cite{ma2021homophily} first reveal that strong homophily is not a necessity, and vanilla GCN can perform well on heterophilic graphs under certain conditions.
    Luan et al.~\cite{luan2024graph,luan2022revisiting} further demonstrate that heterophily is not always harmful to model, and mid-level homophily is the main culprit of bad performance, a phenomenon referred to as the mid-homophily pitfall.
    Apart from heterophily distribution~\cite{wang2024understanding}, some studies provide crucial insights into GNN performance in relation to conditional shift~\cite{zhu2023explaining}, structural disparity~\cite{mao2024demystifying,yang2024leveraging}, and other factors~\cite{lee2024feature,loveland2024performance,chen2023exploiting}.
    We expect more theoretical research on the relationship between model performance and heterophily, laying a solid theoretical foundation for the development of this field.

    \noindent \textbf{Over-smoothing \& Over-squashing.}
    Two notorious issues regarding GNNs are over-smoothing and over-squashing.
    Over-smoothing refers to the phenomenon where node representations become gradually similiar as the depth of GNNs
    gets deeper, leading to a loss of discriminative power~\cite{li2018deeper,oono2019graph,liu2020towards,rusch2023survey}.
    Yan et al.~\cite{yan2022two} first take a unified perspective of the over-smoothing and heterophily. 
    It is commonly believed that the over-smoothing issue can be addressed along with graph heterophily~\cite{bodnar2022neural,guo2023taming,parkmitigating}.
    Another issue of GNNs is over-squashing, where information from distant neighbors is compressed into fixed-length vectors, leading to a loss of messages from high-order neighbors~\cite{topping2021understanding}.
    Rubin et al.~\cite{rubin2023geodesic} first establish a theoretical framework to understand the combined effect of heterophily and over-squashing.
    Remarkably, Huang et al.~\cite{huang2024universal} devises a polynomial filter-based GNN, named UniFilter, to address graph heterophily from a spectral view and effectively prevent over-smoothing while mitigating over-squashing. 
    Meanwhile, some views suggest that heterophily is the key to solving these two issues~\cite{peimulti}.
    As three major pitfalls affecting GNN performance, we look forward to more explorations uncovering the connections among heterophily, over-smoothing, and over-squashing, and developing further solutions to address all three.

    \noindent \textbf{Other Discoveries.}
    In addition to the above, other theoretical analyses regarding heterophily are also worth attention. 
    Yang et al.~\cite{yang2023graph} study the training dynamics in the function space of GNNs, and establish a strong correlation between generalization and homophily by deriving a data-dependent generalization bound that highly depends on heterophily.
    MGNN~\cite{cui2023mgnn} provides a comprehensive analysis regarding the universality of spatial GNNs from the perspectives of geometry and physics, and finds the proposed framework, inspired by Distance Geometry Problem, handles both homophilic and heterophilic graphs well.
    Inspired by the statistical physics and random matrix theory, Shi et al.~\cite{shi2024homophily} explore the double descent phenomenon in GNNs, and further explain its relationship with heterophily. 
    
    \subsection{Broader Learning Scopes}
    We not only need deeper theoretical insights but also urge researchers to contribute to a broader range of explorations, including but not limited to more learning tasks, learning settings, learning paradigms, real-word applications, and beyond.
    
    \noindent \textbf{More Learning tasks.}
    Most existing studies on learning from heterophilic graphs focus primarily on node classification.
    As listed in our survey, researches on the edge-level or graph-level are still in the early stage.
    In addition to existing tasks, we also encourage more diverse ones to consider heterophilic scenarios, such as graph generation~\cite{zhu2022survey,chanpuriya2021interpretable}, 
    graph condensation~\cite{jin2021graph,gao2024graph}, 
    graph matching~\cite{ling2022graph,yu2023seedgnn}, 
    critical node identification~\cite{munikoti2022scalable}, and influence maximization~\cite{ling2023deep,feng2024influence}.
    
    \noindent \textbf{More Task Settings.}
    Node classification is undoubtedly the most studied topic in heterophilic graphs. 
    However, existing works mostly support only semi-supervised settings and neglect more complex task settings in real scenarios.
    Meanwhile, the performance of existing models can be seriously degraded when labels are extremely limited. 
    To bridge this gap, models designed for few-shot settings~\cite{zhou2019meta,wang2023few} or settings with extremely limited labels~\cite{wan2021contrastive} should be further explored. 
    Moreover, the imbalanced class settings~\cite{zhao2021graphsmote,yu2024graphcbal} and long-tail settings~\cite{yun2022lte4g} that affect homophilic graphs should also be considered in heterophilic graphs. 
    Furthermore, we can rethink learning from heterophilic graphs from the perspective of weakly-supervised settings~\cite{wu2023leveraging,liu2023learning}, such as missing or noisy structures, features, and labels.
    
    \noindent \textbf{More Learning Paradigms.}
    In the field of learning from heterophilic graphs, supervised learning still dominates, while self-supervised learning is rapidly advancing, and prompt learning is still in its early stages of development. 
    Apart from paradigms mentioned above, graph heterophily is expanding into other learning paradigms, including reinforcement learning~\cite{peng2024graphrare}, knowledge distillation~\cite{chen2022sa,wu2022knowledge}, graph self-training~\cite{wang2024hc}, and neural architecture search~\cite{wei4825405searching,wei2022enhancing,wei2022designing,zheng2023auto}.
    Further, we are still curious about whether graph heterophily can spark synergies with meta learning~\cite{liu2022few}, multi-task learning~\cite{zhang2021survey}, positive unlabeled learning~\cite{wu2024unraveling}, and other paradigms.
    
    \noindent \textbf{More Applications.}
    As listed above, current applications of heterophilic graphs primarily focus on social networks.
    In fields such as biology, chemistry, and geography, the understanding of graph heterophily remains superficial, only limited to structural inconsistencies. 
    Therefore, there is an urgent need to further study the utilization of heterophily in these fields to address specific application requirements.
    Moreover, we are eager to see further integration of graph heterophily with fields such as agents, AI4Finance, AI4Science and AI4Health. Domains like task planning~\cite{wu2024can,yang2023foundation}, portfolio management~\cite{soleymani2021deep}, climate change~\cite{bayraktar2023graph}, tectonic movements~\cite{wang2023topological,yang2023novel}, and epidemic modeling~\cite{liu2024review} exhibit typical heterophily in data structure, making them prime candidates for such integration.
    Compared to the outstanding performance of Large Language Models (LLMs) in applications like text and dialogue generation, the ``killer application'' for learning from graphs with heterophily has yet to be discovered.  
    
    \subsection{Advanced Learning Architectures}
    The design of backbone architectures is the key factor in learning from heterophilic graphs.
    In the above text, we elaborate two backbone architectures used for learning from graphs with heterophily: Message Passing Framework and Graph Transformer. 
    Here, we discuss some prospects for advanced architectures in this domain.
    
    \noindent \textbf{Advanced Backbones.}
    Inspired by classical State Space Models (SSM), a novel architecture named Mamba~\cite{gu2023mamba} has emerged as a strong competitor to Transformer due to the effectiveness and efficiency in modeling long-range dependencies in sequential data.
    Inspired by this, GMN~\cite{behrouz2024graph} and Graph-Mamba~\cite{wang2024graph} take the lead in adapting SSMs to graph-structured data and achieve the performance comparable to GTs.
    The state selection mechanism, comparable to global attention, can certainly capture long-range dependencies on graphs and address graph heterophily issues. 
    However, how to naturally adapt SSMs to graph-structured data requires further exploration.
    Moreover, we also hope other advanced architectures, such as Kolmogorov Arnold Network (KAN)~\cite{liu2024kan} and Spiking Neural Networks (SNNs)~\cite{yin2024continuous}, to be explored under graph heterophily.

    \noindent \textbf{LLM for Heterophily.}
    With great advancements in LLMs, it is a promising direction to enhance graph learning by extensive knowledge within LLMs.
    Existing methods have demonstrated the capability of LLMs in empowering learning for Text-Attributed Graphs (TAGs)~\cite{yu2023empower,he2023harnessing,wang2024bridging,huang2024gnns,fang2024gaugllm,chen2024exploring,mao2024advancing}. 
    However, recent advances primarily focus on homophilic graphs, leaving graphs with heterophily largely unexplored.
    Notably, LLM4HeG~\cite{wu2024exploring} is the first to integrate LLMs into learning from graph heterophily, providing new insights for future development. 
    The following questions have become current research hotpots and urgently need to be addressed:
    Can LLMs effectively identify graph heterophily? 
    How to reduce the cost and efficiently use LLMs to enhance heterophilic graph learning?
    Should we follow the paradigms of LLM4GNN and GNN4LLM, and use LLMs to replace GNNs or integrate GNNs and LLMs into a unified architecture~\cite{ren2024survey,liu2023towards,fan2024graph} under graph heterophily setting?

    \noindent \textbf{Graph Foundation Models.}
    The goal of Graph Foundation Models (GFMs)~\cite{liu2023towards} is to develop graph models trained on massive data from diverse domains to enhance the applicability across different tasks and domains, emerging as a significant topic of graph domain. 
    Current works~\cite{galkin2023towards,maoposition,tang2024graphgpt,xia2024opengraph} indicate that this field is still in its early stages. 
    Remarkably, AnyGraph~\cite{xia2024anygraph} builds a Graph Mixture-of-Experts (MoE) architecture to effectively manage  cross-domain distribution shifts concerning structure-level and feature-level heterogeneity. 
    This is the first attempt to enable graph models to exhibit scaling law behavior~\cite{alabdulmohsin2022revisiting}, where the performance of model scales favorably with the amount of data and parameters.
    Moreover, the heterophily issue is still a challenge that GFMs with strong generalization capabilities cannot avoid. Here, we call for GFM studies to pay more attention to graph heterophily.

    \section{Conclusion}
    In this paper, we present a comprehensive survey of the benchmark datasets, GNN models, learning paradigms, real-word applications and future directions for heterophilic graphs. 
    Through a detailed overview and an in-depth analysis of recent advances, we aim to provide inspiration and insights for this field, thereby promoting further development of learning from graphs with heterophily.

\bibliographystyle{fcs}
\bibliography{ref}

\begin{thebibliography}{100}

\bibitem{shao2018spread}
Shao C, Ciampaglia G~L, Varol O, Yang K~C, Flammini A, Menczer F.
\newblock The spread of low-credibility content by social bots.
\newblock Nature communications, 2018, 9(1): 1--9

\bibitem{lynn2019physics}
Lynn C~W, Bassett D~S.
\newblock The physics of brain network structure, function and control.
\newblock Nature Reviews Physics, 2019, 1(5): 318--332

\bibitem{zheng2014urban}
Zheng Y, Capra L, Wolfson O, Yang H.
\newblock Urban computing: concepts, methodologies, and applications.
\newblock ACM Transactions on Intelligent Systems and Technology (TIST), 2014, 5(3): 1--55

\bibitem{kipf2016semi}
Kipf T~N, Welling M.
\newblock Semi-supervised classification with graph convolutional networks.
\newblock arXiv preprint arXiv:1609.02907, 2016

\bibitem{hamilton2017inductive}
Hamilton W, Ying Z, Leskovec J.
\newblock Inductive representation learning on large graphs.
\newblock Advances in neural information processing systems, 2017, 30

\bibitem{velivckovic2017graph}
Veli{\v{c}}kovi{\'c} P, Cucurull G, Casanova A, Romero A, Lio P, Bengio Y.
\newblock Graph attention networks.
\newblock arXiv preprint arXiv:1710.10903, 2017

\bibitem{zheng2022graph}
Zheng X, Liu Y, Pan S, Zhang M, Jin D, Yu~P~S.
\newblock Graph neural networks for graphs with heterophily: A survey.
\newblock arXiv preprint arXiv:2202.07082, 2022

\bibitem{zhu2023heterophily}
Zhu J, Yan Y, Heimann M, Zhao L, Akoglu L, Koutra D.
\newblock Heterophily and graph neural networks: Past, present and future.
\newblock IEEE Data Engineering Bulletin, 2023

\bibitem{luan2024graph}
Luan S, Hua C, Xu~M, Lu~Q, Zhu J, Chang X~W, Fu~J, Leskovec J, Precup D.
\newblock When do graph neural networks help with node classification? investigating the homophily principle on node distinguishability.
\newblock Advances in Neural Information Processing Systems, 2024, 36

\bibitem{gilmer2017neural}
Gilmer J, Schoenholz S~S, Riley P~F, Vinyals O, Dahl G~E.
\newblock Neural message passing for quantum chemistry.
\newblock In: International conference on machine learning.
\newblock 2017,  1263--1272

\bibitem{li2018deeper}
Li~Q, Han Z, Wu~X~M.
\newblock Deeper insights into graph convolutional networks for semi-supervised learning.
\newblock In: Proceedings of the AAAI conference on artificial intelligence.
\newblock 2018

\bibitem{alon2020bottleneck}
Alon U, Yahav E.
\newblock On the bottleneck of graph neural networks and its practical implications.
\newblock arXiv preprint arXiv:2006.05205, 2020

\bibitem{corso2020principal}
Corso G, Cavalleri L, Beaini D, Li{\`o} P, Veli{\v{c}}kovi{\'c} P.
\newblock Principal neighbourhood aggregation for graph nets.
\newblock Advances in Neural Information Processing Systems, 2020, 33: 13260--13271

\bibitem{zhu2020beyond}
Zhu J, Yan Y, Zhao L, Heimann M, Akoglu L, Koutra D.
\newblock Beyond homophily in graph neural networks: Current limitations and effective designs.
\newblock Advances in neural information processing systems, 2020, 33: 7793--7804

\bibitem{vaswani2017attention}
Vaswani A, Shazeer N, Parmar N, Uszkoreit J, Jones L, Gomez A~N, Kaiser {\L}, Polosukhin I.
\newblock Attention is all you need.
\newblock Advances in neural information processing systems, 2017, 30

\bibitem{devlin2018bert}
Devlin J, Chang M~W, Lee K, Toutanova K.
\newblock Bert: Pre-training of deep bidirectional transformers for language understanding.
\newblock arXiv preprint arXiv:1810.04805, 2018

\bibitem{brown2020language}
Brown T, Mann B, Ryder N, Subbiah M, Kaplan J~D, Dhariwal P, Neelakantan A, Shyam P, Sastry G, Askell A, others .
\newblock Language models are few-shot learners.
\newblock Advances in neural information processing systems, 2020, 33: 1877--1901

\bibitem{patwardhan2023transformers}
Patwardhan N, Marrone S, Sansone C.
\newblock Transformers in the real world: A survey on nlp applications.
\newblock Information, 2023, 14(4): 242

\bibitem{dosovitskiy2020image}
Dosovitskiy A, Beyer L, Kolesnikov A, Weissenborn D, Zhai X, Unterthiner T, Dehghani M, Minderer M, Heigold G, Gelly S, others .
\newblock An image is worth 16x16 words: Transformers for image recognition at scale.
\newblock arXiv preprint arXiv:2010.11929, 2020

\bibitem{liu2021swin}
Liu Z, Lin Y, Cao Y, Hu~H, Wei Y, Zhang Z, Lin S, Guo B.
\newblock Swin transformer: Hierarchical vision transformer using shifted windows.
\newblock In: Proceedings of the IEEE/CVF international conference on computer vision.
\newblock 2021,  10012--10022

\bibitem{han2022survey}
Han K, Wang Y, Chen H, Chen X, Guo J, Liu Z, Tang Y, Xiao A, Xu~C, Xu~Y, others .
\newblock A survey on vision transformer.
\newblock IEEE transactions on pattern analysis and machine intelligence, 2022, 45(1): 87--110

\bibitem{khan2022transformers}
Khan S, Naseer M, Hayat M, Zamir S~W, Khan F~S, Shah M.
\newblock Transformers in vision: A survey.
\newblock ACM computing surveys (CSUR), 2022, 54(10s): 1--41

\bibitem{min2022transformer}
Min E, Chen R, Bian Y, Xu~T, Zhao K, Huang W, Zhao P, Huang J, Ananiadou S, Rong Y.
\newblock Transformer for graphs: An overview from architecture perspective.
\newblock arXiv preprint arXiv:2202.08455, 2022

\bibitem{shehzad2024graph}
Shehzad A, Xia F, Abid S, Peng C, Yu~S, Zhang D, Verspoor K.
\newblock Graph transformers: A survey.
\newblock arXiv preprint arXiv:2407.09777, 2024

\bibitem{ying2021transformers}
Ying C, Cai T, Luo S, Zheng S, Ke~G, He~D, Shen Y, Liu T~Y.
\newblock Do transformers really perform badly for graph representation?
\newblock Advances in neural information processing systems, 2021, 34: 28877--28888

\bibitem{dwivedi2021graph}
Dwivedi V~P, Luu A~T, Laurent T, Bengio Y, Bresson X.
\newblock Graph neural networks with learnable structural and positional representations.
\newblock arXiv preprint arXiv:2110.07875, 2021

\bibitem{bouritsas2022improving}
Bouritsas G, Frasca F, Zafeiriou S, Bronstein M~M.
\newblock Improving graph neural network expressivity via subgraph isomorphism counting.
\newblock IEEE Transactions on Pattern Analysis and Machine Intelligence, 2022, 45(1): 657--668

\bibitem{kreuzer2021rethinking}
Kreuzer D, Beaini D, Hamilton W, L{\'e}tourneau V, Tossou P.
\newblock Rethinking graph transformers with spectral attention.
\newblock Advances in Neural Information Processing Systems, 2021, 34: 21618--21629

\bibitem{ma2023graph}
Ma~L, Lin C, Lim D, Romero-Soriano A, Dokania P~K, Coates M, Torr P, Lim S~N.
\newblock Graph inductive biases in transformers without message passing.
\newblock In: International Conference on Machine Learning.
\newblock 2023,  23321--23337

\bibitem{zhang2020adaptive}
Zhang K, Zhu Y, Wang J, Zhang J.
\newblock Adaptive structural fingerprints for graph attention networks.
\newblock In: International Conference on Learning Representations.
\newblock 2020

\bibitem{chung1997spectral}
Chung F~R.
\newblock Spectral graph theory. volume~92.
\newblock American Mathematical Soc., 1997

\bibitem{luan2020complete}
Luan S, Zhao M, Hua C, Chang X~W, Precup D.
\newblock Complete the missing half: Augmenting aggregation filtering with diversification for graph convolutional networks.
\newblock arXiv preprint arXiv:2008.08844, 2020

\bibitem{dakovic2019local}
Dakovi{\'c} M, Stankovi{\'c} L, Sejdi{\'c} E.
\newblock Local smoothness of graph signals.
\newblock Mathematical Problems in Engineering, 2019, 2019(1): 3208569

\bibitem{liao2024benchmarking}
Liao N, Liu H, Zhu Z, Luo S, Lakshmanan L~V.
\newblock Benchmarking spectral graph neural networks: A comprehensive study on effectiveness and efficiency.
\newblock arXiv preprint arXiv:2406.09675, 2024

\bibitem{liu2021self}
Liu X, Zhang F, Hou Z, Mian L, Wang Z, Zhang J, Tang J.
\newblock Self-supervised learning: Generative or contrastive.
\newblock IEEE transactions on knowledge and data engineering, 2021, 35(1): 857--876

\bibitem{liu2022graph}
Liu Y, Jin M, Pan S, Zhou C, Zheng Y, Xia F, Philip S~Y.
\newblock Graph self-supervised learning: A survey.
\newblock IEEE transactions on knowledge and data engineering, 2022, 35(6): 5879--5900

\bibitem{hu2019strategies}
Hu~W, Liu B, Gomes J, Zitnik M, Liang P, Pande V, Leskovec J.
\newblock Strategies for pre-training graph neural networks.
\newblock arXiv preprint arXiv:1905.12265, 2019

\bibitem{liu2023pre}
Liu P, Yuan W, Fu~J, Jiang Z, Hayashi H, Neubig G.
\newblock Pre-train, prompt, and predict: A systematic survey of prompting methods in natural language processing.
\newblock ACM Computing Surveys, 2023, 55(9): 1--35

\bibitem{sun2023graph}
Sun X, Zhang J, Wu~X, Cheng H, Xiong Y, Li~J.
\newblock Graph prompt learning: A comprehensive survey and beyond.
\newblock arXiv preprint arXiv:2311.16534, 2023

\bibitem{long2024towards}
Long Q, Yan Y, Zhang P, Fang C, Cui W, Ning Z, Xiao M, Cao N, Luo X, Xu~L, others .
\newblock Towards graph prompt learning: A survey and beyond.
\newblock arXiv preprint arXiv:2408.14520, 2024

\bibitem{rozemberczki2021multi}
Rozemberczki B, Allen C, Sarkar R.
\newblock Multi-scale attributed node embedding.
\newblock Journal of Complex Networks, 2021, 9(2): cnab014

\bibitem{pei2020geom}
Pei H, Wei B, Chang K~C~C, Lei Y, Yang B.
\newblock Geom-gcn: Geometric graph convolutional networks.
\newblock arXiv preprint arXiv:2002.05287, 2020

\bibitem{lim2021new}
Lim D, Li~X, Hohne F, Lim S~N.
\newblock New benchmarks for learning on non-homophilous graphs.
\newblock arXiv preprint arXiv:2104.01404, 2021

\bibitem{lim2021large}
Lim D, Hohne F, Li~X, Huang S~L, Gupta V, Bhalerao O, Lim S~N.
\newblock Large scale learning on non-homophilous graphs: New benchmarks and strong simple methods.
\newblock Advances in Neural Information Processing Systems, 2021, 34: 20887--20902

\bibitem{platonov2023critical}
Platonov O, Kuznedelev D, Diskin M, Babenko A, Prokhorenkova L.
\newblock A critical look at the evaluation of gnns under heterophily: Are we really making progress?
\newblock arXiv preprint arXiv:2302.11640, 2023

\bibitem{luan2024heterophily}
Luan S, Lu~Q, Hua C, Wang X, Zhu J, Chang X~W, Wolf G, Tang J.
\newblock Are heterophily-specific gnns and homophily metrics really effective? evaluation pitfalls and new benchmarks.
\newblock arXiv preprint arXiv:2409.05755, 2024

\bibitem{xiao2024simple}
Xiao T, Zhu H, Chen Z, Wang S.
\newblock Simple and asymmetric graph contrastive learning without augmentations.
\newblock Advances in Neural Information Processing Systems, 2024, 36

\bibitem{cavallo20222}
Cavallo A, Grohnfeldt C, Russo M, Lovisotto G, Vassio L.
\newblock 2-hop neighbor class similarity (2ncs): A graph structural metric indicative of graph neural network performance.
\newblock arXiv preprint arXiv:2212.13202, 2022

\bibitem{platonov2024characterizing}
Platonov O, Kuznedelev D, Babenko A, Prokhorenkova L.
\newblock Characterizing graph datasets for node classification: Homophily-heterophily dichotomy and beyond.
\newblock Advances in Neural Information Processing Systems, 2024, 36

\bibitem{newman2003mixing}
Newman M~E.
\newblock Mixing patterns in networks.
\newblock Physical review E, 2003, 67(2): 026126

\bibitem{li2023restructuring}
Li~S, Kim D, Wang Q.
\newblock Restructuring graph for higher homophily via adaptive spectral clustering.
\newblock In: Proceedings of the AAAI Conference on Artificial Intelligence.
\newblock 2023,  8622--8630

\bibitem{gong2023neighborhood}
Gong S, Zhou J, Xie C, Xuan Q.
\newblock Neighborhood homophily-based graph convolutional network.
\newblock In: Proceedings of the 32nd ACM International Conference on Information and Knowledge Management.
\newblock 2023,  3908--3912

\bibitem{luan2022revisiting}
Luan S, Hua C, Lu~Q, Zhu J, Zhao M, Zhang S, Chang X~W, Precup D.
\newblock Revisiting heterophily for graph neural networks.
\newblock Advances in neural information processing systems, 2022, 35: 1362--1375

\bibitem{yang2021diverse}
Yang L, Li~M, Liu L, Wang C, Cao X, Guo Y, others .
\newblock Diverse message passing for attribute with heterophily.
\newblock Advances in Neural Information Processing Systems, 2021, 34: 4751--4763

\bibitem{jin2022raw}
Jin D, Wang R, Ge~M, He~D, Li~X, Lin W, Zhang W.
\newblock Raw-gnn: Random walk aggregation based graph neural network.
\newblock arXiv preprint arXiv:2206.13953, 2022

\bibitem{lee2024feature}
Lee S~Y, Kim S, Bu~F, Yoo J, Tang J, Shin K.
\newblock Feature distribution on graph topology mediates the effect of graph convolution: Homophily perspective.
\newblock arXiv preprint arXiv:2402.04621, 2024

\bibitem{ojha2024affinity}
Ojha I, Bose K, Das S.
\newblock Affinity-based homophily: Can we measure homophily of a graph without using node labels?
\newblock In: The Second Tiny Papers Track at ICLR 2024.
\newblock 2024

\bibitem{zheng2024missing}
Zheng Y, Luan S, Chen L.
\newblock What is missing in homophily? disentangling graph homophily for graph neural networks.
\newblock arXiv preprint arXiv:2406.18854, 2024

\bibitem{ma2021homophily}
Ma~Y, Liu X, Shah N, Tang J.
\newblock Is homophily a necessity for graph neural networks?
\newblock arXiv preprint arXiv:2106.06134, 2021

\bibitem{huang2017label}
Huang X, Li~J, Hu~X.
\newblock Label informed attributed network embedding.
\newblock In: Proceedings of the tenth ACM international conference on web search and data mining.
\newblock 2017,  731--739

\bibitem{zhiyao2024opengsl}
Zhiyao Z, Zhou S, Mao B, Zhou X, Chen J, Tan Q, Zha D, Feng Y, Chen C, Wang C.
\newblock Opengsl: A comprehensive benchmark for graph structure learning.
\newblock Advances in Neural Information Processing Systems, 2024, 36

\bibitem{tang2009social}
Tang J, Sun J, Wang C, Yang Z.
\newblock Social influence analysis in large-scale networks.
\newblock In: Proceedings of the 15th ACM SIGKDD international conference on Knowledge discovery and data mining.
\newblock 2009,  807--816

\bibitem{garcia2016using}
Garc{\'\i}a-Plaza A~P, Fresno V, Unanue R~M, Zubiaga A.
\newblock Using fuzzy logic to leverage html markup for web page representation.
\newblock IEEE Transactions on Fuzzy Systems, 2016, 25(4): 919--933

\bibitem{dwivedi2023benchmarking}
Dwivedi V~P, Joshi C~K, Luu A~T, Laurent T, Bengio Y, Bresson X.
\newblock Benchmarking graph neural networks.
\newblock Journal of Machine Learning Research, 2023, 24(43): 1--48

\bibitem{traud2012social}
Traud A~L, Mucha P~J, Porter M~A.
\newblock Social structure of facebook networks.
\newblock Physica A: Statistical Mechanics and its Applications, 2012, 391(16): 4165--4180

\bibitem{jure2014snap}
Jure L.
\newblock Snap datasets: Stanford large network dataset collection.
\newblock Retrieved December 2021 from http://snap. stanford. edu/data, 2014

\bibitem{lim2021expertise}
Lim D, Benson A~R.
\newblock Expertise and dynamics within crowdsourced musical knowledge curation: A case study of the genius platform.
\newblock In: Proceedings of the International AAAI Conference on Web and Social Media.
\newblock 2021,  373--384

\bibitem{rozemberczki2021twitch}
Rozemberczki B, Sarkar R.
\newblock Twitch gamers: a dataset for evaluating proximity preserving and structural role-based node embeddings.
\newblock arXiv preprint arXiv:2101.03091, 2021

\bibitem{rozemberczki2020characteristic}
Rozemberczki B, Sarkar R.
\newblock Characteristic functions on graphs: Birds of a feather, from statistical descriptors to parametric models.
\newblock In: Proceedings of the 29th ACM international conference on information \& knowledge management.
\newblock 2020,  1325--1334

\bibitem{hu2020open}
Hu~W, Fey M, Zitnik M, Dong Y, Ren H, Liu B, Catasta M, Leskovec J.
\newblock Open graph benchmark: Datasets for machine learning on graphs.
\newblock Advances in neural information processing systems, 2020, 33: 22118--22133

\bibitem{leskovec2005graphs}
Leskovec J, Kleinberg J, Faloutsos C.
\newblock Graphs over time: densification laws, shrinking diameters and possible explanations.
\newblock In: Proceedings of the eleventh ACM SIGKDD international conference on Knowledge discovery in data mining.
\newblock 2005,  177--187

\bibitem{lhoest2021datasets}
Lhoest Q, Del~Moral A~V, Jernite Y, Thakur A, Von~Platen P, Patil S, Chaumond J, Drame M, Plu J, Tunstall L, others .
\newblock Datasets: A community library for natural language processing.
\newblock arXiv preprint arXiv:2109.02846, 2021

\bibitem{luo2024classic}
Luo Y, Shi L, Wu~X~M.
\newblock Classic gnns are strong baselines: Reassessing gnns for node classification.
\newblock arXiv preprint arXiv:2406.08993, 2024

\bibitem{srivastava2014dropout}
Srivastava N, Hinton G, Krizhevsky A, Sutskever I, Salakhutdinov R.
\newblock Dropout: a simple way to prevent neural networks from overfitting.
\newblock The journal of machine learning research, 2014, 15(1): 1929--1958

\bibitem{shu2022understanding}
Shu J, Xi~B, Li~Y, Wu~F, Kamhoua C, Ma~J.
\newblock Understanding dropout for graph neural networks.
\newblock In: Companion Proceedings of the Web Conference 2022.
\newblock 2022,  1128--1138

\bibitem{cai2021graphnorm}
Cai T, Luo S, Xu~K, He~D, Liu T~y, Wang L.
\newblock Graphnorm: A principled approach to accelerating graph neural network training.
\newblock In: International Conference on Machine Learning.
\newblock 2021,  1204--1215

\bibitem{he2016deep}
He~K, Zhang X, Ren S, Sun J.
\newblock Deep residual learning for image recognition.
\newblock In: Proceedings of the IEEE conference on computer vision and pattern recognition.
\newblock 2016,  770--778

\bibitem{li2019deepgcns}
Li~G, Muller M, Thabet A, Ghanem B.
\newblock Deepgcns: Can gcns go as deep as cnns?
\newblock In: Proceedings of the IEEE/CVF international conference on computer vision.
\newblock 2019,  9267--9276

\bibitem{li2020deepergcn}
Li~G, Xiong C, Thabet A, Ghanem B.
\newblock Deepergcn: All you need to train deeper gcns.
\newblock arXiv preprint arXiv:2006.07739, 2020

\bibitem{xu2018representation}
Xu~K, Li~C, Tian Y, Sonobe T, Kawarabayashi K~i, Jegelka S.
\newblock Representation learning on graphs with jumping knowledge networks.
\newblock In: International conference on machine learning.
\newblock 2018,  5453--5462

\bibitem{luan2024heterophilic}
Luan S, Hua C, Lu~Q, Ma~L, Wu~L, Wang X, Xu~M, Chang X~W, Precup D, Ying R, others .
\newblock The heterophilic graph learning handbook: Benchmarks, models, theoretical analysis, applications and challenges.
\newblock arXiv preprint arXiv:2407.09618, 2024

\bibitem{chen2020simple}
Chen M, Wei Z, Huang Z, Ding B, Li~Y.
\newblock Simple and deep graph convolutional networks.
\newblock In: International conference on machine learning.
\newblock 2020,  1725--1735

\bibitem{chien2020adaptive}
Chien E, Peng J, Li~P, Milenkovic O.
\newblock Adaptive universal generalized pagerank graph neural network.
\newblock arXiv preprint arXiv:2006.07988, 2020

\bibitem{chanpuriya2022simplified}
Chanpuriya S, Musco C.
\newblock Simplified graph convolution with heterophily.
\newblock Advances in Neural Information Processing Systems, 2022, 35: 27184--27197

\bibitem{he2022convolutional}
He~M, Wei Z, Wen J~R.
\newblock Convolutional neural networks on graphs with chebyshev approximation, revisited.
\newblock Advances in neural information processing systems, 2022, 35: 7264--7276

\bibitem{guo2023clenshaw}
Guo Y, Wei Z.
\newblock Clenshaw graph neural networks.
\newblock In: Proceedings of the 29th ACM SIGKDD Conference on Knowledge Discovery and Data Mining.
\newblock 2023,  614--625

\bibitem{he2021bernnet}
He~M, Wei Z, Xu~H, others .
\newblock Bernnet: Learning arbitrary graph spectral filters via bernstein approximation.
\newblock Advances in Neural Information Processing Systems, 2021, 34: 14239--14251

\bibitem{chen2023improved}
Chen J, Xu~L.
\newblock Improved modeling and generalization capabilities of graph neural networks with legendre polynomials.
\newblock IEEE Access, 2023, 11: 63442--63450

\bibitem{wang2022powerful}
Wang T, Jin D, Wang R, He~D, Huang Y.
\newblock Powerful graph convolutional networks with adaptive propagation mechanism for homophily and heterophily.
\newblock In: Proceedings of the AAAI conference on artificial intelligence.
\newblock 2022,  4210--4218

\bibitem{guo2023graph}
Guo Y, Wei Z.
\newblock Graph neural networks with learnable and optimal polynomial bases.
\newblock In: International Conference on Machine Learning.
\newblock 2023,  12077--12097

\bibitem{du2024graph}
Du~Z, Liang J, Liang J, Yao K, Cao F.
\newblock Graph regulation network for point cloud segmentation.
\newblock IEEE Transactions on Pattern Analysis and Machine Intelligence, 2024

\bibitem{li2022g}
Li~M, Guo X, Wang Y, Wang Y, Lin Z.
\newblock G$^2$cn: Graph gaussian convolution networks with concentrated graph filters.
\newblock In: International Conference on Machine Learning.
\newblock 2022,  12782--12796

\bibitem{ekbote2024figure}
Ekbote C, Deshpande A, Iyer A, Sellamanickam S, Bairi R.
\newblock Figure: Simple and efficient unsupervised node representations with filter augmentations.
\newblock Advances in Neural Information Processing Systems, 2024, 36

\bibitem{geng2023pyramid}
Geng H, Chen C, He~Y, Zeng G, Han Z, Chai H, Yan J.
\newblock Pyramid graph neural network: A graph sampling and filtering approach for multi-scale disentangled representations.
\newblock In: Proceedings of the 29th ACM SIGKDD Conference on Knowledge Discovery and Data Mining.
\newblock 2023,  518--530

\bibitem{lu2024flexible}
Lu~Q, Zhu J, Luan S, Chang X~W.
\newblock Flexible diffusion scopes with parameterized laplacian for heterophilic graph learning.
\newblock arXiv preprint arXiv:2409.09888, 2024

\bibitem{li2024pc}
Li~B, Pan E, Kang Z.
\newblock Pc-conv: Unifying homophily and heterophily with two-fold filtering.
\newblock In: Proceedings of the AAAI Conference on Artificial Intelligence.
\newblock 2024,  13437--13445

\bibitem{xu2023shape}
Xu~J, Dai E, Luo D, Zhang X, Wang S.
\newblock Shape-aware graph spectral learning.
\newblock Openreview, 2023

\bibitem{huang2024optimizing}
Huang K, Cao W, Ta~H, Xiao X, Li{\`o} P.
\newblock Optimizing polynomial graph filters: A novel adaptive krylov subspace approach.
\newblock In: Proceedings of the ACM on Web Conference 2024.
\newblock 2024,  1057--1068

\bibitem{huang2024universal}
Huang K, Wang Y~G, Li~M, others .
\newblock How universal polynomial bases enhance spectral graph neural networks: Heterophily, over-smoothing, and over-squashing.
\newblock arXiv preprint arXiv:2405.12474, 2024

\bibitem{han2024node}
Han H, Li~J, Huang W, Tang X, Lu~H, Luo C, Liu H, Tang J.
\newblock Node-wise filtering in graph neural networks: A mixture of experts approach.
\newblock arXiv preprint arXiv:2406.03464, 2024

\bibitem{abu2019mixhop}
Abu-El-Haija S, Perozzi B, Kapoor A, Alipourfard N, Lerman K, Harutyunyan H, Ver~Steeg G, Galstyan A.
\newblock Mixhop: Higher-order graph convolutional architectures via sparsified neighborhood mixing.
\newblock In: international conference on machine learning.
\newblock 2019,  21--29

\bibitem{jin2021universal}
Jin D, Yu~Z, Huo C, Wang R, Wang X, He~D, Han J.
\newblock Universal graph convolutional networks.
\newblock Advances in Neural Information Processing Systems, 2021, 34: 10654--10664

\bibitem{maurya2022simplifying}
Maurya S~K, Liu X, Murata T.
\newblock Simplifying approach to node classification in graph neural networks.
\newblock Journal of Computational Science, 2022, 62: 101695

\bibitem{xu2022hp}
Xu~J, Dai E, Zhang X, Wang S.
\newblock Hp-gmn: Graph memory networks for heterophilous graphs.
\newblock In: 2022 IEEE International Conference on Data Mining (ICDM).
\newblock 2022,  1263--1268

\bibitem{huang2022local}
Huang N, Villar S, Priebe C~E, Zheng D, Huang C, Yang L, Braverman V.
\newblock From local to global: Spectral-inspired graph neural networks.
\newblock arXiv preprint arXiv:2209.12054, 2022

\bibitem{choi2023node}
Choi S, Kim G, Yun S~Y.
\newblock Node mutual information: Enhancing graph neural networks for heterophily.
\newblock In: NeurIPS 2023 Workshop: New Frontiers in Graph Learning.
\newblock 2023

\bibitem{wang2021tree}
Wang Y, Derr T.
\newblock Tree decomposed graph neural network.
\newblock In: Proceedings of the 30th ACM international conference on information \& knowledge management.
\newblock 2021,  2040--2049

\bibitem{song2023ordered}
Song Y, Zhou C, Wang X, Lin Z.
\newblock Ordered gnn: Ordering message passing to deal with heterophily and over-smoothing.
\newblock arXiv preprint arXiv:2302.01524, 2023

\bibitem{sun2022beyond}
Sun Y, Deng H, Yang Y, Wang C, Xu~J, Huang R, Cao L, Wang Y, Chen L.
\newblock Beyond homophily: Structure-aware path aggregation graph neural network.
\newblock In: IJCAI.
\newblock 2022,  2233--2240

\bibitem{xie2023pathmlp}
Xie C, Zhou J, Gong S, Wan J, Qian J, Yu~S, Xuan Q, Yang X.
\newblock Pathmlp: Smooth path towards high-order homophily.
\newblock arXiv preprint arXiv:2306.13532, 2023

\bibitem{jin2021node}
Jin W, Derr T, Wang Y, Ma~Y, Liu Z, Tang J.
\newblock Node similarity preserving graph convolutional networks.
\newblock In: Proceedings of the 14th ACM international conference on web search and data mining.
\newblock 2021,  148--156

\bibitem{suresh2021breaking}
Suresh S, Budde V, Neville J, Li~P, Ma~J.
\newblock Breaking the limit of graph neural networks by improving the assortativity of graphs with local mixing patterns.
\newblock In: Proceedings of the 27th ACM SIGKDD Conference on Knowledge Discovery \& Data Mining.
\newblock 2021,  1541--1551

\bibitem{ai2024neighbors}
Ai~G, Gao Y, Wang H, Li~X, Wang J, Yan H.
\newblock Neighbors selective graph convolutional network for homophily and heterophily.
\newblock Pattern Recognition Letters, 2024

\bibitem{liu2025beyond}
Liu S, He~D, Yu~Z, Jin D, Feng Z.
\newblock Beyond homophily: Neighborhood distribution-guided graph convolutional networks.
\newblock Expert Systems with Applications, 2025, 259: 125274

\bibitem{wang2023improving}
Wang Y, Xiang S, Pan C.
\newblock Improving the homophily of heterophilic graphs for semi-supervised node classification.
\newblock In: 2023 IEEE International Conference on Multimedia and Expo (ICME).
\newblock 2023,  1865--1870

\bibitem{liu2021non}
Liu M, Wang Z, Ji~S.
\newblock Non-local graph neural networks.
\newblock TPAMI, 2021, 44(12): 10270--10276

\bibitem{yang2022graph}
Yang T, Wang Y, Yue Z, Yang Y, Tong Y, Bai J.
\newblock Graph pointer neural networks.
\newblock In: Proceedings of the AAAI conference on artificial intelligence.
\newblock 2022,  8832--8839

\bibitem{zou2023similarity}
Zou M, Gan Z, Cao R, Guan C, Leng S.
\newblock Similarity-navigated graph neural networks for node classification.
\newblock Information Sciences, 2023, 633: 41--69

\bibitem{li2024knn}
Li~L, Yang W, Bai S, Ma~Z.
\newblock Knn-gnn: A powerful graph neural network enhanced by aggregating k-nearest neighbors in common subspace.
\newblock Expert Systems with Applications, 2024, 253: 124217

\bibitem{dong2024differentiable}
Dong Y, Dupty M~H, Deng L, Liu Z, Goh Y~L, Lee W~S.
\newblock Differentiable cluster graph neural network.
\newblock arXiv preprint arXiv:2405.16185, 2024

\bibitem{park2022deformable}
Park J, Yun S, Park H, Kang J, Jeong J, Kim K~M, Ha~J~W, Kim H~J.
\newblock Deformable graph transformer.
\newblock arXiv preprint arXiv:2206.14337, 2022

\bibitem{zhu2021graph}
Zhu J, Rossi R~A, Rao A, Mai T, Lipka N, Ahmed N~K, Koutra D.
\newblock Graph neural networks with heterophily.
\newblock In: Proceedings of the AAAI conference on artificial intelligence.
\newblock 2021,  11168--11176

\bibitem{zhong2022simplifying}
Zhong Z, Ivanov S, Pang J.
\newblock Simplifying node classification on heterophilous graphs with compatible label propagation.
\newblock arXiv preprint arXiv:2205.09389, 2022

\bibitem{zheng2024revisiting}
Zheng Z, Bei Y, Zhou S, Ma~Y, Gu~M, Xu~H, Lai C, Chen J, Bu~J.
\newblock Revisiting the message passing in heterophilous graph neural networks.
\newblock arXiv preprint arXiv:2405.17768, 2024

\bibitem{li2022finding}
Li~X, Zhu R, Cheng Y, Shan C, Luo S, Li~D, Qian W.
\newblock Finding global homophily in graph neural networks when meeting heterophily.
\newblock In: International Conference on Machine Learning.
\newblock 2022,  13242--13256

\bibitem{he2022block}
He~D, Liang C, Liu H, Wen M, Jiao P, Feng Z.
\newblock Block modeling-guided graph convolutional neural networks.
\newblock In: Proceedings of the AAAI conference on artificial intelligence.
\newblock 2022,  4022--4029

\bibitem{yu2024lggnn}
Yu~Z, Feng B, He~D, Wang Z, Huang Y, Feng Z.
\newblock Lg-gnn: Local-global adaptive graph neural network for modeling both homophily and heterophily.
\newblock In: Proceedings of the Thirty-Third International Joint Conference on Artificial Intelligence.
\newblock 2024

\bibitem{liu2023simga}
Liu H, Liao N, Luo S.
\newblock Simga: A simple and effective heterophilous graph neural network with efficient global aggregation.
\newblock arXiv preprint arXiv:2305.09958, 2023

\bibitem{liu2022uplifting}
Liu X, Zhang L, Guan H.
\newblock Uplifting message passing neural network with graph original information.
\newblock arXiv preprint arXiv:2210.05382, 2022

\bibitem{bo2021beyond}
Bo~D, Wang X, Shi C, Shen H.
\newblock Beyond low-frequency information in graph convolutional networks.
\newblock In: Proceedings of the AAAI conference on artificial intelligence.
\newblock 2021,  3950--3957

\bibitem{wu2023signed}
Wu~Y, Hu~L, Wang Y.
\newblock Signed attention based graph neural network for graphs with heterophily.
\newblock Neurocomputing, 2023, 557: 126731

\bibitem{lai2023self}
Lai Y, Zhang T, Fan R.
\newblock Self-attention dual embedding for graphs with heterophily.
\newblock arXiv preprint arXiv:2305.18385, 2023

\bibitem{yan2022two}
Yan Y, Hashemi M, Swersky K, Yang Y, Koutra D.
\newblock Two sides of the same coin: Heterophily and oversmoothing in graph convolutional neural networks.
\newblock In: 2022 IEEE International Conference on Data Mining (ICDM).
\newblock 2022,  1287--1292

\bibitem{choi2023signed}
Choi Y, Choi J, Ko~T, Kim C~K.
\newblock Is signed message essential for graph neural networks.
\newblock arXiv preprint arXiv:2301.08918, 2023

\bibitem{liang2024sign}
Liang L, Kim S, Shin K, Xu~Z, Pan S, Qi~Y.
\newblock Sign is not a remedy: Multiset-to-multiset message passing for learning on heterophilic graphs.
\newblock arXiv preprint arXiv:2405.20652, 2024

\bibitem{sun2024breaking}
Sun H, Li~X, Wu~Z, Su~D, Li~R~H, Wang G.
\newblock Breaking the entanglement of homophily and heterophily in semi-supervised node classification.
\newblock In: 2024 IEEE 40th International Conference on Data Engineering (ICDE).
\newblock 2024,  2379--2392

\bibitem{chaudhary2024gnndld}
Chaudhary C, Boran N~K, Sangeeth N, Singh V.
\newblock Gnndld: Graph neural network with directional label distribution.
\newblock In: ICAART (2).
\newblock 2024,  165--176

\bibitem{rossi2024edge}
Rossi E, Charpentier B, Di~Giovanni F, Frasca F, G{\"u}nnemann S, Bronstein M~M.
\newblock Edge directionality improves learning on heterophilic graphs.
\newblock In: Learning on Graphs Conference.
\newblock 2024,  25--1

\bibitem{koke2023holonets}
Koke C, Cremers D.
\newblock Holonets: Spectral convolutions do extend to directed graphs.
\newblock arXiv preprint arXiv:2310.02232, 2023

\bibitem{zhuo2024commute}
Zhuo W, Tan G.
\newblock Commute graph neural networks.
\newblock arXiv preprint arXiv:2407.01635, 2024

\bibitem{du2022gbk}
Du~L, Shi X, Fu~Q, Ma~X, Liu H, Han S, Zhang D.
\newblock Gbk-gnn: Gated bi-kernel graph neural networks for modeling both homophily and heterophily.
\newblock In: Proceedings of the ACM Web Conference 2022.
\newblock 2022,  1550--1558

\bibitem{wang2024heterophilic}
Wang K, Zhang G, Zhang X, Fang J, Wu~X, Li~G, Pan S, Huang W, Liang Y.
\newblock The heterophilic snowflake hypothesis: Training and empowering gnns for heterophilic graphs.
\newblock In: Proceedings of the 30th ACM SIGKDD Conference on Knowledge Discovery and Data Mining.
\newblock 2024,  3164--3175

\bibitem{cheng2023prioritized}
Cheng Y, Chen M, Li~X, Shan C, Gao M.
\newblock Prioritized propagation in graph neural networks.
\newblock arXiv preprint arXiv:2311.02832, 2023

\bibitem{deng2024learning}
Deng G, Zhou H, Kannan R, Prasanna V.
\newblock Learning personalized scoping for graph neural networks under heterophily.
\newblock arXiv preprint arXiv:2409.06998, 2024

\bibitem{wang2024heterophily}
Wang J, Guo Y, Yang L, Wang Y.
\newblock Heterophily-aware graph attention network.
\newblock Pattern Recognition, 2024, 156: 110738

\bibitem{rusch2022gradient}
Rusch T~K, Chamberlain B~P, Mahoney M~W, Bronstein M~M, Mishra S.
\newblock Gradient gating for deep multi-rate learning on graphs.
\newblock In: The Eleventh International Conference on Learning Representations.
\newblock 2022

\bibitem{finkelshtein2023cooperative}
Finkelshtein B, Huang X, Bronstein M, Ceylan I~I.
\newblock Cooperative graph neural networks.
\newblock arXiv preprint arXiv:2310.01267, 2023

\bibitem{ma2024polyformer}
Ma~J, He~M, Wei Z.
\newblock Polyformer: Scalable node-wise filters via polynomial graph transformer.
\newblock arXiv preprint arXiv:2407.14459, 2024

\bibitem{deng2024polynormer}
Deng C, Yue Z, Zhang Z.
\newblock Polynormer: Polynomial-expressive graph transformer in linear time.
\newblock arXiv preprint arXiv:2403.01232, 2024

\bibitem{chen2023signgt}
Chen J, Li~G, Hopcroft J~E, He~K.
\newblock Signgt: Signed attention-based graph transformer for graph representation learning.
\newblock arXiv preprint arXiv:2310.11025, 2023

\bibitem{chen2024sigformer}
Chen S, Chen J, Zhou S, Wang B, Han S, Su~C, Yuan Y, Wang C.
\newblock Sigformer: Sign-aware graph transformer for recommendation.
\newblock arXiv preprint arXiv:2404.11982, 2024

\bibitem{kuang2024transformer}
Kuang W, Wang Z, Wei Z, Li~Y, Ding B.
\newblock When transformer meets large graphs: An expressive and efficient two-view architecture.
\newblock IEEE Transactions on Knowledge and Data Engineering, 2024

\bibitem{liu2023gapformer}
Liu C, Zhan Y, Ma~X, Ding L, Tao D, Wu~J, Hu~W.
\newblock Gapformer: Graph transformer with graph pooling for node classification.
\newblock In: IJCAI.
\newblock 2023,  2196--2205

\bibitem{li2024learning}
Li~W, Chen K, Liu S, Zheng T, Huang W, Song M.
\newblock Learning a mini-batch graph transformer via two-stage interaction augmentation.
\newblock arXiv preprint arXiv:2407.09904, 2024

\bibitem{fu2024vcr}
Fu~D, Hua Z, Xie Y, Fang J, Zhang S, Sancak K, Wu~H, Malevich A, He~J, Long B.
\newblock Vcr-graphormer: A mini-batch graph transformer via virtual connections.
\newblock arXiv preprint arXiv:2403.16030, 2024

\bibitem{xing2024less}
Xing Y, Wang X, Li~Y, Huang H, Shi C.
\newblock Less is more: on the over-globalizing problem in graph transformers.
\newblock arXiv preprint arXiv:2405.01102, 2024

\bibitem{zhang2022hierarchical}
Zhang Z, Liu Q, Hu~Q, Lee C~K.
\newblock Hierarchical graph transformer with adaptive node sampling.
\newblock Advances in Neural Information Processing Systems, 2022, 35: 21171--21183

\bibitem{chen2024ntformer}
Chen J, Jiang S, He~K.
\newblock Ntformer: A composite node tokenized graph transformer for node classification.
\newblock arXiv preprint arXiv:2406.19249, 2024

\bibitem{limpformer}
Li~D, Qi~B, Gao J, Xiong H, Gu~B, Chen X.
\newblock Mpformer: Advancing graph modeling through heterophily relationship-based position encoding.
\newblock Openreview, 2024

\bibitem{ma2023rethinking}
Ma~X, Chen Q, Wu~Y, Song G, Wang L, Zheng B.
\newblock Rethinking structural encodings: Adaptive graph transformer for node classification task.
\newblock In: Proceedings of the ACM Web Conference 2023.
\newblock 2023,  533--544

\bibitem{muller2023attending}
M{\"u}ller L, Galkin M, Morris C, Ramp{\'a}{\v{s}}ek L.
\newblock Attending to graph transformers.
\newblock arXiv preprint arXiv:2302.04181, 2023

\bibitem{bo2023specformer}
Bo~D, Shi C, Wang L, Liao R.
\newblock Specformer: Spectral graph neural networks meet transformers.
\newblock arXiv preprint arXiv:2303.01028, 2023

\bibitem{wang2024graph}
Wang X, Zhu Y, Shi H, Liu Y, Hong C.
\newblock Graph triple attention network: A decoupled perspective.
\newblock arXiv preprint arXiv:2408.07654, 2024

\bibitem{chen2022optimization}
Chen Q, Wang Y, Wang Y, Yang J, Lin Z.
\newblock Optimization-induced graph implicit nonlinear diffusion.
\newblock In: International Conference on Machine Learning.
\newblock 2022,  3648--3661

\bibitem{eliasof2021pde}
Eliasof M, Haber E, Treister E.
\newblock Pde-gcn: Novel architectures for graph neural networks motivated by partial differential equations.
\newblock Advances in neural information processing systems, 2021, 34: 3836--3849

\bibitem{rusch2022graph}
Rusch T~K, Chamberlain B, Rowbottom J, Mishra S, Bronstein M.
\newblock Graph-coupled oscillator networks.
\newblock In: International Conference on Machine Learning.
\newblock 2022,  18888--18909

\bibitem{zhao2023graph}
Zhao K, Kang Q, Song Y, She R, Wang S, Tay W~P.
\newblock Graph neural convection-diffusion with heterophily.
\newblock arXiv preprint arXiv:2305.16780, 2023

\bibitem{wang2022acmp}
Wang Y, Yi~K, Liu X, Wang Y~G, Jin S.
\newblock Acmp: Allen-cahn message passing with attractive and repulsive forces for graph neural networks.
\newblock In: The Eleventh International Conference on Learning Representations.
\newblock 2022

\bibitem{choi2023gread}
Choi J, Hong S, Park N, Cho S~B.
\newblock Gread: Graph neural reaction-diffusion networks.
\newblock In: International Conference on Machine Learning.
\newblock 2023,  5722--5747

\bibitem{eliasof2023adr}
Eliasof M, Haber E, Treister E.
\newblock Adr-gnn: Advection-diffusion-reaction graph neural networks.
\newblock arXiv preprint arXiv:2307.16092, 2023

\bibitem{maskey2024fractional}
Maskey S, Paolino R, Bacho A, Kutyniok G.
\newblock A fractional graph laplacian approach to oversmoothing.
\newblock Advances in Neural Information Processing Systems, 2024, 36

\bibitem{zhang2024unleashing}
Zhang A, Li~P.
\newblock Unleashing the power of high-pass filtering in continuous graph neural networks.
\newblock In: Asian Conference on Machine Learning.
\newblock 2024,  1683--1698

\bibitem{shao2023unifying}
Shao Z, Shi D, Han A, Guo Y, Zhao Q, Gao J.
\newblock Unifying over-smoothing and over-squashing in graph neural networks: A physics informed approach and beyond.
\newblock arXiv preprint arXiv:2309.02769, 2023

\bibitem{gravina2022anti}
Gravina A, Bacciu D, Gallicchio C.
\newblock Anti-symmetric dgn: a stable architecture for deep graph networks.
\newblock arXiv preprint arXiv:2210.09789, 2022

\bibitem{bodnar2022neural}
Bodnar C, Di~Giovanni F, Chamberlain B, Lio P, Bronstein M.
\newblock Neural sheaf diffusion: A topological perspective on heterophily and oversmoothing in gnns.
\newblock Advances in Neural Information Processing Systems, 2022, 35: 18527--18541

\bibitem{barbero2022sheaf}
Barbero F, Bodnar C, Oc{\'a}riz~Borde d~H~S, Bronstein M, Veli{\v{c}}kovi{\'c} P, Li{\`o} P.
\newblock Sheaf neural networks with connection laplacians.
\newblock In: Topological, Algebraic and Geometric Learning Workshops 2022.
\newblock 2022,  28--36

\bibitem{markovich2023qdc}
Markovich T.
\newblock Qdc: Quantum diffusion convolution kernels on graphs.
\newblock arXiv preprint arXiv:2307.11234, 2023

\bibitem{di2022understanding}
Di~Giovanni F, Rowbottom J, Chamberlain B~P, Markovich T, Bronstein M~M.
\newblock Understanding convolution on graphs via energies.
\newblock arXiv preprint arXiv:2206.10991, 2022

\bibitem{zhang2023steering}
Zhang A, Li~P, Chen G.
\newblock Steering graph neural networks with pinning control.
\newblock arXiv preprint arXiv:2303.01265, 2023

\bibitem{wan2024flexible}
Wan L, Han H, Sun L, Zhang Z, Ning Z, Yan X, Xia F.
\newblock Flexible graph neural diffusion with latent class representation learning.
\newblock In: Proceedings of the 30th ACM SIGKDD Conference on Knowledge Discovery and Data Mining.
\newblock 2024,  2936--2947

\bibitem{li2024generalized}
Li~Y, Wang X, Liu H, Shi C.
\newblock A generalized neural diffusion framework on graphs.
\newblock In: Proceedings of the AAAI Conference on Artificial Intelligence.
\newblock 2024,  8707--8715

\bibitem{ortega2018graph}
Ortega A, Frossard P, Kova{\v{c}}evi{\'c} J, Moura J~M, Vandergheynst P.
\newblock Graph signal processing: Overview, challenges, and applications.
\newblock Proceedings of the IEEE, 2018, 106(5): 808--828

\bibitem{nt2019revisiting}
Nt~H, Maehara T.
\newblock Revisiting graph neural networks: All we have is low-pass filters.
\newblock arXiv preprint arXiv:1905.09550, 2019

\bibitem{gasteiger2018predict}
Gasteiger J, Bojchevski A, G{\"u}nnemann S.
\newblock Predict then propagate: Graph neural networks meet personalized pagerank.
\newblock arXiv preprint arXiv:1810.05997, 2018

\bibitem{li2019label}
Li~Q, Wu~X~M, Liu H, Zhang X, Guan Z.
\newblock Label efficient semi-supervised learning via graph filtering.
\newblock In: Proceedings of the IEEE/CVF conference on computer vision and pattern recognition.
\newblock 2019,  9582--9591

\bibitem{gasteiger2019diffusion}
Gasteiger J, Wei{\ss}enberger S, G{\"u}nnemann S.
\newblock Diffusion improves graph learning.
\newblock Advances in neural information processing systems, 2019, 32

\bibitem{li2019optimizing}
Li~P, Chien I, Milenkovic O.
\newblock Optimizing generalized pagerank methods for seed-expansion community detection.
\newblock Advances in Neural Information Processing Systems, 2019, 32

\bibitem{defferrard2016convolutional}
Defferrard M, Bresson X, Vandergheynst P.
\newblock Convolutional neural networks on graphs with fast localized spectral filtering.
\newblock Advances in neural information processing systems, 2016, 29

\bibitem{gil2007numerical}
Gil A, Segura J, Temme N~M.
\newblock Numerical methods for special functions.
\newblock SIAM, 2007

\bibitem{weber2008analysis}
Weber A.
\newblock Analysis of the physical laplacian and the heat flow on a locally finite graph.
\newblock arXiv preprint arXiv:0801.0812, 2008

\bibitem{kroeker1977wiener}
Kroeker J.
\newblock Wiener analysis of nonlinear systems using poisson-charlier crosscorrelation.
\newblock Biological Cybernetics, 1977, 27(4): 221--227

\bibitem{hildebrand1987introduction}
Hildebrand F.
\newblock Introduction to numerical analysis: Courier corporation.
\newblock Courier Corporation, Chelmsford, MA, USA, 1987

\bibitem{liesen2013krylov}
Liesen J, Strakos Z.
\newblock Krylov subspace methods: principles and analysis.
\newblock Numerical Mathematics and Scie, 2013

\bibitem{cai2024survey}
Cai W, Jiang J, Wang F, Tang J, Kim S, Huang J.
\newblock A survey on mixture of experts.
\newblock arXiv preprint arXiv:2407.06204, 2024

\bibitem{zheng2023node}
Zheng S, Zhu Z, Liu Z, Li~Y, Zhao Y.
\newblock Node-oriented spectral filtering for graph neural networks.
\newblock IEEE Transactions on Pattern Analysis and Machine Intelligence, 2023

\bibitem{xuslog}
Xu~H, Yan Y, Wang D, Xu~Z, Zeng Z, Abdelzaher T~F, Han J, Tong H.
\newblock Slog: An inductive spectral graph neural network beyond polynomial filter.
\newblock In: Forty-first International Conference on Machine Learning.
\newblock 2024

\bibitem{tenenbaum2000global}
Tenenbaum J~B, Silva V~d, Langford J~C.
\newblock A global geometric framework for nonlinear dimensionality reduction.
\newblock science, 2000, 290(5500): 2319--2323

\bibitem{nickel2017poincare}
Nickel M, Kiela D.
\newblock Poincar{\'e} embeddings for learning hierarchical representations.
\newblock Advances in neural information processing systems, 2017, 30

\bibitem{ribeiro2017struc2vec}
Ribeiro L~F, Saverese P~H, Figueiredo D~R.
\newblock struc2vec: Learning node representations from structural identity.
\newblock In: Proceedings of the 23rd ACM SIGKDD international conference on knowledge discovery and data mining.
\newblock 2017,  385--394

\bibitem{grover2016node2vec}
Grover A, Leskovec J.
\newblock node2vec: Scalable feature learning for networks.
\newblock In: Proceedings of the 22nd ACM SIGKDD international conference on Knowledge discovery and data mining.
\newblock 2016,  855--864

\bibitem{vinyals2015pointer}
Vinyals O, Fortunato M, Jaitly N.
\newblock Pointer networks.
\newblock Advances in neural information processing systems, 2015, 28

\bibitem{liang2024predicting}
Liang L, Hu~X, Xu~Z, Song Z, King I.
\newblock Predicting global label relationship matrix for graph neural networks under heterophily.
\newblock Advances in Neural Information Processing Systems, 2024, 36

\bibitem{jeh2002simrank}
Jeh G, Widom J.
\newblock Simrank: a measure of structural-context similarity.
\newblock In: Proceedings of the eighth ACM SIGKDD international conference on Knowledge discovery and data mining.
\newblock 2002,  538--543

\bibitem{cao2007learning}
Cao Z, Qin T, Liu T~Y, Tsai M~F, Li~H.
\newblock Learning to rank: from pairwise approach to listwise approach.
\newblock In: Proceedings of the 24th international conference on Machine learning.
\newblock 2007,  129--136

\bibitem{wang2023snowflake}
Wang K, Li~G, Wang S, Zhang G, Wang K, You Y, Peng X, Liang Y, Wang Y.
\newblock The snowflake hypothesis: Training deep gnn with one node one receptive field.
\newblock arXiv preprint arXiv:2308.10051, 2023

\bibitem{katz1953new}
Katz L.
\newblock A new status index derived from sociometric analysis.
\newblock Psychometrika, 1953, 18(1): 39--43

\bibitem{chamberlain2021grand}
Chamberlain B, Rowbottom J, Gorinova M~I, Bronstein M, Webb S, Rossi E.
\newblock Grand: Graph neural diffusion.
\newblock In: International conference on machine learning.
\newblock 2021,  1407--1418

\bibitem{han2023continuous}
Han A, Shi D, Lin L, Gao J.
\newblock From continuous dynamics to graph neural networks: Neural diffusion and beyond.
\newblock arXiv preprint arXiv:2310.10121, 2023

\bibitem{chamberlain2021beltrami}
Chamberlain B, Rowbottom J, Eynard D, Di~Giovanni F, Dong X, Bronstein M.
\newblock Beltrami flow and neural diffusion on graphs.
\newblock Advances in Neural Information Processing Systems, 2021, 34: 1594--1609

\bibitem{thorpe2022grand++}
Thorpe M, Nguyen T, Xia H, Strohmer T, Bertozzi A, Osher S, Wang B.
\newblock Grand++: Graph neural diffusion with a source term.
\newblock ICLR, 2022

\bibitem{allen1979microscopic}
Allen S~M, Cahn J~W.
\newblock A microscopic theory for antiphase boundary motion and its application to antiphase domain coarsening.
\newblock Acta metallurgica, 1979, 27(6): 1085--1095

\bibitem{fisher1937wave}
Fisher R~A.
\newblock The wave of advance of advantageous genes.
\newblock Annals of eugenics, 1937, 7(4): 355--369

\bibitem{gilding2004travelling}
Gilding B~H, Kersner R.
\newblock Travelling waves in nonlinear diffusion-convection reaction. volume~60.
\newblock Springer Science \& Business Media, 2004

\bibitem{hansen2019toward}
Hansen J, Ghrist R.
\newblock Toward a spectral theory of cellular sheaves.
\newblock Journal of Applied and Computational Topology, 2019, 3(4): 315--358

\bibitem{yu2013synchronization}
Yu~W, Chen G, Lu~J, Kurths J.
\newblock Synchronization via pinning control on general complex networks.
\newblock SIAM Journal on Control and Optimization, 2013, 51(2): 1395--1416

\bibitem{parkmitigating}
Park M, Heo J, Kim D.
\newblock Mitigating oversmoothing through reverse process of gnns for heterophilic graphs.
\newblock In: Forty-first International Conference on Machine Learning.
\newblock 2024

\bibitem{jaiswal2020survey}
Jaiswal A, Babu A~R, Zadeh M~Z, Banerjee D, Makedon F.
\newblock A survey on contrastive self-supervised learning.
\newblock Technologies, 2020, 9(1): 2

\bibitem{jing2020self}
Jing L, Tian Y.
\newblock Self-supervised visual feature learning with deep neural networks: A survey.
\newblock IEEE transactions on pattern analysis and machine intelligence, 2020, 43(11): 4037--4058

\bibitem{gui2024survey}
Gui J, Chen T, Zhang J, Cao Q, Sun Z, Luo H, Tao D.
\newblock A survey on self-supervised learning: Algorithms, applications, and future trends.
\newblock IEEE Transactions on Pattern Analysis and Machine Intelligence, 2024

\bibitem{guo2023architecture}
Guo X, Wang Y, Wei Z, Wang Y.
\newblock Architecture matters: Uncovering implicit mechanisms in graph contrastive learning.
\newblock Advances in Neural Information Processing Systems, 2023, 36: 28585--28610

\bibitem{yanggraph}
Yang W, Mirzasoleiman B.
\newblock Graph contrastive learning under heterophily via graph filters.
\newblock In: The 40th Conference on Uncertainty in Artificial Intelligence.
\newblock 2024

\bibitem{liu2023beyond}
Liu Y, Zheng Y, Zhang D, Lee V~C, Pan S.
\newblock Beyond smoothing: Unsupervised graph representation learning with edge heterophily discriminating.
\newblock In: Proceedings of the AAAI conference on artificial intelligence.
\newblock 2023,  4516--4524

\bibitem{chen2024polygcl}
Chen J, Lei R, Wei Z.
\newblock Polygcl: Graph contrastive learning via learnable spectral polynomial filters.
\newblock In: The Twelfth International Conference on Learning Representations.
\newblock 2024

\bibitem{velivckovic2018deep}
Veli{\v{c}}kovi{\'c} P, Fedus W, Hamilton W~L, Li{\`o} P, Bengio Y, Hjelm R~D.
\newblock Deep graph infomax.
\newblock arXiv preprint arXiv:1809.10341, 2018

\bibitem{you2020graph}
You Y, Chen T, Sui Y, Chen T, Wang Z, Shen Y.
\newblock Graph contrastive learning with augmentations.
\newblock Advances in neural information processing systems, 2020, 33: 5812--5823

\bibitem{zhu2020deep}
Zhu Y, Xu~Y, Yu~F, Liu Q, Wu~S, Wang L.
\newblock Deep graph contrastive representation learning.
\newblock arXiv preprint arXiv:2006.04131, 2020

\bibitem{liu2024simgcl}
Liu C, Yu~C, Gui N, Yu~Z, Deng S.
\newblock Simgcl: graph contrastive learning by finding homophily in heterophily.
\newblock Knowledge and Information Systems, 2024, 66(3): 2089--2114

\bibitem{chen2022towards}
Chen J, Zhu G, Qi~Y, Yuan C, Huang Y.
\newblock Towards self-supervised learning on graphs with heterophily.
\newblock In: Proceedings of the 31st ACM International Conference on Information \& Knowledge Management.
\newblock 2022,  201--211

\bibitem{wanghetergcl}
Wang C, Liu Y, Yang Y, Li~W.
\newblock Hetergcl: Graph contrastive learning framework on heterophilic graph.
\newblock In: Proceedings of the Thirty-Third International Joint Conference on Artificial Intelligence.
\newblock 2024

\bibitem{yang2024gauss}
Yang L, Hu~W, Xu~J, Shi R, He~D, Wang C, Cao X, Wang Z, Niu B, Guo Y.
\newblock Gauss: Graph-customized universal self-supervised learning.
\newblock In: Proceedings of the ACM on Web Conference 2024.
\newblock 2024,  582--593

\bibitem{wang2022single}
Wang H, Zhang J, Zhu Q, Huang W, Kawaguchi K, Xiao X.
\newblock Single-pass contrastive learning can work for both homophilic and heterophilic graph.
\newblock arXiv preprint arXiv:2211.10890, 2022

\bibitem{wang2022augmentation}
Wang H, Zhang J, Zhu Q, Huang W.
\newblock Augmentation-free graph contrastive learning with performance guarantee.
\newblock arXiv preprint arXiv:2204.04874, 2022

\bibitem{khan2023contrastive}
Khan A, Storkey A.
\newblock Contrastive learning for non-local graphs with multi-resolution structural views.
\newblock arXiv preprint arXiv:2308.10077, 2023

\bibitem{coifman2006diffusion}
Coifman R~R, Maggioni M.
\newblock Diffusion wavelets.
\newblock Applied and computational harmonic analysis, 2006, 21(1): 53--94

\bibitem{yuan2023muse}
Yuan M, Chen M, Li~X.
\newblock Muse: Multi-view contrastive learning for heterophilic graphs.
\newblock In: Proceedings of the 32nd ACM International Conference on Information and Knowledge Management.
\newblock 2023,  3094--3103

\bibitem{altenburger2018monophily}
Altenburger K~M, Ugander J.
\newblock Monophily in social networks introduces similarity among friends-of-friends.
\newblock Nature human behaviour, 2018, 2(4): 284--290

\bibitem{xiaoefficient}
Xiao T, Zhu H, Zhang Z, Guo Z, Aggarwal C~C, Wang S, Honavar V~G.
\newblock Efficient contrastive learning for fast and accurate inference on graphs.
\newblock In: Forty-first International Conference on Machine Learning.
\newblock 2024

\bibitem{wans3gcl}
Wan G, Tian Y, Huang W, Chawla N~V, Ye~M.
\newblock S3gcl: Spectral, swift, spatial graph contrastive learning.
\newblock In: Forty-first International Conference on Machine Learning.
\newblock 2024

\bibitem{he2023contrastive}
He~D, Zhao J, Guo R, Feng Z, Jin D, Huang Y, Wang Z, Zhang W.
\newblock Contrastive learning meets homophily: two birds with one stone.
\newblock In: International Conference on Machine Learning.
\newblock 2023,  12775--12789

\bibitem{li2023homogcl}
Li~W~Z, Wang C~D, Xiong H, Lai J~H.
\newblock Homogcl: Rethinking homophily in graph contrastive learning.
\newblock In: Proceedings of the 29th ACM SIGKDD Conference on Knowledge Discovery and Data Mining.
\newblock 2023,  1341--1352

\bibitem{zhuo2024improving}
Zhuo J, Qin F, Cui C, Fu~K, Niu B, Wang M, Guo Y, Wang C, Wang Z, Cao X, others .
\newblock Improving graph contrastive learning via adaptive positive sampling.
\newblock In: Proceedings of the IEEE/CVF Conference on Computer Vision and Pattern Recognition.
\newblock 2024,  23179--23187

\bibitem{zhuo2024graph}
Zhuo J, Cui C, Fu~K, Niu B, He~D, Wang C, Guo Y, Wang Z, Cao X, Yang L.
\newblock Graph contrastive learning reimagined: Exploring universality.
\newblock In: Proceedings of the ACM on Web Conference 2024.
\newblock 2024,  641--651

\bibitem{zhao2024disambiguated}
Zhao T, Zhang X, Wang S.
\newblock Disambiguated node classification with graph neural networks.
\newblock In: Proceedings of the ACM on Web Conference 2024.
\newblock 2024,  914--923

\bibitem{kipf2016variational}
Kipf T~N, Welling M.
\newblock Variational graph auto-encoders.
\newblock arXiv preprint arXiv:1611.07308, 2016

\bibitem{li2023s}
Li~J, Wu~R, Sun W, Chen L, Tian S, Zhu L, Meng C, Zheng Z, Wang W.
\newblock What's behind the mask: Understanding masked graph modeling for graph autoencoders.
\newblock In: Proceedings of the 29th ACM SIGKDD Conference on Knowledge Discovery and Data Mining.
\newblock 2023,  1268--1279

\bibitem{zhong2022unsupervised}
Zhong Z, Gonzalez G, Grattarola D, Pang J.
\newblock Unsupervised network embedding beyond homophily.
\newblock arXiv preprint arXiv:2203.10866, 2022

\bibitem{lin2023multi}
Lin B, Li~Y, Gui N, Xu~Z, Yu~Z.
\newblock Multi-view graph representation learning beyond homophily.
\newblock ACM Transactions on Knowledge Discovery from Data, 2023, 17(8): 1--21

\bibitem{li2024redundancy}
Li~M, Zhang Y, Wang S, Hu~Y, Yin B.
\newblock Redundancy is not what you need: An embedding fusion graph auto-encoder for self-supervised graph representation learning.
\newblock IEEE Transactions on Neural Networks and Learning Systems, 2024

\bibitem{li2022graph}
Li~Y, Lin B, Luo B, Gui N.
\newblock Graph representation learning beyond node and homophily.
\newblock IEEE Transactions on Knowledge and Data Engineering, 2022, 35(5): 4880--4893

\bibitem{tang2022graph}
Tang M, Yang C, Li~P.
\newblock Graph auto-encoder via neighborhood wasserstein reconstruction.
\newblock arXiv preprint arXiv:2202.09025, 2022

\bibitem{tian2024ugmae}
Tian Y, Zhang C, Kou Z, Liu Z, Zhang X, Chawla N~V.
\newblock Ugmae: A unified framework for graph masked autoencoders.
\newblock arXiv preprint arXiv:2402.08023, 2024

\bibitem{luo2024masked}
Luo Y, Li~S, Sui Y, Wu~J, Wu~J, Wang X.
\newblock Masked graph modeling with multi-view contrast.
\newblock In: 2024 IEEE 40th International Conference on Data Engineering (ICDE).
\newblock 2024

\bibitem{fang2024masked}
Fang D, Zhu F, Xie D, Min W.
\newblock Masked graph autoencoders with contrastive augmentation for spatially resolved transcriptomics data.
\newblock arXiv preprint arXiv:2408.06377, 2024

\bibitem{yang2023cmgae}
Yang W, Zhou L.
\newblock Cmgae: Enhancing graph masked autoencoders through the use of contrastive learning.
\newblock In: 2023 2nd International Conference on Machine Learning, Control, and Robotics (MLCR).
\newblock 2023,  42--47

\bibitem{xiao2022decoupled}
Xiao T, Chen Z, Guo Z, Zhuang Z, Wang S.
\newblock Decoupled self-supervised learning for graphs.
\newblock Advances in Neural Information Processing Systems, 2022, 35: 620--634

\bibitem{wei2021finetuned}
Wei J, Bosma M, Zhao V~Y, Guu K, Yu~A~W, Lester B, Du~N, Dai A~M, Le~Q~V.
\newblock Finetuned language models are zero-shot learners.
\newblock arXiv preprint arXiv:2109.01652, 2021

\bibitem{jia2021scaling}
Jia C, Yang Y, Xia Y, Chen Y~T, Parekh Z, Pham H, Le~Q, Sung Y~H, Li~Z, Duerig T.
\newblock Scaling up visual and vision-language representation learning with noisy text supervision.
\newblock In: International conference on machine learning.
\newblock 2021,  4904--4916

\bibitem{jia2022visual}
Jia M, Tang L, Chen B~C, Cardie C, Belongie S, Hariharan B, Lim S~N.
\newblock Visual prompt tuning.
\newblock In: European Conference on Computer Vision.
\newblock 2022,  709--727

\bibitem{wang2021afec}
Wang L, Zhang M, Jia Z, Li~Q, Bao C, Ma~K, Zhu J, Zhong Y.
\newblock Afec: Active forgetting of negative transfer in continual learning.
\newblock Advances in Neural Information Processing Systems, 2021, 34: 22379--22391

\bibitem{sun2022gppt}
Sun M, Zhou K, He~X, Wang Y, Wang X.
\newblock Gppt: Graph pre-training and prompt tuning to generalize graph neural networks.
\newblock In: Proceedings of the 28th ACM SIGKDD Conference on Knowledge Discovery and Data Mining.
\newblock 2022,  1717--1727

\bibitem{liu2023graphprompt}
Liu Z, Yu~X, Fang Y, Zhang X.
\newblock Graphprompt: Unifying pre-training and downstream tasks for graph neural networks.
\newblock In: Proceedings of the ACM Web Conference 2023.
\newblock 2023,  417--428

\bibitem{yu2024generalized}
Yu~X, Liu Z, Fang Y, Liu Z, Chen S, Zhang X.
\newblock Generalized graph prompt: Toward a unification of pre-training and downstream tasks on graphs.
\newblock IEEE Transactions on Knowledge and Data Engineering, 2024

\bibitem{huang2024prodigy}
Huang Q, Ren H, Chen P, Kr{\v{z}}manc G, Zeng D, Liang P~S, Leskovec J.
\newblock Prodigy: Enabling in-context learning over graphs.
\newblock Advances in Neural Information Processing Systems, 2024, 36

\bibitem{zhu2023sgl}
Zhu Y, Guo J, Tang S.
\newblock Sgl-pt: A strong graph learner with graph prompt tuning.
\newblock arXiv preprint arXiv:2302.12449, 2023

\bibitem{sun2023all}
Sun X, Cheng H, Li~J, Liu B, Guan J.
\newblock All in one: Multi-task prompting for graph neural networks.
\newblock In: Proceedings of the 29th ACM SIGKDD Conference on Knowledge Discovery and Data Mining.
\newblock 2023,  2120--2131

\bibitem{zi2024prog}
Zi~C, Zhao H, Sun X, Lin Y, Cheng H, Li~J.
\newblock Prog: A graph prompt learning benchmark.
\newblock arXiv preprint arXiv:2406.05346, 2024

\bibitem{liu2023one}
Liu H, Feng J, Kong L, Liang N, Tao D, Chen Y, Zhang M.
\newblock One for all: Towards training one graph model for all classification tasks.
\newblock arXiv preprint arXiv:2310.00149, 2023

\bibitem{fang2024universal}
Fang T, Zhang Y, Yang Y, Wang C, Chen L.
\newblock Universal prompt tuning for graph neural networks.
\newblock Advances in Neural Information Processing Systems, 2024, 36

\bibitem{tan2023virtual}
Tan Z, Guo R, Ding K, Liu H.
\newblock Virtual node tuning for few-shot node classification.
\newblock In: Proceedings of the 29th ACM SIGKDD Conference on Knowledge Discovery and Data Mining.
\newblock 2023,  2177--2188

\bibitem{lee2024subgraph}
Lee J, Yang W, Kang J.
\newblock Subgraph-level universal prompt tuning.
\newblock arXiv preprint arXiv:2402.10380, 2024

\bibitem{yu2024multigprompt}
Yu~X, Zhou C, Fang Y, Zhang X.
\newblock Multigprompt for multi-task pre-training and prompting on graphs.
\newblock In: Proceedings of the ACM on Web Conference 2024.
\newblock 2024,  515--526

\bibitem{jiang2024unified}
Jiang B, Wu~H, Zhang Z, Wang B, Tang J.
\newblock A unified graph selective prompt learning for graph neural networks.
\newblock arXiv preprint arXiv:2406.10498, 2024

\bibitem{yan2024inductive}
Yan Y, Zhang P, Fang Z, Long Q.
\newblock Inductive graph alignment prompt: Bridging the gap between graph pre-training and inductive fine-tuning from spectral perspective.
\newblock In: Proceedings of the ACM on Web Conference 2024.
\newblock 2024,  4328--4339

\bibitem{chen2023ultra}
Chen M, Liu Z, Liu C, Li~J, Mao Q, Sun J.
\newblock Ultra-dp: Unifying graph pre-training with multi-task graph dual prompt.
\newblock arXiv preprint arXiv:2310.14845, 2023

\bibitem{wang2024novel}
Wang J, Deng Z, Lin T, Li~W, Ling S.
\newblock A novel prompt tuning for graph transformers: Tailoring prompts to graph topologies.
\newblock In: Proceedings of the 30th ACM SIGKDD Conference on Knowledge Discovery and Data Mining.
\newblock 2024,  3116--3127

\bibitem{ma2024hetgpt}
Ma~Y, Yan N, Li~J, Mortazavi M, Chawla N~V.
\newblock Hetgpt: Harnessing the power of prompt tuning in pre-trained heterogeneous graph neural networks.
\newblock In: Proceedings of the ACM on Web Conference 2024.
\newblock 2024,  1015--1023

\bibitem{yu2024hgprompt}
Yu~X, Fang Y, Liu Z, Zhang X.
\newblock Hgprompt: Bridging homogeneous and heterogeneous graphs for few-shot prompt learning.
\newblock In: Proceedings of the AAAI Conference on Artificial Intelligence.
\newblock 2024,  16578--16586

\bibitem{yu2024dygprompt}
Yu~X, Liu Z, Fang Y, Zhang X.
\newblock Dygprompt: Learning feature and time prompts on dynamic graphs.
\newblock arXiv preprint arXiv:2405.13937, 2024

\bibitem{song2024krait}
Song Y, Singh R, Palanisamy B.
\newblock Krait: A backdoor attack against graph prompt tuning.
\newblock arXiv preprint arXiv:2407.13068, 2024

\bibitem{lyu2024cross}
Lyu X, Han Y, Wang W, Qian H, Tsang I, Zhang X.
\newblock Cross-context backdoor attacks against graph prompt learning.
\newblock In: Proceedings of the 30th ACM SIGKDD Conference on Knowledge Discovery and Data Mining.
\newblock 2024,  2094--2105

\bibitem{wang2024ddiprompt}
Wang Y, Xiong Y, Wu~X, Sun X, Zhang J.
\newblock Ddiprompt: Drug-drug interaction event prediction based on graph prompt learning.
\newblock arXiv preprint arXiv:2402.11472, 2024

\bibitem{ye2023natural}
Ye~R, Zhang C, Wang R, Xu~S, Zhang Y, others .
\newblock Natural language is all a graph needs.
\newblock arXiv preprint arXiv:2308.07134, 2023, 4(5): 7

\bibitem{duan2024g}
Duan Y, Liu J, Chen S, Chen L, Wu~J.
\newblock G-prompt: Graphon-based prompt tuning for graph classification.
\newblock Information Processing \& Management, 2024, 61(3): 103639

\bibitem{fang2024gaugllm}
Fang Y, Fan D, Zha D, Tan Q.
\newblock Gaugllm: Improving graph contrastive learning for text-attributed graphs with large language models.
\newblock In: Proceedings of the 30th ACM SIGKDD Conference on Knowledge Discovery and Data Mining.
\newblock 2024,  747--758

\bibitem{jiang2024killing}
Jiang W, Wu~W, Zhang L, Yuan Z, Xiang J, Zhou J, Xiong H.
\newblock Killing two birds with one stone: Cross-modal reinforced prompting for graph and language tasks.
\newblock In: Proceedings of the 30th ACM SIGKDD Conference on Knowledge Discovery and Data Mining.
\newblock 2024,  1301--1312

\bibitem{jin2024urban}
Jin J, Song Y, Kan D, Zhu H, Sun X, Li~Z, Sun X, Zhang J.
\newblock Urban region pre-training and prompting: A graph-based approach.
\newblock arXiv preprint arXiv:2408.05920, 2024

\bibitem{zhang2024gpt4rec}
Zhang P, Yan Y, Zhang X, Kang L, Li~C, Huang F, Wang S, Kim S.
\newblock Gpt4rec: Graph prompt tuning for streaming recommendation.
\newblock In: Proceedings of the 47th International ACM SIGIR Conference on Research and Development in Information Retrieval.
\newblock 2024,  1774--1784

\bibitem{gong2024self}
Gong C, Li~X, Yu~J, Cheng Y, Tan J, Yu~C.
\newblock Self-pro: A self-prompt and tuning framework for graph neural networks.
\newblock In: Joint European Conference on Machine Learning and Knowledge Discovery in Databases.
\newblock 2024,  197--215

\bibitem{yu2024non}
Yu~X, Zhang J, Fang Y, Jiang R.
\newblock Non-homophilic graph pre-training and prompt learning.
\newblock arXiv preprint arXiv:2408.12594, 2024

\bibitem{zhou2022conditional}
Zhou K, Yang J, Loy C~C, Liu Z.
\newblock Conditional prompt learning for vision-language models.
\newblock In: Proceedings of the IEEE/CVF conference on computer vision and pattern recognition.
\newblock 2022,  16816--16825

\bibitem{ge2023enhancing}
Ge~Q, Zhao Z, others .
\newblock Enhancing graph neural networks with structure-based prompt.
\newblock arXiv preprint arXiv:2310.17394, 2023

\bibitem{xie2023contrastive}
Xie X, Chen W, Kang Z, Peng C.
\newblock Contrastive graph clustering with adaptive filter.
\newblock Expert Systems with Applications, 2023, 219: 119645

\bibitem{wen2024homophily}
Wen Z, Ling Y, Ren Y, Wu~T, Chen J, Pu~X, Hao Z, He~L.
\newblock Homophily-related: Adaptive hybrid graph filter for multi-view graph clustering.
\newblock In: Proceedings of the AAAI Conference on Artificial Intelligence.
\newblock 2024,  15841--15849

\bibitem{pan2023beyond}
Pan E, Kang Z.
\newblock Beyond homophily: Reconstructing structure for graph-agnostic clustering.
\newblock In: International Conference on Machine Learning.
\newblock 2023,  26868--26877

\bibitem{xie2024provable}
Xie X, Pan E, Kang Z, Chen W, Li~B.
\newblock Provable filter for real-world graph clustering.
\newblock arXiv preprint arXiv:2403.03666, 2024

\bibitem{zhu2024boosting}
Zhu P, Li~J, Wang Y, Xiao B, Zhang J, Lin W, Hu~Q.
\newblock Boosting pseudo-labeling with curriculum self-reflection for attributed graph clustering.
\newblock IEEE Transactions on Neural Networks and Learning Systems, 2024

\bibitem{gu2023homophily}
Gu~M, Yang G, Zhou S, Ma~N, Chen J, Tan Q, Liu M, Bu~J.
\newblock Homophily-enhanced structure learning for graph clustering.
\newblock In: Proceedings of the 32nd ACM International Conference on Information and Knowledge Management.
\newblock 2023,  577--586

\bibitem{zhou2022link}
Zhou S, Guo Z, Aggarwal C, Zhang X, Wang S.
\newblock Link prediction on heterophilic graphs via disentangled representation learning.
\newblock arXiv preprint arXiv:2208.01820, 2022

\bibitem{di2024link}
Di~Francesco A~G, Caso F, Bucarelli M~S, Silvestri F.
\newblock Link prediction under heterophily: A physics-inspired graph neural network approach.
\newblock arXiv preprint arXiv:2402.14802, 2024

\bibitem{yang2022incorporating}
Yang J, Medya S, Ye~W.
\newblock Incorporating heterophily into graph neural networks for graph classification.
\newblock arXiv preprint arXiv:2203.07678, 2022

\bibitem{papachristou2022glinkx}
Papachristou M, Goel R, Portman F, Miller M, Jin R.
\newblock Glinkx: A scalable unified framework for homophilous and heterophilous graphs.
\newblock arXiv preprint arXiv:2211.00550, 2022

\bibitem{liao2024ld2}
Liao N, Luo S, Li~X, Shi J.
\newblock Ld2: Scalable heterophilous graph neural network with decoupled embeddings.
\newblock Advances in Neural Information Processing Systems, 2024, 36

\bibitem{chen2023node}
Chen J, Li~Z, Zhu Y, Zhang J, Pu~J.
\newblock From node interaction to hop interaction: New effective and scalable graph learning paradigm.
\newblock In: Proceedings of the IEEE/CVF Conference on Computer Vision and Pattern Recognition.
\newblock 2023,  7876--7885

\bibitem{das2024ags}
Das S~S, Ferdous S, Halappanavar M~M, Serra E, Pothen A.
\newblock Ags-gnn: Attribute-guided sampling for graph neural networks.
\newblock In: Proceedings of the 30th ACM SIGKDD Conference on Knowledge Discovery and Data Mining.
\newblock 2024,  538--549

\bibitem{wu2022nodeformer}
Wu~Q, Zhao W, Li~Z, Wipf D~P, Yan J.
\newblock Nodeformer: A scalable graph structure learning transformer for node classification.
\newblock Advances in Neural Information Processing Systems, 2022, 35: 27387--27401

\bibitem{wu2023difformer}
Wu~Q, Yang C, Zhao W, He~Y, Wipf D, Yan J.
\newblock Difformer: Scalable (graph) transformers induced by energy constrained diffusion.
\newblock arXiv preprint arXiv:2301.09474, 2023

\bibitem{wu2024simplifying}
Wu~Q, Zhao W, Yang C, Zhang H, Nie F, Jiang H, Bian Y, Yan J.
\newblock Simplifying and empowering transformers for large-graph representations.
\newblock Advances in Neural Information Processing Systems, 2024, 36

\bibitem{kong2023goat}
Kong K, Chen J, Kirchenbauer J, Ni~R, Bruss C~B, Goldstein T.
\newblock Goat: A global transformer on large-scale graphs.
\newblock In: International Conference on Machine Learning.
\newblock 2023,  17375--17390

\bibitem{sun2024spikegraphormer}
Sun Y, Zhu D, Wang Y, Tian Z, Cao N, O'Hared G.
\newblock Spikegraphormer: A high-performance graph transformer with spiking graph attention.
\newblock arXiv preprint arXiv:2403.15480, 2024

\bibitem{zhu2022does}
Zhu J, Jin J, Loveland D, Schaub M~T, Koutra D.
\newblock How does heterophily impact the robustness of graph neural networks? theoretical connections and practical implications.
\newblock In: Proceedings of the 28th ACM SIGKDD Conference on Knowledge Discovery and Data Mining.
\newblock 2022,  2637--2647

\bibitem{huang2023robust}
Huang J, Du~L, Chen X, Fu~Q, Han S, Zhang D.
\newblock Robust mid-pass filtering graph convolutional networks.
\newblock In: Proceedings of the ACM Web Conference 2023.
\newblock 2023,  328--338

\bibitem{zhu2024universally}
Zhu Y, Lai Y, Ai~X, Zhou K.
\newblock Universally robust graph neural networks by preserving neighbor similarity.
\newblock arXiv preprint arXiv:2401.09754, 2024

\bibitem{lei2022evennet}
Lei R, Wang Z, Li~Y, Ding B, Wei Z.
\newblock Evennet: Ignoring odd-hop neighbors improves robustness of graph neural networks.
\newblock Advances in Neural Information Processing Systems, 2022, 35: 4694--4706

\bibitem{qiu2024refining}
Qiu C, Nan G, Xiong T, Deng W, Wang D, Teng Z, Sun L, Cui Q, Tao X.
\newblock Refining latent homophilic structures over heterophilic graphs for robust graph convolution networks.
\newblock In: Proceedings of the AAAI Conference on Artificial Intelligence.
\newblock 2024,  8930--8938

\bibitem{dengprospect}
Deng B, Chen J, Hu~Y, Xu~Z, Chen C, Tao Z.
\newblock Prospect: Learn mlps robust against graph adversarial structure attacks.
\newblock Openreview, 2023

\bibitem{cheng2024resurrecting}
Cheng Y, Shan C, Shen Y, Li~X, Luo S, Li~D.
\newblock Resurrecting label propagation for graphs with heterophily and label noise.
\newblock In: Proceedings of the 30th ACM SIGKDD Conference on Knowledge Discovery and Data Mining.
\newblock 2024,  433--444

\bibitem{yuan2024unveiling}
Yuan H, Xu~J, Wang C, Yang Z, Wang C, Yin K, Yang Y.
\newblock Unveiling privacy vulnerabilities: Investigating the role of structure in graph data.
\newblock In: Proceedings of the 30th ACM SIGKDD Conference on Knowledge Discovery and Data Mining.
\newblock 2024,  4059--4070

\bibitem{wu2024provable}
Wu~R, Fang G, Pan Q, Zhang M, Liu T, Wang W, Zhao W.
\newblock On provable privacy vulnerabilities of graph representations.
\newblock arXiv preprint arXiv:2402.04033, 2024

\bibitem{mueller2023privacy}
Mueller T~T, Chevli M, Daigavane A, Rueckert D, Kaissis G.
\newblock Privacy-utility trade-offs in neural networks for medical population graphs: Insights from differential privacy and graph structure.
\newblock arXiv preprint arXiv:2307.06760, 2023

\bibitem{loveland2022graph}
Loveland D, Zhu J, Heimann M, Fish B, Schaub M~T, Koutra D.
\newblock On graph neural network fairness in the presence of heterophilous neighborhoods.
\newblock arXiv preprint arXiv:2207.04376, 2022

\bibitem{xue2024data}
Xue Y, Jin Z, Gao W.
\newblock A data-centric graph neural network for node classification of heterophilic networks.
\newblock International Journal of Machine Learning and Cybernetics, 2024,  1--11

\bibitem{yang2024graph}
Yang Y, Sun Y, Wang S, Guo J, Gao J, Ju~F, Yin B.
\newblock Graph neural networks with soft association between topology and attribute.
\newblock In: Proceedings of the AAAI Conference on Artificial Intelligence.
\newblock 2024,  9260--9268

\bibitem{deac2023evolving}
Deac A, Tang J.
\newblock Evolving computation graphs.
\newblock arXiv preprint arXiv:2306.12943, 2023

\bibitem{zheng2023finding}
Zheng Y, Zhang H, Lee V, Zheng Y, Wang X, Pan S.
\newblock Finding the missing-half: Graph complementary learning for homophily-prone and heterophily-prone graphs.
\newblock In: International Conference on Machine Learning.
\newblock 2023,  42492--42505

\bibitem{xu2023node}
Xu~Z, Chen Y, Zhou Q, Wu~Y, Pan M, Yang H, Tong H.
\newblock Node classification beyond homophily: Towards a general solution.
\newblock In: Proceedings of the 29th ACM SIGKDD Conference on Knowledge Discovery and Data Mining.
\newblock 2023,  2862--2873

\bibitem{bi2024make}
Bi~W, Du~L, Fu~Q, Wang Y, Han S, Zhang D.
\newblock Make heterophilic graphs better fit gnn: A graph rewiring approach.
\newblock IEEE Transactions on Knowledge and Data Engineering, 2024

\bibitem{choi2022finding}
Choi Y, Choi J, Ko~T, Byun H, Kim C~K.
\newblock Finding heterophilic neighbors via confidence-based subgraph matching for semi-supervised node classification.
\newblock In: Proceedings of the 31st ACM International Conference on Information \& Knowledge Management.
\newblock 2022,  283--292

\bibitem{jiang2021gcn}
Jiang M, Liu G, Su~Y, Wu~X.
\newblock Gcn-sl: Graph convolutional networks with structure learning for graphs under heterophily.
\newblock arXiv preprint arXiv:2105.13795, 2021

\bibitem{wang2021graph}
Wang R, Mou S, Wang X, Xiao W, Ju~Q, Shi C, Xie X.
\newblock Graph structure estimation neural networks.
\newblock In: Proceedings of the web conference 2021.
\newblock 2021,  342--353

\bibitem{wu2023learning}
Wu~L, Tan C, Liu Z, Gao Z, Lin H, Li~S~Z.
\newblock Learning to augment graph structure for both homophily and heterophily graphs.
\newblock In: Joint European Conference on Machine Learning and Knowledge Discovery in Databases.
\newblock 2023,  3--18

\bibitem{dong2023towards}
Dong M, Kluger Y.
\newblock Towards understanding and reducing graph structural noise for gnns.
\newblock In: International Conference on Machine Learning.
\newblock 2023,  8202--8226

\bibitem{wu2023homophily}
Wu~L, Lin H, Liu Z, Liu Z, Huang Y, Li~S~Z.
\newblock Homophily-enhanced self-supervision for graph structure learning: Insights and directions.
\newblock IEEE Transactions on Neural Networks and Learning Systems, 2023

\bibitem{liu2022towards}
Liu Y, Zheng Y, Zhang D, Chen H, Peng H, Pan S.
\newblock Towards unsupervised deep graph structure learning.
\newblock In: Proceedings of the ACM Web Conference 2022.
\newblock 2022,  1392--1403

\bibitem{wu2024learning}
Wu~L, Lin H, Zhao G, Tan C, Li~S~Z.
\newblock Learning to model graph structural information on mlps via graph structure self-contrasting.
\newblock arXiv preprint arXiv:2409.05573, 2024

\bibitem{li2024gslb}
Li~Z, Sun X, Luo Y, Zhu Y, Chen D, Luo Y, Zhou X, Liu Q, Wu~S, Wang L, others .
\newblock Gslb: the graph structure learning benchmark.
\newblock Advances in Neural Information Processing Systems, 2024, 36

\bibitem{wang2023overview}
Wang S, Yang J, Yao J, Bai Y, Zhu W.
\newblock An overview of advanced deep graph node clustering.
\newblock IEEE Transactions on Computational Social Systems, 2023, 11(1): 1302--1314

\bibitem{tian2014learning}
Tian F, Gao B, Cui Q, Chen E, Liu T~Y.
\newblock Learning deep representations for graph clustering.
\newblock In: Proceedings of the AAAI conference on artificial intelligence.
\newblock 2014

\bibitem{pan2021multi}
Pan E, Kang Z.
\newblock Multi-view contrastive graph clustering.
\newblock Advances in neural information processing systems, 2021, 34: 2148--2159

\bibitem{zhao2021graph}
Zhao H, Yang X, Wang Z, Yang E, Deng C.
\newblock Graph debiased contrastive learning with joint representation clustering.
\newblock In: IJCAI.
\newblock 2021,  3434--3440

\bibitem{lu2011link}
L{\"u}~L, Zhou T.
\newblock Link prediction in complex networks: A survey.
\newblock Physica A: statistical mechanics and its applications, 2011, 390(6): 1150--1170

\bibitem{zhang2018link}
Zhang M, Chen Y.
\newblock Link prediction based on graph neural networks.
\newblock Advances in neural information processing systems, 2018, 31

\bibitem{li2023seegera}
Li~X, Ye~T, Shan C, Li~D, Gao M.
\newblock Seegera: Self-supervised semi-implicit graph variational auto-encoders with masking.
\newblock In: Proceedings of the ACM web conference 2023.
\newblock 2023,  143--153

\bibitem{tan2023s2gae}
Tan Q, Liu N, Huang X, Choi S~H, Li~L, Chen R, Hu~X.
\newblock S2gae: Self-supervised graph autoencoders are generalizable learners with graph masking.
\newblock In: Proceedings of the sixteenth ACM international conference on web search and data mining.
\newblock 2023,  787--795

\bibitem{ding2023self}
Ding Y, Liu Z, Hao H.
\newblock Self-supervised learning and graph classification under heterophily.
\newblock In: Proceedings of the 32nd ACM International Conference on Information and Knowledge Management.
\newblock 2023,  3849--3853

\bibitem{ghosh2009spiking}
Ghosh-Dastidar S, Adeli H.
\newblock Spiking neural networks.
\newblock International journal of neural systems, 2009, 19(04): 295--308

\bibitem{tavanaei2019deep}
Tavanaei A, Ghodrati M, Kheradpisheh S~R, Masquelier T, Maida A.
\newblock Deep learning in spiking neural networks.
\newblock Neural networks, 2019, 111: 47--63

\bibitem{chakraborty2018adversarial}
Chakraborty A, Alam M, Dey V, Chattopadhyay A, Mukhopadhyay D.
\newblock Adversarial attacks and defences: A survey.
\newblock arXiv preprint arXiv:1810.00069, 2018

\bibitem{silva2020opportunities}
Silva S~H, Najafirad P.
\newblock Opportunities and challenges in deep learning adversarial robustness: A survey.
\newblock arXiv preprint arXiv:2007.00753, 2020

\bibitem{xu2021robustness}
Xu~J, Chen J, You S, Xiao Z, Yang Y, Lu~J.
\newblock Robustness of deep learning models on graphs: A survey.
\newblock AI Open, 2021, 2: 69--78

\bibitem{zugner2018adversarial}
Z{\"u}gner D, Akbarnejad A, G{\"u}nnemann S.
\newblock Adversarial attacks on neural networks for graph data.
\newblock In: Proceedings of the 24th ACM SIGKDD international conference on knowledge discovery \& data mining.
\newblock 2018,  2847--2856

\bibitem{dai2018adversarial}
Dai H, Li~H, Tian T, Huang X, Wang L, Zhu J, Song L.
\newblock Adversarial attack on graph structured data.
\newblock In: International conference on machine learning.
\newblock 2018,  1115--1124

\bibitem{he2021node}
He~X, Wen R, Wu~Y, Backes M, Shen Y, Zhang Y.
\newblock Node-level membership inference attacks against graph neural networks.
\newblock arXiv preprint arXiv:2102.05429, 2021

\bibitem{liao2021information}
Liao P, Zhao H, Xu~K, Jaakkola T, Gordon G~J, Jegelka S, Salakhutdinov R.
\newblock Information obfuscation of graph neural networks.
\newblock In: International conference on machine learning.
\newblock 2021,  6600--6610

\bibitem{hu2022learning}
Hu~H, Cheng L, Vap J~P, Borowczak M.
\newblock Learning privacy-preserving graph convolutional network with partially observed sensitive attributes.
\newblock In: Proceedings of the ACM Web Conference 2022.
\newblock 2022,  3552--3561

\bibitem{dai2021say}
Dai E, Wang S.
\newblock Say no to the discrimination: Learning fair graph neural networks with limited sensitive attribute information.
\newblock In: Proceedings of the 14th ACM International Conference on Web Search and Data Mining.
\newblock 2021,  680--688

\bibitem{li2021dyadic}
Li~P, Wang Y, Zhao H, Hong P, Liu H.
\newblock On dyadic fairness: Exploring and mitigating bias in graph connections.
\newblock In: International Conference on Learning Representations.
\newblock 2021

\bibitem{zhu2021deep}
Zhu Y, Xu~W, Zhang J, Liu Q, Wu~S, Wang L.
\newblock Deep graph structure learning for robust representations: A survey.
\newblock arXiv preprint arXiv:2103.03036, 2021, 14: 1--1

\bibitem{ye2021sparse}
Ye~Y, Ji~S.
\newblock Sparse graph attention networks.
\newblock IEEE Transactions on Knowledge and Data Engineering, 2021, 35(1): 905--916

\bibitem{nie2011spectral}
Nie F, Zeng Z, Tsang I~W, Xu~D, Zhang C.
\newblock Spectral embedded clustering: A framework for in-sample and out-of-sample spectral clustering.
\newblock IEEE Transactions on Neural Networks, 2011, 22(11): 1796--1808

\bibitem{vincent2008extracting}
Vincent P, Larochelle H, Bengio Y, Manzagol P~A.
\newblock Extracting and composing robust features with denoising autoencoders.
\newblock In: Proceedings of the 25th international conference on Machine learning.
\newblock 2008,  1096--1103

\bibitem{gong2023beyond}
Gong Z, Wang G, Sun Y, Liu Q, Ning Y, Xiong H, Peng J.
\newblock Beyond homophily: Robust graph anomaly detection via neural sparsification.
\newblock In: IJCAI.
\newblock 2023,  2104--2113

\bibitem{wen2024ta}
Wen J, Jiang N, Li~L, Zhou J, Li~Y, Zhan H, Kou G, Gu~W, Zhao J.
\newblock Ta-detector: A gnn-based anomaly detector via trust relationship.
\newblock ACM Transactions on Multimedia Computing, Communications and Applications, 2024

\bibitem{gao2023addressing}
Gao Y, Wang X, He~X, Liu Z, Feng H, Zhang Y.
\newblock Addressing heterophily in graph anomaly detection: A perspective of graph spectrum.
\newblock In: Proceedings of the ACM Web Conference 2023.
\newblock 2023,  1528--1538

\bibitem{qiao2024truncated}
Qiao H, Pang G.
\newblock Truncated affinity maximization: One-class homophily modeling for graph anomaly detection.
\newblock Advances in Neural Information Processing Systems, 2024, 36

\bibitem{zhang2024generation}
Zhang R, Cheng D, Liu X, Yang J, Ouyang Y, Wu~X, Zheng Y.
\newblock Generation is better than modification: Combating high class homophily variance in graph anomaly detection.
\newblock arXiv preprint arXiv:2403.10339, 2024

\bibitem{tang2022rethinking}
Tang J, Li~J, Gao Z, Li~J.
\newblock Rethinking graph neural networks for anomaly detection.
\newblock In: International Conference on Machine Learning.
\newblock 2022,  21076--21089

\bibitem{chai2022can}
Chai Z, You S, Yang Y, Pu~S, Xu~J, Cai H, Jiang W.
\newblock Can abnormality be detected by graph neural networks?
\newblock In: IJCAI.
\newblock 2022,  1945--1951

\bibitem{jin2024multi}
Jin W, Ma~H, Zhang Y, Li~Z, Chang L.
\newblock Multi-view discriminative edge heterophily contrastive learning network for attributed graph anomaly detection.
\newblock Expert Systems with Applications, 2024,  124460

\bibitem{roy2024gad}
Roy A, Shu J, Li~J, Yang C, Elshocht O, Smeets J, Li~P.
\newblock Gad-nr: Graph anomaly detection via neighborhood reconstruction.
\newblock In: Proceedings of the 17th ACM International Conference on Web Search and Data Mining.
\newblock 2024,  576--585

\bibitem{gao2023alleviating}
Gao Y, Wang X, He~X, Liu Z, Feng H, Zhang Y.
\newblock Alleviating structural distribution shift in graph anomaly detection.
\newblock In: Proceedings of the sixteenth ACM international conference on web search and data mining.
\newblock 2023,  357--365

\bibitem{pourhabibi2020fraud}
Pourhabibi T, Ong K~L, Kam B~H, Boo Y~L.
\newblock Fraud detection: A systematic literature review of graph-based anomaly detection approaches.
\newblock Decision Support Systems, 2020, 133: 113303

\bibitem{shi2022h2}
Shi F, Cao Y, Shang Y, Zhou Y, Zhou C, Wu~J.
\newblock H2-fdetector: A gnn-based fraud detector with homophilic and heterophilic connections.
\newblock In: Proceedings of the ACM Web Conference 2022.
\newblock 2022,  1486--1494

\bibitem{kim2023dynamic}
Kim H, Choi J, Whang J~J.
\newblock Dynamic relation-attentive graph neural networks for fraud detection.
\newblock In: 2023 IEEE International Conference on Data Mining Workshops (ICDMW).
\newblock 2023,  1092--1096

\bibitem{wang2023label}
Wang Y, Zhang J, Huang Z, Li~W, Feng S, Ma~Z, Sun Y, Yu~D, Dong F, Jin J, others .
\newblock Label information enhanced fraud detection against low homophily in graphs.
\newblock In: Proceedings of the ACM Web Conference 2023.
\newblock 2023,  406--416

\bibitem{duan2024dga}
Duan M, Zheng T, Gao Y, Wang G, Feng Z, Wang X.
\newblock Dga-gnn: Dynamic grouping aggregation gnn for fraud detection.
\newblock In: Proceedings of the AAAI Conference on Artificial Intelligence.
\newblock 2024,  11820--11828

\bibitem{wu2023splitgnn}
Wu~B, Yao X, Zhang B, Chao K~M, Li~Y.
\newblock Splitgnn: Spectral graph neural network for fraud detection against heterophily.
\newblock In: Proceedings of the 32nd ACM International Conference on Information and Knowledge Management.
\newblock 2023,  2737--2746

\bibitem{xu2024revisiting}
Xu~F, Wang N, Wu~H, Wen X, Zhao X, Wan H.
\newblock Revisiting graph-based fraud detection in sight of heterophily and spectrum.
\newblock In: Proceedings of the AAAI Conference on Artificial Intelligence.
\newblock 2024,  9214--9222

\bibitem{bergstrom2019information}
Bergstrom C~T, Bak-Coleman J~B.
\newblock Information gerrymandering in social networks skews collective decision-making, 2019

\bibitem{deb2019perils}
Deb A, Luceri L, Badaway A, Ferrara E.
\newblock Perils and challenges of social media and election manipulation analysis: The 2018 us midterms.
\newblock In: Companion proceedings of the 2019 world wide web conference.
\newblock 2019,  237--247

\bibitem{ferrara2017disinformation}
Ferrara E.
\newblock Disinformation and social bot operations in the run up to the 2017 french presidential election.
\newblock arXiv preprint arXiv:1707.00086, 2017

\bibitem{ashmore2023hover}
Ashmore B, Chen L.
\newblock Hover: Homophilic oversampling via edge removal for class-imbalanced bot detection on graphs.
\newblock In: Proceedings of the 32nd ACM International Conference on Information and Knowledge Management.
\newblock 2023,  3728--3732

\bibitem{zhou2023semi}
Zhou M, Feng W, Zhu Y, Zhang D, Dong Y, Tang J.
\newblock Semi-supervised social bot detection with initial residual relation attention networks.
\newblock In: Joint European Conference on Machine Learning and Knowledge Discovery in Databases.
\newblock 2023,  207--224

\bibitem{li2023multi}
Li~S, Qiao B, Li~K, Lu~Q, Lin M, Zhou W.
\newblock Multi-modal social bot detection: Learning homophilic and heterophilic connections adaptively.
\newblock In: Proceedings of the 31st ACM International Conference on Multimedia.
\newblock 2023,  3908--3916

\bibitem{ye2023hofa}
Ye~S, Tan Z, Lei Z, He~R, Wang H, Zheng Q, Luo M.
\newblock Hofa: Twitter bot detection with homophily-oriented augmentation and frequency adaptive attention.
\newblock arXiv preprint arXiv:2306.12870, 2023

\bibitem{shi2023muti}
Shi S, Qiao K, Wang Z, Yang J, Song B, Chen J, Yan B.
\newblock Muti-scale graph neural network with signed-attention for social bot detection: A frequency perspective.
\newblock arXiv preprint arXiv:2307.01968, 2023

\bibitem{wu2023heterophily}
Wu~Q, Yang Y, He~B, Liu H, Wang X, Liao Y, Yang R, Zhou P.
\newblock Heterophily-aware social bot detection with supervised contrastive learning.
\newblock arXiv preprint arXiv:2306.07478, 2023

\bibitem{chen2022multi}
Chen X, Zhou F, Trajcevski G, Bonsangue M.
\newblock Multi-view learning with distinguishable feature fusion for rumor detection.
\newblock Knowledge-Based Systems, 2022, 240: 108085

\bibitem{yang2021rumor}
Yang X, Lyu Y, Tian T, Liu Y, Liu Y, Zhang X.
\newblock Rumor detection on social media with graph structured adversarial learning.
\newblock In: Proceedings of the twenty-ninth international conference on international joint conferences on artificial intelligence.
\newblock 2021,  1417--1423

\bibitem{yan2023graph}
Yan Y, Wang Y, Zheng P.
\newblock A graph-based pivotal semantic mining framework for rumor detection.
\newblock Engineering Applications of Artificial Intelligence, 2023, 118: 105613

\bibitem{nguyen2024portable}
Nguyen T~T, Ren Z, Nguyen T~T, Jo~J, Nguyen Q~V~H, Yin H.
\newblock Portable graph-based rumour detection against multi-modal heterophily.
\newblock Knowledge-Based Systems, 2024, 284: 111310

\bibitem{wang2022hagen}
Wang C, Lin Z, Yang X, Sun J, Yue M, Shahabi C.
\newblock Hagen: Homophily-aware graph convolutional recurrent network for crime forecasting.
\newblock In: Proceedings of the AAAI Conference on Artificial Intelligence.
\newblock 2022,  4193--4200

\bibitem{he2020lightgcn}
He~X, Deng K, Wang X, Li~Y, Zhang Y, Wang M.
\newblock Lightgcn: Simplifying and powering graph convolution network for recommendation.
\newblock In: Proceedings of the 43rd International ACM SIGIR conference on research and development in Information Retrieval.
\newblock 2020,  639--648

\bibitem{wang2019neural}
Wang X, He~X, Wang M, Feng F, Chua T~S.
\newblock Neural graph collaborative filtering.
\newblock In: Proceedings of the 42nd international ACM SIGIR conference on Research and development in Information Retrieval.
\newblock 2019,  165--174

\bibitem{sun2020framework}
Sun J, Guo W, Zhang D, Zhang Y, Regol F, Hu~Y, Guo H, Tang R, Yuan H, He~X, others .
\newblock A framework for recommending accurate and diverse items using bayesian graph convolutional neural networks.
\newblock In: Proceedings of the 26th ACM SIGKDD international conference on knowledge discovery \& data mining.
\newblock 2020,  2030--2039

\bibitem{wu2021self}
Wu~J, Wang X, Feng F, He~X, Chen L, Lian J, Xie X.
\newblock Self-supervised graph learning for recommendation.
\newblock In: Proceedings of the 44th international ACM SIGIR conference on research and development in information retrieval.
\newblock 2021,  726--735

\bibitem{mcpherson2001birds}
McPherson M, Smith-Lovin L, Cook J~M.
\newblock Birds of a feather: Homophily in social networks.
\newblock Annual review of sociology, 2001, 27(1): 415--444

\bibitem{jiang2022triangle}
Jiang W, Jiao Y, Wang Q, Liang C, Guo L, Zhang Y, Sun Z, Xiong Y, Zhu Y.
\newblock Triangle graph interest network for click-through rate prediction.
\newblock In: Proceedings of the fifteenth ACM international conference on web search and data mining.
\newblock 2022,  401--409

\bibitem{jiang2024challenging}
Jiang W, Gao X, Xu~G, Chen T, Yin H.
\newblock Challenging low homophily in social recommendation.
\newblock In: Proceedings of the ACM on Web Conference 2024.
\newblock 2024,  3476--3484

\bibitem{gholinejad2024heterophily}
Gholinejad N, Chehreghani M~H.
\newblock Heterophily-aware fair recommendation using graph convolutional networks.
\newblock arXiv preprint arXiv:2402.03365, 2024

\bibitem{xiao2023spatial}
Xiao C, Zhou J, Huang J, Xu~T, Xiong H.
\newblock Spatial heterophily aware graph neural networks.
\newblock In: Proceedings of the 29th ACM SIGKDD Conference on Knowledge Discovery and Data Mining.
\newblock 2023,  2752--2763

\bibitem{zoidi2015graph}
Zoidi O, Fotiadou E, Nikolaidis N, Pitas I.
\newblock Graph-based label propagation in digital media: A review.
\newblock ACM Computing Surveys (CSUR), 2015, 47(3): 1--35

\bibitem{taelman2021exploitation}
Taelman C, Chlaily S, Khachatrian E, Sommen v.~d F, Marinoni A.
\newblock On the exploitation of heterophily in graph-based multimodal remote sensing data analysis.
\newblock CEUR Workshop Proceedings, 2022

\bibitem{eswaran2017zoobp}
Eswaran D, G{\"u}nnemann S, Faloutsos C, Makhija D, Kumar M.
\newblock Zoobp: Belief propagation for heterogeneous networks.
\newblock Proceedings of the VLDB Endowment, 2017, 10(5): 625--636

\bibitem{han2022vision}
Han K, Wang Y, Guo J, Tang Y, Wu~E.
\newblock Vision gnn: An image is worth graph of nodes.
\newblock Advances in neural information processing systems, 2022, 35: 8291--8303

\bibitem{peng2020learning}
Peng W, Hong X, Chen H, Zhao G.
\newblock Learning graph convolutional network for skeleton-based human action recognition by neural searching.
\newblock In: Proceedings of the AAAI conference on artificial intelligence.
\newblock 2020,  2669--2676

\bibitem{chen2019graph}
Chen Y, Rohrbach M, Yan Z, Shuicheng Y, Feng J, Kalantidis Y.
\newblock Graph-based global reasoning networks.
\newblock In: Proceedings of the IEEE/CVF conference on computer vision and pattern recognition.
\newblock 2019,  433--442

\bibitem{yang2018graph}
Yang J, Lu~J, Lee S, Batra D, Parikh D.
\newblock Graph r-cnn for scene graph generation.
\newblock In: Proceedings of the European conference on computer vision (ECCV).
\newblock 2018,  670--685

\bibitem{zhu2022scene}
Zhu G, Zhang L, Jiang Y, Dang Y, Hou H, Shen P, Feng M, Zhao X, Miao Q, Shah S~A~A, others .
\newblock Scene graph generation: A comprehensive survey.
\newblock arXiv preprint arXiv:2201.00443, 2022

\bibitem{chang2021comprehensive}
Chang X, Ren P, Xu~P, Li~Z, Chen X, Hauptmann A.
\newblock A comprehensive survey of scene graphs: Generation and application.
\newblock IEEE Transactions on Pattern Analysis and Machine Intelligence, 2021, 45(1): 1--26

\bibitem{lin2022hl}
Lin X, Ding C, Zhan Y, Li~Z, Tao D.
\newblock Hl-net: Heterophily learning network for scene graph generation.
\newblock In: proceedings of the IEEE/CVF conference on computer vision and pattern recognition.
\newblock 2022,  19476--19485

\bibitem{shuman2013emerging}
Shuman D~I, Narang S~K, Frossard P, Ortega A, Vandergheynst P.
\newblock The emerging field of signal processing on graphs: Extending high-dimensional data analysis to networks and other irregular domains.
\newblock IEEE signal processing magazine, 2013, 30(3): 83--98

\bibitem{chen2024kumaraswamy}
Chen L, Song Y, Lin S, Wang C, He~G.
\newblock Kumaraswamy wavelet for heterophilic scene graph generation.
\newblock In: Proceedings of the AAAI Conference on Artificial Intelligence.
\newblock 2024,  1138--1146

\bibitem{kumaraswamy1980generalized}
Kumaraswamy P.
\newblock A generalized probability density function for double-bounded random processes.
\newblock Journal of hydrology, 1980, 46(1-2): 79--88

\bibitem{wang2019graph}
Wang L, Huang Y, Hou Y, Zhang S, Shan J.
\newblock Graph attention convolution for point cloud semantic segmentation.
\newblock In: Proceedings of the IEEE/CVF conference on computer vision and pattern recognition.
\newblock 2019,  10296--10305

\bibitem{lei2020spherical}
Lei H, Akhtar N, Mian A.
\newblock Spherical kernel for efficient graph convolution on 3d point clouds.
\newblock IEEE transactions on pattern analysis and machine intelligence, 2020, 43(10): 3664--3680

\bibitem{chen2024joint}
Chen S, Wei L, Liang L, Lang C.
\newblock Joint homophily and heterophily relational knowledge distillation for efficient and compact 3d object detection.
\newblock In: ACM Multimedia.
\newblock 2024

\bibitem{wang2022molecular}
Wang Y, Wang J, Cao Z, Barati~Farimani A.
\newblock Molecular contrastive learning of representations via graph neural networks.
\newblock Nature Machine Intelligence, 2022, 4(3): 279--287

\bibitem{merchant2023scaling}
Merchant A, Batzner S, Schoenholz S~S, Aykol M, Cheon G, Cubuk E~D.
\newblock Scaling deep learning for materials discovery.
\newblock Nature, 2023, 624(7990): 80--85

\bibitem{fang2022geometry}
Fang X, Liu L, Lei J, He~D, Zhang S, Zhou J, Wang F, Wu~H, Wang H.
\newblock Geometry-enhanced molecular representation learning for property prediction.
\newblock Nature Machine Intelligence, 2022, 4(2): 127--134

\bibitem{li2024cgmega}
Li~H, Han Z, Sun Y, Wang F, Hu~P, Gao Y, Bai X, Peng S, Ren C, Xu~X, others .
\newblock Cgmega: explainable graph neural network framework with attention mechanisms for cancer gene module dissection.
\newblock Nature Communications, 2024, 15(1): 5997

\bibitem{wu2022machine}
Wu~L, Wen Y, Leng D, Zhang Q, Dai C, Wang Z, Liu Z, Yan B, Zhang Y, Wang J, others .
\newblock Machine learning methods, databases and tools for drug combination prediction.
\newblock Briefings in bioinformatics, 2022, 23(1): bbab355

\bibitem{cheng2019network}
Cheng F, Kov{\'a}cs I~A, Barab{\'a}si A~L.
\newblock Network-based prediction of drug combinations.
\newblock Nature communications, 2019, 10(1): 1197

\bibitem{jia2016overcoming}
Jia Y, Yun C~H, Park E, Ercan D, Manuia M, Juarez J, Xu~C, Rhee K, Chen T, Zhang H, others .
\newblock Overcoming egfr (t790m) and egfr (c797s) resistance with mutant-selective allosteric inhibitors.
\newblock Nature, 2016, 534(7605): 129--132

\bibitem{chen2022drug}
Chen H, Lu~Y, Yang Y, Rao Y.
\newblock A drug combination prediction framework based on graph convolutional network and heterogeneous information.
\newblock IEEE/ACM Transactions on Computational Biology and Bioinformatics, 2022

\bibitem{liu2024slgcn}
Liu B~M, Gao Y~L, Li~F, Zheng C~H, Liu J~X.
\newblock Slgcn: Structure-enhanced line graph convolutional network for predicting drug--disease associations.
\newblock Knowledge-Based Systems, 2024, 283: 111187

\bibitem{xue2018review}
Xue H, Li~J, Xie H, Wang Y.
\newblock Review of drug repositioning approaches and resources.
\newblock International journal of biological sciences, 2018, 14(10): 1232

\bibitem{jarada2020review}
Jarada T~N, Rokne J~G, Alhajj R.
\newblock A review of computational drug repositioning: strategies, approaches, opportunities, challenges, and directions.
\newblock Journal of cheminformatics, 2020, 12: 1--23

\bibitem{lotfi2018review}
Lotfi~Shahreza M, Ghadiri N, Mousavi S~R, Varshosaz J, Green J~R.
\newblock A review of network-based approaches to drug repositioning.
\newblock Briefings in bioinformatics, 2018, 19(5): 878--892

\bibitem{kang2018conditional}
Kang S, Cho K.
\newblock Conditional molecular design with deep generative models.
\newblock Journal of chemical information and modeling, 2018, 59(1): 43--52

\bibitem{yang2023cmgn}
Yang M, Sun H, Liu X, Xue X, Deng Y, Wang X.
\newblock Cmgn: a conditional molecular generation net to design target-specific molecules with desired properties.
\newblock Briefings in bioinformatics, 2023, 24(4): bbad185

\bibitem{rigoni2020conditional}
Rigoni D, Navarin N, Sperduti A.
\newblock Conditional constrained graph variational autoencoders for molecule design.
\newblock In: 2020 IEEE Symposium Series on Computational Intelligence (SSCI).
\newblock 2020,  729--736

\bibitem{wangmolecule}
Wang H, Solin A, Garg V.
\newblock Molecule generation by heterophilious triple flows.
\newblock Openreview, 2024

\bibitem{wang2017multi}
Wang Z, Zhu X, Adeli E, Zhu Y, Nie F, Munsell B, Wu~G, others .
\newblock Multi-modal classification of neurodegenerative disease by progressive graph-based transductive learning.
\newblock Medical image analysis, 2017, 39: 218--230

\bibitem{zhu2018dynamic}
Zhu Y, Zhu X, Kim M, Yan J, Kaufer D, Wu~G.
\newblock Dynamic hyper-graph inference framework for computer-assisted diagnosis of neurodegenerative diseases.
\newblock IEEE transactions on medical imaging, 2018, 38(2): 608--616

\bibitem{wang2023hypergraph}
Wang M, Shao W, Huang S, Zhang D.
\newblock Hypergraph-regularized multimodal learning by graph diffusion for imaging genetics based alzheimer’s disease diagnosis.
\newblock Medical Image Analysis, 2023, 89: 102883

\bibitem{qu2023graph}
Qu~Z, Yao T, Liu X, Wang G.
\newblock A graph convolutional network based on univariate neurodegeneration biomarker for alzheimer’s disease diagnosis.
\newblock IEEE Journal of Translational Engineering in Health and Medicine, 2023, 11: 405--416

\bibitem{xu2024data}
Xu~J, Yang Y, Huang D, Gururajapathy S~S, Ke~Y, Qiao M, Wang A, Kumar H, McGeown J, Kwon E.
\newblock Data-driven network neuroscience: On data collection and benchmark.
\newblock Advances in Neural Information Processing Systems, 2024, 36

\bibitem{choneurodegenerative}
Cho H, Sim J, Wu~G, Kim W~H.
\newblock Neurodegenerative brain network classification via adaptive diffusion with temporal regularization.
\newblock In: Forty-first International Conference on Machine Learning.
\newblock 2024

\bibitem{hammond2011wavelets}
Hammond D~K, Vandergheynst P, Gribonval R.
\newblock Wavelets on graphs via spectral graph theory.
\newblock Applied and Computational Harmonic Analysis, 2011, 30(2): 129--150

\bibitem{xu2019graph}
Xu~B, Shen H, Cao Q, Qiu Y, Cheng X.
\newblock Graph wavelet neural network.
\newblock arXiv preprint arXiv:1904.07785, 2019

\bibitem{xu2024heterophily}
Xu~S, Shen J, Li~Y, Yao Y, Yu~P, Xu~F, Ma~X.
\newblock On the heterophily of program graphs: A case study of graph-based type inference.
\newblock In: Proceedings of the 15th Asia-Pacific Symposium on Internetware.
\newblock 2024,  1--10

\bibitem{yan2020just}
Yan M, Xia X, Fan Y, Hassan A~E, Lo~D, Li~S.
\newblock Just-in-time defect identification and localization: A two-phase framework.
\newblock IEEE Transactions on Software Engineering, 2020, 48(1): 82--101

\bibitem{qiu2020jito}
Qiu F, Yan M, Xia X, Wang X, Fan Y, Hassan A~E, Lo~D.
\newblock Jito: a tool for just-in-time defect identification and localization.
\newblock In: Proceedings of the 28th ACM joint meeting on european software engineering conference and symposium on the foundations of software engineering.
\newblock 2020,  1586--1590

\bibitem{qiu2021deep}
Qiu F, Gao Z, Xia X, Lo~D, Grundy J, Wang X.
\newblock Deep just-in-time defect localization.
\newblock IEEE Transactions on Software Engineering, 2021, 48(12): 5068--5086

\bibitem{zhang2024just}
Zhang H, Min W, Wei Z, Kuang L, Gao H, Miao H.
\newblock A just-in-time software defect localization method based on code graph representation.
\newblock In: Proceedings of the 32nd IEEE/ACM International Conference on Program Comprehension.
\newblock 2024,  293--303

\bibitem{shi2016survey}
Shi C, Li~Y, Zhang J, Sun Y, Philip S~Y.
\newblock A survey of heterogeneous information network analysis.
\newblock IEEE Transactions on Knowledge and Data Engineering, 2016, 29(1): 17--37

\bibitem{wang2019heterogeneous}
Wang X, Ji~H, Shi C, Wang B, Ye~Y, Cui P, Yu~P~S.
\newblock Heterogeneous graph attention network.
\newblock In: The world wide web conference.
\newblock 2019,  2022--2032

\bibitem{li2017semi}
Li~X, Wu~Y, Ester M, Kao B, Wang X, Zheng Y.
\newblock Semi-supervised clustering in attributed heterogeneous information networks.
\newblock In: Proceedings of the 26th international conference on world wide web.
\newblock 2017,  1621--1629

\bibitem{li2021leveraging}
Li~X, Ding D, Kao B, Sun Y, Mamoulis N.
\newblock Leveraging meta-path contexts for classification in heterogeneous information networks.
\newblock In: 2021 IEEE 37th International Conference on Data Engineering (ICDE).
\newblock 2021,  912--923

\bibitem{li2016transductive}
Li~X, Kao B, Zheng Y, Huang Z.
\newblock On transductive classification in heterogeneous information networks.
\newblock In: Proceedings of the 25th ACM International on Conference on Information and Knowledge Management.
\newblock 2016,  811--820

\bibitem{zhao2021heterogeneous}
Zhao J, Wang X, Shi C, Hu~B, Song G, Ye~Y.
\newblock Heterogeneous graph structure learning for graph neural networks.
\newblock In: Proceedings of the AAAI conference on artificial intelligence.
\newblock 2021,  4697--4705

\bibitem{hu2019adversarial}
Hu~B, Fang Y, Shi C.
\newblock Adversarial learning on heterogeneous information networks.
\newblock In: Proceedings of the 25th ACM SIGKDD international conference on knowledge discovery \& data mining.
\newblock 2019,  120--129

\bibitem{ji2021heterogeneous}
Ji~H, Wang X, Shi C, Wang B, Philip S~Y.
\newblock Heterogeneous graph propagation network.
\newblock IEEE Transactions on Knowledge and Data Engineering, 2021, 35(1): 521--532

\bibitem{guo2023homophily}
Guo J, Du~L, Bi~W, Fu~Q, Ma~X, Chen X, Han S, Zhang D, Zhang Y.
\newblock Homophily-oriented heterogeneous graph rewiring.
\newblock In: Proceedings of the ACM Web Conference 2023.
\newblock 2023,  511--522

\bibitem{li2023hetero}
Li~J, Wei Z, Dan J, Zhou J, Zhu Y, Wu~R, Wang B, Zhen Z, Meng C, Jin H, others .
\newblock Hetero$^2$net: Heterophily-aware representation learning on heterogenerous graphs.
\newblock arXiv preprint arXiv:2310.11664, 2023

\bibitem{shen2024heterophily}
Shen Z, Kang Z.
\newblock When heterophily meets heterogeneous graphs: Latent graphs guided unsupervised representation learning.
\newblock arXiv preprint arXiv:2409.00687, 2024

\bibitem{lin2024heterophily}
Lin J, Guo X, Zhang S, Zhou D, Zhu Y, Shun J.
\newblock When heterophily meets heterogeneity: New graph benchmarks and effective methods.
\newblock arXiv preprint arXiv:2407.10916, 2024

\bibitem{longa2023graph}
Longa A, Lachi V, Santin G, Bianchini M, Lepri B, Lio P, Scarselli F, Passerini A.
\newblock Graph neural networks for temporal graphs: State of the art, open challenges, and opportunities.
\newblock arXiv preprint arXiv:2302.01018, 2023

\bibitem{sahili2023spatio}
Sahili Z~A, Awad M.
\newblock Spatio-temporal graph neural networks: A survey.
\newblock arXiv preprint arXiv:2301.10569, 2023

\bibitem{zhou2022greto}
Zhou Z, Huang Q, Lin G, Yang K, Bai L, Wang Y.
\newblock Greto: remedying dynamic graph topology-task discordance via target homophily.
\newblock In: The eleventh international conference on learning representations.
\newblock 2022

\bibitem{antelmi2023survey}
Antelmi A, Cordasco G, Polato M, Scarano V, Spagnuolo C, Yang D.
\newblock A survey on hypergraph representation learning.
\newblock ACM Computing Surveys, 2023, 56(1): 1--38

\bibitem{veldt2023combinatorial}
Veldt N, Benson A~R, Kleinberg J.
\newblock Combinatorial characterizations and impossibilities for higher-order homophily.
\newblock Science Advances, 2023, 9(1): eabq3200

\bibitem{wang2022equivariant}
Wang P, Yang S, Liu Y, Wang Z, Li~P.
\newblock Equivariant hypergraph diffusion neural operators.
\newblock arXiv preprint arXiv:2207.06680, 2022

\bibitem{nguyen2024sheaf}
Nguyen B, Sani L, Qiu X, Li{\`o} P, Lane N~D.
\newblock Sheaf hypernetworks for personalized federated learning.
\newblock arXiv preprint arXiv:2405.20882, 2024

\bibitem{zou2024unig}
Zou M, Gan Z, Wang Y, Zhang J, Sui D, Guan C, Leng S.
\newblock Unig-encoder: A universal feature encoder for graph and hypergraph node classification.
\newblock Pattern Recognition, 2024, 147: 110115

\bibitem{wang2024understanding}
Wang J, Guo Y, Yang L, Wang Y.
\newblock Understanding heterophily for graph neural networks.
\newblock arXiv preprint arXiv:2401.09125, 2024

\bibitem{zhu2023explaining}
Zhu Q, Jiao Y, Ponomareva N, Han J, Perozzi B.
\newblock Explaining and adapting graph conditional shift.
\newblock arXiv preprint arXiv:2306.03256, 2023

\bibitem{mao2024demystifying}
Mao H, Chen Z, Jin W, Han H, Ma~Y, Zhao T, Shah N, Tang J.
\newblock Demystifying structural disparity in graph neural networks: Can one size fit all?
\newblock Advances in neural information processing systems, 2024, 36

\bibitem{yang2024leveraging}
Yang J, Chen Z, Xiao T, Zhang W, Lin Y, Kuang K.
\newblock Leveraging invariant principle for heterophilic graph structure distribution shifts.
\newblock arXiv preprint arXiv:2408.09490, 2024

\bibitem{loveland2024performance}
Loveland D, Zhu J, Heimann M, Fish B, Schaub M~T, Koutra D.
\newblock On performance discrepancies across local homophily levels in graph neural networks.
\newblock In: Learning on Graphs Conference.
\newblock 2024,  6--1

\bibitem{chen2023exploiting}
Chen J, Chen S, Gao J, Huang Z, Zhang J, Pu~J.
\newblock Exploiting neighbor effect: Conv-agnostic gnn framework for graphs with heterophily.
\newblock IEEE Transactions on Neural Networks and Learning Systems, 2023

\bibitem{oono2019graph}
Oono K, Suzuki T.
\newblock Graph neural networks exponentially lose expressive power for node classification.
\newblock arXiv preprint arXiv:1905.10947, 2019

\bibitem{liu2020towards}
Liu M, Gao H, Ji~S.
\newblock Towards deeper graph neural networks.
\newblock In: Proceedings of the 26th ACM SIGKDD international conference on knowledge discovery \& data mining.
\newblock 2020,  338--348

\bibitem{rusch2023survey}
Rusch T~K, Bronstein M~M, Mishra S.
\newblock A survey on oversmoothing in graph neural networks.
\newblock arXiv preprint arXiv:2303.10993, 2023

\bibitem{guo2023taming}
Guo K, Cao X, Liu Z, Chang Y.
\newblock Taming over-smoothing representation on heterophilic graphs.
\newblock Information Sciences, 2023, 647: 119463

\bibitem{topping2021understanding}
Topping J, Di~Giovanni F, Chamberlain B~P, Dong X, Bronstein M~M.
\newblock Understanding over-squashing and bottlenecks on graphs via curvature.
\newblock arXiv preprint arXiv:2111.14522, 2021

\bibitem{rubin2023geodesic}
Rubin J, Loomba S, Jones N~S.
\newblock Geodesic distributions reveal how heterophily and bottlenecks limit the expressive power of message passing neural networks.
\newblock In: The Second Learning on Graphs Conference.
\newblock 2023

\bibitem{peimulti}
Pei H, Li~Y, Deng H, Hai J, Wang P, Ma~J, Tao J, Xiong Y, Guan X.
\newblock Multi-track message passing: Tackling oversmoothing and oversquashing in graph learning via preventing heterophily mixing.
\newblock In: Forty-first International Conference on Machine Learning.
\newblock 2024

\bibitem{yang2023graph}
Yang C, Wu~Q, Wipf D, Sun R, Yan J.
\newblock How graph neural networks learn: Lessons from training dynamics in function space.
\newblock arXiv preprint arXiv:2310.05105, 2023

\bibitem{cui2023mgnn}
Cui G, Wei Z.
\newblock Mgnn: Graph neural networks inspired by distance geometry problem.
\newblock In: Proceedings of the 29th ACM SIGKDD Conference on Knowledge Discovery and Data Mining.
\newblock 2023,  335--347

\bibitem{shi2024homophily}
Shi C, Pan L, Hu~H, Dokmani{\'c} I.
\newblock Homophily modulates double descent generalization in graph convolution networks.
\newblock Proceedings of the National Academy of Sciences, 2024, 121(8): e2309504121

\bibitem{zhu2022survey}
Zhu Y, Du~Y, Wang Y, Xu~Y, Zhang J, Liu Q, Wu~S.
\newblock A survey on deep graph generation: Methods and applications.
\newblock In: Learning on Graphs Conference.
\newblock 2022,  47--1

\bibitem{chanpuriya2021interpretable}
Chanpuriya S, Rossi R, Rao A, Mai T, Lipka N, Song Z, Musco C~N.
\newblock An interpretable graph generative model with heterophily.
\newblock Openreview, 2021

\bibitem{jin2021graph}
Jin W, Zhao L, Zhang S, Liu Y, Tang J, Shah N.
\newblock Graph condensation for graph neural networks.
\newblock arXiv preprint arXiv:2110.07580, 2021

\bibitem{gao2024graph}
Gao X, Yu~J, Jiang W, Chen T, Zhang W, Yin H.
\newblock Graph condensation: A survey.
\newblock arXiv preprint arXiv:2401.11720, 2024

\bibitem{ling2022graph}
Ling X, Wu~L, Wu~C, Ji~S.
\newblock Graph neural networks: Graph matching.
\newblock Graph Neural Networks: Foundations, Frontiers, and Applications, 2022,  277--295

\bibitem{yu2023seedgnn}
Yu~L, Xu~J, Lin X.
\newblock Seedgnn: graph neural network for supervised seeded graph matching.
\newblock In: International Conference on Machine Learning.
\newblock 2023,  40390--40411

\bibitem{munikoti2022scalable}
Munikoti S, Das L, Natarajan B.
\newblock Scalable graph neural network-based framework for identifying critical nodes and links in complex networks.
\newblock Neurocomputing, 2022, 468: 211--221

\bibitem{ling2023deep}
Ling C, Jiang J, Wang J, Thai M~T, Xue R, Song J, Qiu M, Zhao L.
\newblock Deep graph representation learning and optimization for influence maximization.
\newblock In: International Conference on Machine Learning.
\newblock 2023,  21350--21361

\bibitem{feng2024influence}
Feng Y, Tan V~Y, Cautis B.
\newblock Influence maximization via graph neural bandits.
\newblock In: Proceedings of the 30th ACM SIGKDD Conference on Knowledge Discovery and Data Mining.
\newblock 2024,  771--781

\bibitem{zhou2019meta}
Zhou F, Cao C, Zhang K, Trajcevski G, Zhong T, Geng J.
\newblock Meta-gnn: On few-shot node classification in graph meta-learning.
\newblock In: Proceedings of the 28th ACM International Conference on Information and Knowledge Management.
\newblock 2019,  2357--2360

\bibitem{wang2023few}
Wang S, Dong Y, Ding K, Chen C, Li~J.
\newblock Few-shot node classification with extremely weak supervision.
\newblock In: Proceedings of the Sixteenth ACM International Conference on Web Search and Data Mining.
\newblock 2023,  276--284

\bibitem{wan2021contrastive}
Wan S, Zhan Y, Liu L, Yu~B, Pan S, Gong C.
\newblock Contrastive graph poisson networks: Semi-supervised learning with extremely limited labels.
\newblock Advances in Neural Information Processing Systems, 2021, 34: 6316--6327

\bibitem{zhao2021graphsmote}
Zhao T, Zhang X, Wang S.
\newblock Graphsmote: Imbalanced node classification on graphs with graph neural networks.
\newblock In: Proceedings of the 14th ACM international conference on web search and data mining.
\newblock 2021,  833--841

\bibitem{yu2024graphcbal}
Yu~C, Zhu J, Li~X.
\newblock Graphcbal: Class-balanced active learning for graph neural networks via reinforcement learning.
\newblock arXiv preprint arXiv:2402.10074, 2024

\bibitem{yun2022lte4g}
Yun S, Kim K, Yoon K, Park C.
\newblock Lte4g: Long-tail experts for graph neural networks.
\newblock In: Proceedings of the 31st ACM International Conference on Information \& Knowledge Management.
\newblock 2022,  2434--2443

\bibitem{wu2023leveraging}
Wu~X, Wu~H, Wang R, Li~D, Zhou X, Lu~K.
\newblock Leveraging free labels to power up heterophilic graph learning in weakly-supervised settings: An empirical study.
\newblock In: Joint European Conference on Machine Learning and Knowledge Discovery in Databases.
\newblock 2023,  140--156

\bibitem{liu2023learning}
Liu Y, Ding K, Wang J, Lee V, Liu H, Pan S.
\newblock Learning strong graph neural networks with weak information.
\newblock In: Proceedings of the 29th ACM SIGKDD Conference on Knowledge Discovery and Data Mining.
\newblock 2023,  1559--1571

\bibitem{peng2024graphrare}
Peng T, Wu~W, Yuan H, Bao Z, Pengru Z, Yu~X, Lin X, Liang Y, Pu~Y.
\newblock Graphrare: Reinforcement learning enhanced graph neural network with relative entropy.
\newblock In: 2024 IEEE 40th International Conference on Data Engineering (ICDE).
\newblock 2024,  2489--2502

\bibitem{chen2022sa}
Chen J, Chen S, Bai M, Gao J, Zhang J, Pu~J.
\newblock Sa-mlp: Distilling graph knowledge from gnns into structure-aware mlp.
\newblock arXiv preprint arXiv:2210.09609, 2022

\bibitem{wu2022knowledge}
Wu~L, Lin H, Huang Y, Li~S~Z.
\newblock Knowledge distillation improves graph structure augmentation for graph neural networks.
\newblock Advances in Neural Information Processing Systems, 2022, 35: 11815--11827

\bibitem{wang2024hc}
Wang F, Zhao T, Xu~J, Wang S.
\newblock Hc-gst: Heterophily-aware distribution consistency based graph self-training.
\newblock arXiv preprint arXiv:2407.17787, 2024

\bibitem{wei4825405searching}
Wei L, He~Z, Zhao H, Yao Q.
\newblock Searching heterophily-agnostic graph neural networks.
\newblock Available at SSRN 4825405, 2024

\bibitem{wei2022enhancing}
Wei L, He~Z, Zhao H, Yao Q.
\newblock Enhancing intra-class information extraction for heterophilous graphs: One neural architecture search approach.
\newblock arXiv preprint arXiv:2211.10990, 2022

\bibitem{wei2022designing}
Wei L, Zhao H, He~Z.
\newblock Designing the topology of graph neural networks: A novel feature fusion perspective.
\newblock In: Proceedings of the ACM Web Conference 2022.
\newblock 2022,  1381--1391

\bibitem{zheng2023auto}
Zheng X, Zhang M, Chen C, Zhang Q, Zhou C, Pan S.
\newblock Auto-heg: Automated graph neural network on heterophilic graphs.
\newblock In: Proceedings of the ACM Web Conference 2023.
\newblock 2023,  611--620

\bibitem{liu2022few}
Liu Y, Li~M, Li~X, Giunchiglia F, Feng X, Guan R.
\newblock Few-shot node classification on attributed networks with graph meta-learning.
\newblock In: Proceedings of the 45th international ACM SIGIR conference on research and development in information retrieval.
\newblock 2022,  471--481

\bibitem{zhang2021survey}
Zhang Y, Yang Q.
\newblock A survey on multi-task learning.
\newblock IEEE transactions on knowledge and data engineering, 2021, 34(12): 5586--5609

\bibitem{wu2024unraveling}
Wu~Y, Yao J, Han B, Yao L, Liu T.
\newblock Unraveling the impact of heterophilic structures on graph positive-unlabeled learning.
\newblock arXiv preprint arXiv:2405.19919, 2024

\bibitem{wu2024can}
Wu~X, Shen Y, Shan C, Song K, Wang S, Zhang B, Feng J, Cheng H, Chen W, Xiong Y, others .
\newblock Can graph learning improve task planning?
\newblock arXiv preprint arXiv:2405.19119, 2024

\bibitem{yang2023foundation}
Yang S, Nachum O, Du~Y, Wei J, Abbeel P, Schuurmans D.
\newblock Foundation models for decision making: Problems, methods, and opportunities.
\newblock arXiv preprint arXiv:2303.04129, 2023

\bibitem{soleymani2021deep}
Soleymani F, Paquet E.
\newblock Deep graph convolutional reinforcement learning for financial portfolio management--deeppocket.
\newblock Expert Systems with Applications, 2021, 182: 115127

\bibitem{bayraktar2023graph}
Bayraktar Z, Molla S, Mahavadi S.
\newblock Graph neural network generated metal-organic frameworks for carbon capture.
\newblock In: ICLR 2023 Workshop on Tackling Climate Change with Machine Learning.
\newblock 2023

\bibitem{wang2023topological}
Wang M, Wang E, Liu X, Wang C.
\newblock Topological graph representation of stratigraphic properties of spatial-geological characteristics and compression modulus prediction by mechanism-driven learning.
\newblock Computers and Geotechnics, 2023, 153: 105112

\bibitem{yang2023novel}
Yang Q, Wang X, Zhang X, Zheng J, Ke~Y, Wang L, Guo H.
\newblock A novel deep learning method for automatic recognition of coseismic landslides.
\newblock Remote Sensing, 2023, 15(4): 977

\bibitem{liu2024review}
Liu Z, Wan G, Prakash B~A, Lau M~S, Jin W.
\newblock A review of graph neural networks in epidemic modeling.
\newblock In: Proceedings of the 30th ACM SIGKDD Conference on Knowledge Discovery and Data Mining.
\newblock 2024,  6577--6587

\bibitem{gu2023mamba}
Gu~A, Dao T.
\newblock Mamba: Linear-time sequence modeling with selective state spaces.
\newblock arXiv preprint arXiv:2312.00752, 2023

\bibitem{behrouz2024graph}
Behrouz A, Hashemi F.
\newblock Graph mamba: Towards learning on graphs with state space models.
\newblock In: Proceedings of the 30th ACM SIGKDD Conference on Knowledge Discovery and Data Mining.
\newblock 2024,  119--130

\bibitem{liu2024kan}
Liu Z, Wang Y, Vaidya S, Ruehle F, Halverson J, Solja{\v{c}}i{\'c} M, Hou T~Y, Tegmark M.
\newblock Kan: Kolmogorov-arnold networks.
\newblock arXiv preprint arXiv:2404.19756, 2024

\bibitem{yin2024continuous}
Yin N, Wan M, Shen L, Patel H~L, Li~B, Gu~B, Xiong H.
\newblock Continuous spiking graph neural networks.
\newblock arXiv preprint arXiv:2404.01897, 2024

\bibitem{yu2023empower}
Yu~J, Ren Y, Gong C, Tan J, Li~X, Zhang X.
\newblock Empower text-attributed graphs learning with large language models (llms).
\newblock arXiv preprint arXiv:2310.09872, 2023

\bibitem{he2023harnessing}
He~X, Bresson X, Laurent T, Perold A, LeCun Y, Hooi B.
\newblock Harnessing explanations: Llm-to-lm interpreter for enhanced text-attributed graph representation learning.
\newblock arXiv preprint arXiv:2305.19523, 2023

\bibitem{wang2024bridging}
Wang Y, Zhu Y, Zhang W, Zhuang Y, Li~Y, Tang S.
\newblock Bridging local details and global context in text-attributed graphs.
\newblock arXiv preprint arXiv:2406.12608, 2024

\bibitem{huang2024gnns}
Huang X, Han K, Yang Y, Bao D, Tao Q, Chai Z, Zhu Q.
\newblock Gnns as adapters for llms on text-attributed graphs.
\newblock In: The Web Conference 2024.
\newblock 2024

\bibitem{chen2024exploring}
Chen Z, Mao H, Li~H, Jin W, Wen H, Wei X, Wang S, Yin D, Fan W, Liu H, others .
\newblock Exploring the potential of large language models (llms) in learning on graphs.
\newblock ACM SIGKDD Explorations Newsletter, 2024, 25(2): 42--61

\bibitem{mao2024advancing}
Mao Q, Liu Z, Liu C, Li~Z, Sun J.
\newblock Advancing graph representation learning with large language models: A comprehensive survey of techniques.
\newblock arXiv preprint arXiv:2402.05952, 2024

\bibitem{wu2024exploring}
Wu~Y, Li~S, Fang Y, Shi C.
\newblock Exploring the potential of large language models for heterophilic graphs.
\newblock arXiv preprint arXiv:2408.14134, 2024

\bibitem{ren2024survey}
Ren X, Tang J, Yin D, Chawla N, Huang C.
\newblock A survey of large language models for graphs.
\newblock In: Proceedings of the 30th ACM SIGKDD Conference on Knowledge Discovery and Data Mining.
\newblock 2024,  6616--6626

\bibitem{liu2023towards}
Liu J, Yang C, Lu~Z, Chen J, Li~Y, Zhang M, Bai T, Fang Y, Sun L, Yu~P~S, others .
\newblock Towards graph foundation models: A survey and beyond.
\newblock arXiv preprint arXiv:2310.11829, 2023

\bibitem{fan2024graph}
Fan W, Wang S, Huang J, Chen Z, Song Y, Tang W, Mao H, Liu H, Liu X, Yin D, others .
\newblock Graph machine learning in the era of large language models (llms).
\newblock arXiv preprint arXiv:2404.14928, 2024

\bibitem{galkin2023towards}
Galkin M, Yuan X, Mostafa H, Tang J, Zhu Z.
\newblock Towards foundation models for knowledge graph reasoning.
\newblock arXiv preprint arXiv:2310.04562, 2023

\bibitem{maoposition}
Mao H, Chen Z, Tang W, Zhao J, Ma~Y, Zhao T, Shah N, Galkin M, Tang J.
\newblock Position: Graph foundation models are already here.
\newblock In: Forty-first International Conference on Machine Learning.
\newblock 2024

\bibitem{tang2024graphgpt}
Tang J, Yang Y, Wei W, Shi L, Su~L, Cheng S, Yin D, Huang C.
\newblock Graphgpt: Graph instruction tuning for large language models.
\newblock In: Proceedings of the 47th International ACM SIGIR Conference on Research and Development in Information Retrieval.
\newblock 2024,  491--500

\bibitem{xia2024opengraph}
Xia L, Kao B, Huang C.
\newblock Opengraph: Towards open graph foundation models.
\newblock arXiv preprint arXiv:2403.01121, 2024

\bibitem{xia2024anygraph}
Xia L, Huang C.
\newblock Anygraph: Graph foundation model in the wild.
\newblock arXiv preprint arXiv:2408.10700, 2024

\bibitem{alabdulmohsin2022revisiting}
Alabdulmohsin I~M, Neyshabur B, Zhai X.
\newblock Revisiting neural scaling laws in language and vision.
\newblock Advances in Neural Information Processing Systems, 2022, 35: 22300--22312

\end{thebibliography}
\begin{biography}{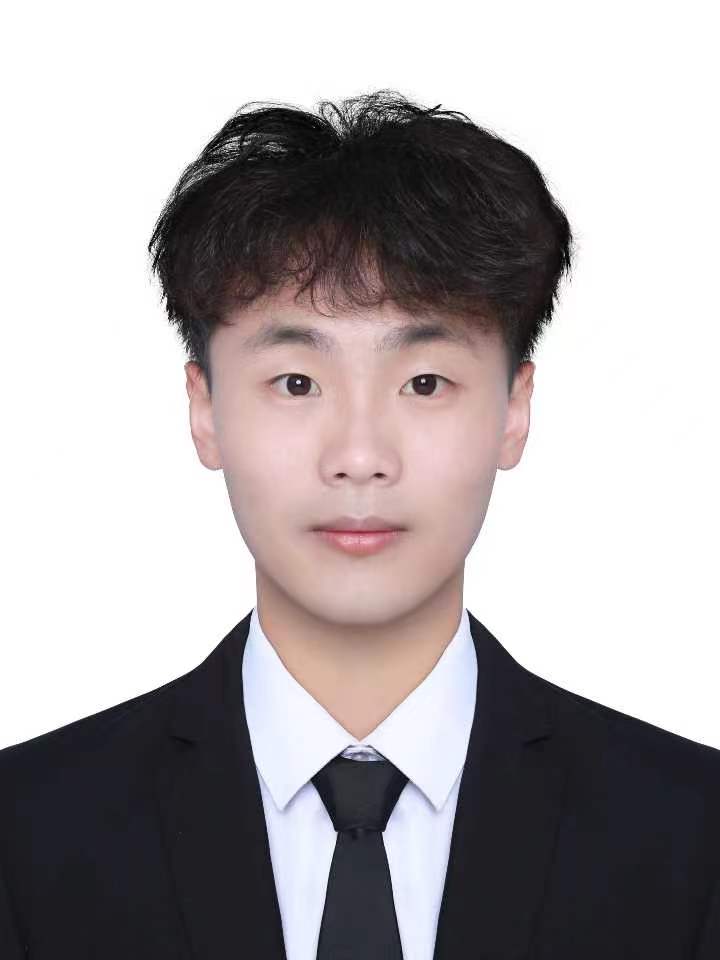}
{Chenghua~Gong} is currently working toward his M.S. degree at East China Normal University, Shanghai, China. His research interests include graph neural networks and graph prompt learning.
\end{biography}
\vfill
\begin{biography}{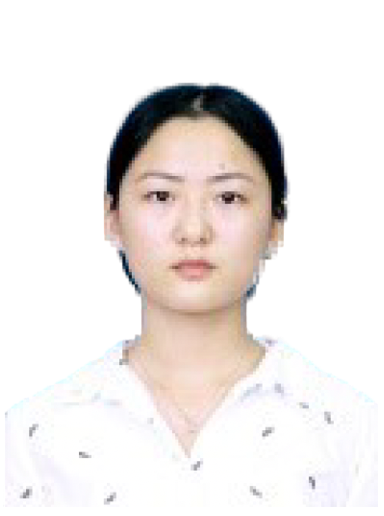}
{Yao~Cheng} is currently working toward her Ph.D. degree at East China Normal University, Shanghai, China. Her general research interests include graph neural networks, robust graph learning, and graph foundation models.
\end{biography}
\vfill
\begin{biography}{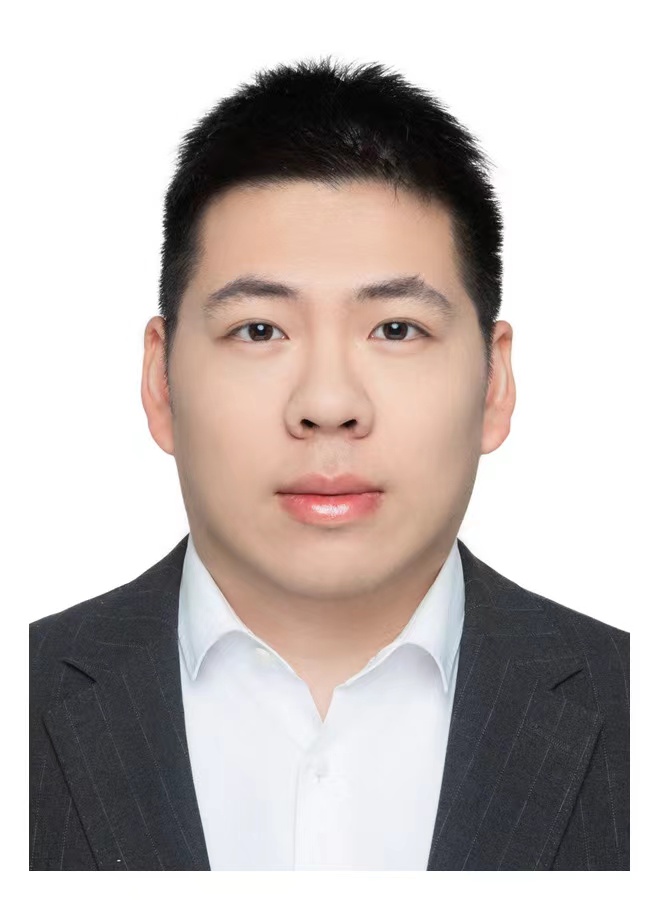}
{Jianxiang~Yu} is currently working toward his Ph.D. degree at East China Normal University, Shanghai, China. His research interests include graph neural networks and natural language processing.
\end{biography}
\vfill
\begin{biography}{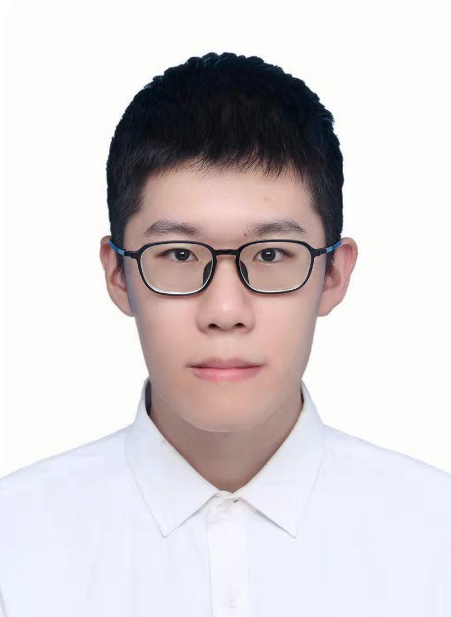}
{Can~Xu} is currently pursuing his Ph.D. degree at East China Normal University. His research interests include graph neural networks and generative models.
\end{biography}
\vfill
\begin{biography}{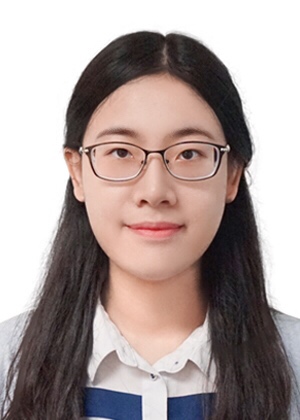}
{Shan~Caihua} is a Senior Researcher at Microsoft Research Asia (Shanghai). She received her Ph.D. degree in Computer Science from the University of Hong Kong in 2020. Her primary research interests include graph neural network and molecular generation. 
She has published over 30 papers at top-tier conferences such as ICML, NeurIPS, ICLR, KDD, WWW, SIGMOD, VLDB, and ICDE.
\end{biography}
\vfill
\begin{biography}{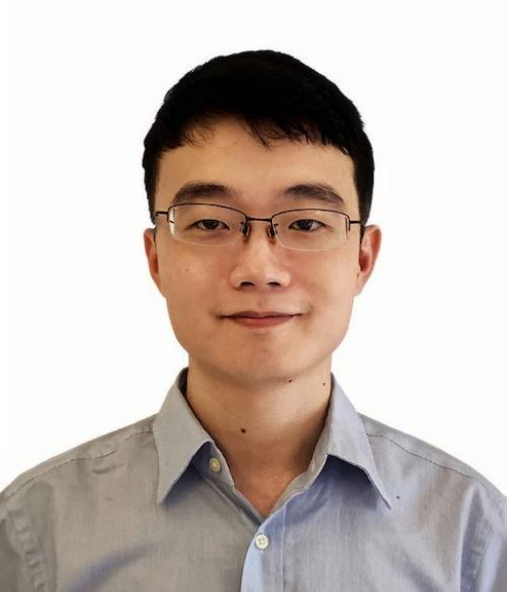}
{Siqiang~Luo} is a Nanyang Assistant Professor at the College of Computing and Data Science, NTU, Singapore. He received his Ph.D. degree in computer science from HKU in 2019. His research interest lies in graph data management, such as graph neural networks, PageRanks and community search. 
His research has been
regularly published in top venues such as SIGMOD, PVLDB and
ICDE.
\end{biography}
\vfill
\begin{biography}{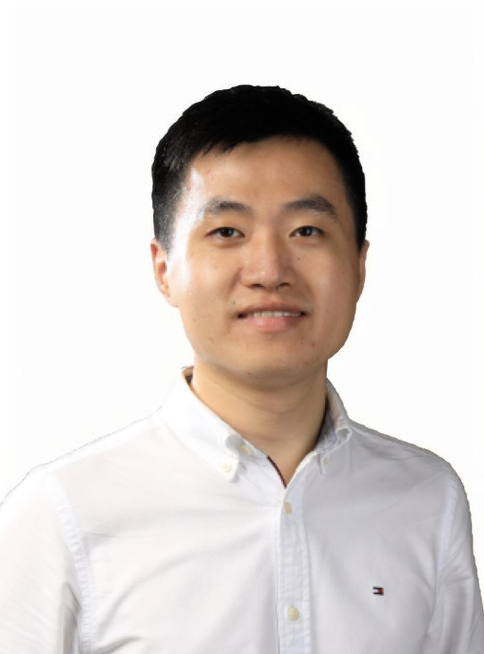}
{Xiang~Li} received his Ph.D. degree from the University of Hong Kong in 2018. From 2018 to 2020, he worked as a research scientist in the Data Science Lab at JD.com and a research associate at The University of Hong Kong, respectively. 
He is currently a research professor at the School of Data Science and Engineering in East China Normal University. 
His general research interests include Data Mining and Machine Learning.
\end{biography}
\end{document}